\documentclass[12pt]{article}
\usepackage[utf8]{inputenc}
\usepackage[english]{babel}
\usepackage[centertags]{amsmath}
\usepackage{amssymb}
\usepackage{bm}
\usepackage{amsbsy}
\usepackage{comment}
\usepackage{graphicx}
\usepackage{appendix}
\usepackage{mathrsfs}
\usepackage{epsfig}
\usepackage{color}
\usepackage{euscript}
\usepackage{ulem}
\usepackage{url}
\usepackage{multirow}
\usepackage{cite}
\usepackage{textcomp}
\usepackage{bigints}
\usepackage{lipsum}
\usepackage{setspace}
\usepackage{cellspace}
\usepackage{booktabs}
\usepackage{mathtools}
\usepackage{float}
\DeclareMathOperator{\const}{const}

\definecolor{darkblue}{rgb}{0,0,1}
\definecolor{darkraspberry}{rgb}{0.53,0.15,0.34}
\definecolor{dgreen}{rgb}{0,0.6,0}
\definecolor{customgreen}{rgb}{0, 0.5, 0}
\definecolor{customgray}{rgb}{0.5, 0.5, 0.5}
\definecolor{dgreen}{rgb}{0,0.6,0}
\definecolor{darkblue}{rgb}{0., 0, 1}
\definecolor{purple}{rgb}{0.65,0.,0.78}

\usepackage{jheppubm}
\newcommand{\nn}{\nonumber}
\newcommand{\be}{\begin{equation}}
\newcommand{\ee}{\end{equation}}
\newcommand{\bea}{\begin{eqnarray}}
\newcommand{\eea}{\end{eqnarray}}

\newcommand{\fb}{\mathfrak{b}}

\newcommand{\ff}{\mathfrak{f}}
\newcommand{\fg}{\mathfrak{g}}

\newcommand{\fq}{\mathfrak{q}}

\newcommand{\fs}{\mathfrak{s}}

\newcommand{\fB}{\mathfrak{B}}

\newcommand{\fF}{\mathfrak{F}}
\newcommand{\fG}{\mathfrak{G}}
\newcommand{\fJ}{\mathfrak{J}}
\newcommand{\fK}{\mathfrak{K}}
\newcommand{\fL}{\mathfrak{L}}

\newcommand{\bldf}{\mathbf{f}}

\newcommand{\Gg}[1]{\ensuremath{\displaystyle \frac{G_{#1 #1}}{g_{#1 #1}}}}

\newcommand{\cL}{\cal{L}} 
\newcommand{\cN}{\mathcal{N}}
\newcommand{\cK}{\mathcal{K}}

\newcommand{\cM}{\mathcal{M}}

\makeatletter
    \newcommand{\hlineb}[1]{\hline
      \noalign{\kern -#1}
      \noalign{\hrule height #1}
      \noalign{\kern -#1}}
\makeatother
\makeatletter
    \newcommand{\clineb}[2]{%
      \noalign{\kern -#1}%
      \multicolumn{#2}{c}{\rule[-#1]{\dimexpr\linewidth/#2}{#1}}%
      \noalign{\kern -#1}%
    }
\makeatother
\usepackage{tabularray}
\usepackage{calc} 
\usepackage{accents}

\UseTblrLibrary{booktabs}
\usepackage{tikz}
\usetikzlibrary{calc}

\allowdisplaybreaks

\numberwithin{equation}{section}
\title{Lifshitz-like Magnetic Black Branes: \\Third Law of Thermodynamics \\and Null Energy Condition
}

\author{Irina Ya. Aref'eva$^{a,b}$, Kristina Rannu$^{c}$ and Viktor Zlobin$^{d}$}

\affiliation{$^a$Steklov Mathematical Institute, Russian Academy of Sciences, Gubkina str. 8, 119333,  Moscow, Russia \\
 $^b$Moscow State University, Physical Department, Moscow, Russia\\
$^c$Peoples Friendship University of Russia, Miklukho-Maklaya str. 6, 117198, Moscow, Russia \\
$^d$Institute for Theorerical and Mathematical Physics, Moscow State University, Moscow, Russia
}

\emailAdd{arefeva@mi-ras.ru}
\emailAdd{rannu-ka@rudn.ru}
\emailAdd{zlobin.vo@phystech.edu}

\abstract{We develop a procedure to solve Einstein-dilaton-Maxwell models in quadratures using the potential reconstruction approach. We then apply this procedure to three distinct models, examining both the null energy condition (NEC) and the validity of the third law of thermodynamics in each case. The explicit knowledge of the blackening function --- as opposed to relying solely on numerical data --- allows us to discuss the validity of the third law in detail. The three models considered are:  (I) a 5D model with two Maxwell fields, featuring anisotropy specified by a Gaussian function and a Lifshitz function; (II) the same 5D model as in (I), but with anisotropy parametrized by two Lifshitz parameters; and (III)  a 6D model with one 2-form and one 3-form field, with the metric parametrized by two Lifshitz parameters. We show that for models I and II the parameter regions, where both the NEC and the third law are satisfied, exhibit no correlation between the two conditions. In contrast, for model III the validity of the NEC implies the validity of the third law.
}


\begin{document}

\maketitle

\section{Introduction}
The study of black branes in multidimensional spaces is a key ingredient in constructing realistic holographic quantum chromodynamics (HQCD), which generalizes the simplest case of the AdS/CFT correspondence. The HQCD framework has made it possible to examine the thermodynamics of quark-gluon plasma (QGP) models \cite{Arefeva:2014ufn, Arefeva:2021kku}.
\\

In attempting to construct five-dimensional gravity duals that reproduce the experimentally observed properties of the QGP, one widely used class of models is based on the Einstein-dilaton-Maxwell action. Anisotropic models are of particular interest, since QGP is a highly anisotropic medium in the early stages after its formation in heavy-ion collisions (HIC), occurring within $10^{-24}$ s \cite{Arefeva:2014ufn}. In such models, the AdS metric is modified by a warp factor and spatial anisotropy \cite{DHoker:2009ixq,Giataganas:2013lga,Arefeva:2014vjl,Taylor:2015glc,Rougemont:2015oea,Arefeva:2016phb,Giataganas:2017koz,Arefeva:2018hyo,Bohra:2019ebj, He:2020fdi,Arefeva:2022bhx,Arefeva:2023jjh,Shukla:2023pbp,Arefeva:2024xmg,Rannu:2024vrq,Arefeva:2025okg,Giataganas:2025ing,Arefeva:2026yms}. The early stages of HIC, as experiments suggest, show signs of a very strong magnetic field \cite{Skokov:2009qp}, which is incorporated via Maxwell fields in the gravity dual \cite{Rougemont:2015oea}. Another Maxwell field gives rise to the quark chemical potential \cite{Nakamura:2007nk, GaugeGravity}.
\\

Holographic anisotropic setups are rich gravitational and thermodynamic systems in their own right and can be studied independently of their direct application to QGP phenomenology. Analyzing these anisotropic models is far from a purely abstract exercise; considered in isolation, it is a crucial step toward a better understanding of their fundamental features and capabilities. Thus, new aspects of their wider and more complete application to QGP studies can be investigated.
\\ 

Motivated by these circumstances, this paper examines the third law of thermodynamics and the null energy condition (NEC) for several asymptotically AdS black brane models. In this context the third law of thermodynamics has been discussed in \cite{DHoker:2009ixq,Arefeva:2023kpu,Arefeva:2024wqb}, NEC has been discussed in \cite{Giataganas:2025ing}. By enforcing these constraints in the bulk, we can hope that the boundary models --- which, according to the holographic principle, are supposed to describe realistic systems --- exhibit physically meaningful behavior.
\\

A further more technical question is whether Einstein-dilaton-Maxwell models considered in the potential reconstruction approach \cite{DeWolfe:1999cp}
 can be solved by quadratures. In the existing literature, the AdS metric is usually modified by Lifshitz \cite{Pang:2009ad, Goldstein:2009cv, Azeyanagi:2009pr,Taylor:2015glc} and Gauss \cite{Bohra:2019ebj,
He:2020fdi,Arefeva:2022bhx} anisotropic factors, while the warp factor is chosen as an exponential of a polynomial; analytic expressions for the solutions are provided, though they can be considerably complicated \cite{Arefeva:2022bhx, Arefeva:2023jjh,Giataganas:2025ing}. As always, the role of explicit solutions should not be overestimated, but their advantage over solutions obtainable only numerically is that they allow a more complete and confident description of the properties of the models under consideration.
\\

In this paper, we systematically generalize a procedure --- found previously for a certain class of models \cite{Arefeva:2018hyo, Arefeva:2022bhx} --- to solve for the blackening function in quadratures, extending it to a broad class of Einstein-dilaton-Maxwell models. It turns out, that for the class of models with a diagonal metric in the form
\begin{gather}
  ds^2 = \fB^2(z) \left( 
    - \, g(z)dt^2 + \sum^{D-2}_{i=1} \fg_{i}(z) dx_i^2 
    + \cfrac{dz^2}{g(z)} \right), \label{intro-metric}
\end{gather}
there is a nice formula, that represents the blackening function $g$ equation in the form
\be
  \left(
\fL\left(\fK\,g\right)^{\prime} \right)
  ^{\prime} =  \fJ,\ee
where $\fL$ and $\fK$ depend on all the metric coefficients except for the blackening function, and $\fJ$ depends also on the interaction of the Maxwell fields with dilaton. If functions $\fB$ and $\fg_i$ along with some coupling functions $f_i$ are specified, this equation lets us straightforwardly solve for the blackening function $g$. We also extend this procedure to models with 3-form magnetic fields introduced into the action. We illustrate this procedure with two models: first, a $D=6$ model that includes both a 2-form and a 3-form magnetic field; and second, the fully anisotropic Einstein-dilaton-four-Maxwell model considered in \cite{Arefeva:2024mtl}, for which we write down the blackening function solution in quadratures.
\\

This paper is organized as follows. In Section~\ref{sec:actionmetric}, we present the setup of a $D$-dimensional Einstein-dilaton-Maxwell model with $3$-form fields additionally introduced into the action. In Section~\ref{sec:quadratures}, the quadrature derivation procedure is laid out and applied to two models, including the fully anisotropic Einstein-dilaton-four-Maxwell model \cite{Arefeva:2024mtl}. In Sections~\ref{sec:model1}, \ref{sec:model2}, and \ref{sec:model3}, we consider three different models and discuss in detail the NEC and the third law of thermodynamics for each one. In Section~\ref{sec:discussion}, we briefly outline the main results. The paper is supplemented by Appendices~\ref{app::6D-EOM}, \ref{app::A}, \ref{app::B}, and \ref{app::C}, where we provide necessary comments on the process of solving for the blackening function in quadratures and write down EOM for 6-dimensional models.

\section{Action and Metric}\label{sec:actionmetric}

Section~\ref{sec:quadratures} treats the broad class of Einstein-dilaton-Maxwell models extended by 3-form fields, deriving the general quadrature procedure. In Sections~\ref{sec:model1}--\ref{sec:model2}, we examine particular models from this class. The action for the general $D$-dimensional Einstein-dilaton-Maxwell theory in the Einstein frame, supplemented by the 3-form fields, takes the form:
\begin{align}
  S &= \int d^D x\, {\cL} = \int d^D x \left({\cL}_g - {\cL}_m \right) \label{action}\\
  & = \int d^D x \, \sqrt{- g_D} \left[ R 
  - \sum_{\cM} \cfrac{f_{\cM}(\phi)}{2 \cdot 2!} \, F^2_{\cM}
  - \sum_{\cN} \cfrac{h_{\cN}(\phi) }{2 \cdot 3!} \, H^2_{\cN}
  - \cfrac{1}{2} \, \partial_{\mu} \phi \, \partial^{\,\mu} \phi 
  - V(\phi) \right], \nn
\end{align}
where the symbolic expressions $F^2_{\cM}$ and $H^2_{\cN}$ denote the 2- and 3-form field contributions to the action, respectively: one has $F^2_{\cM} \equiv F^{(\cM)}_{\mu\nu}F^{(\cM)\mu\nu}$ and $H^2_{\cN} \equiv H^{(\cN)}_{\mu\nu\rho}H^{(\cN)\mu\nu\rho}$. The $p$-form fields are coupled to the dilaton $\phi$ by coupling functions $f_{\cM}(\phi)$ and $h_{\cN}(\phi)$; $V(\phi)$ is the potential of the dilaton; $g_D$ is the determinant of the metric and $R$ is its Ricci scalar. 

The metric is always assumed to be diagonal and dependent on one of the coordinates --- namely, the holographic coordinate --- only. In the holographic approach to the quark-gluon plasma study it is customary to represent the general metric (\ref{intro-metric}) in the following way:
\begin{gather}
  ds^2 = \cfrac{L^2\,\fb(z)}{z^2} \left( 
    - \, g(z)dt^2 + \sum^{D-2}_{i=1} \fg_{i}(z) dx_i^2 
    + \cfrac{dz^2}{g(z)} \right), \label{metric}
\end{gather}
where $\fb(z)$ and $g(z)$ are the warp function and the blackening function, correspondingly, while $\fg_{i}(z)$ are anisotropic factors. Setting the warp factor and the anisotropic factors to be trivial $\fb(z)=\fg_i(z)\equiv1$, one obtains the AdS metric, thus, $L$ is called the radius of the AdS space.

For the material fields, the following ansatzes are employed:
\begin{itemize}
  \item $\phi = \phi(z)$;
  \item magnetic ansatz: 
  $F = q \, dx^{\mu} \wedge dx^{\nu}$ for 2-form fields, 
  $H = Q \, dx^{\mu} \wedge dx^{\nu} \wedge dx^{\rho}$ for 3-form fields;
  \item electric ansatz: $A = A_t(z) \, dt$ for a $2$-form field. 
\end{itemize}

The variational principle applied to the action \eqref{action} leads to the following equations of motion (EOM):
\begin{gather}
  G_{\mu\nu} = T_{\mu\nu}, \label{einstein-eq} \\
  \partial_{\mu} \Big( 
    \sqrt{-g_D} \, f_{\cM} \, F_{\cM}^{\mu\nu} \Big) = 0, \label{maxwell-1-5d} \\
  \partial_{\mu} \Big( 
    \sqrt{-g_D} \, h_{\cN} \, H_{\cN}^{\mu\nu\rho} \Big) = 0, \\
  D_\mu D^\mu \phi 
  - \cfrac{\partial \, V(\phi)}{\partial \phi} 
  - \sum_{\cM} \cfrac{1}{4} \, \cfrac{\partial \, f_{\cM}(\phi)}{\partial \phi} \, F_{\cM}^2 
  - \sum_{\cN} \cfrac{1}{12} \, \cfrac{\partial \, h_{\cN}(\phi)}{\partial \phi} \, H_{\cN}^2  = 0, \label{dilaton-5d}
\end{gather}
where the stress-energy tensor is defined as
\begin{gather}
  T_{\mu\nu} = \cfrac{1}{\sqrt{-g_D}} \, \cfrac{\delta (\sqrt{-g_D} \, {\cL}_m)}{\delta g^{\mu\nu}}. \label{Tmunu-def}
\end{gather}

\subsection{On Notation for 2- and 3-Form Fields in D = 6}\label{subsec::6D-notation}

In Section~\ref{sec:model3} we consider a $D = 6$ model with magnetic 2- and 3-form fields, which differs from the widely studied $D = 5$ models with 2-form fields \cite{Rougemont:2015oea,
Arefeva:2018hyo,Bohra:2019ebj, He:2020fdi,Arefeva:2022bhx, Arefeva:2023jjh,Shukla:2023pbp, Arefeva:2024xmg,  Rannu:2024vrq,Arefeva:2025okg,Giataganas:2025ing,Arefeva:2026yms}. Thus, this motivates us to set efficient and intuitive rules for the notation of the form fields.
\begin{itemize}
  \item For 2-form fields, that do not propagate along the $x^4$ coordinate, the notation is inherited from some existing works on Einstein-dilaton-Maxwell models \cite{Arefeva:2024mtl}:
  \begin{gather}
    F_1 = q_1 \, dx^2 \wedge dx^3, \qquad 
    F_2 = q_2 \, dx^1 \wedge dx^3, \qquad 
    F_3 = q_3 \, dx^1 \wedge dx^2,
  \end{gather}
  with the corresponding coupling functions denoted as $f_1, f_2$, and $f_3$.
  \item For 2-form fields, that do propagate along the $x^4$ coordinate, we introduce
  \begin{gather}
    \fF_1 = \fq_1 \, dx^1 \wedge dx^4, \qquad 
    \fF_2 = \fq_2 \, dx^2 \wedge dx^4, \qquad 
    \fF_3 = \fq_3 \, dx^3 \wedge dx^4,
  \end{gather}
  with the corresponding coupling functions denoted as $\ff_1$, $\ff_2$, and $\ff_3$.
  \item For 3-form fields, we introduce a rule analogous to that of the four-Maxwell~\cite{Arefeva:2024mtl} case: the index is inherited from the $x^i$ coordinate the field does not propagate along ($123 \leftrightarrow 4$, $134 \leftrightarrow 2$, and so on). We therefore introduce
  \begin{gather}
    H_1 = Q_1 \, dx^2 \wedge dx^3 \wedge dx^4, \quad
    H_2 = Q_2 \, dx^1 \wedge dx^3 \wedge dx^4, \\
    H_3 = Q_3 \, dx^1 \wedge dx^2 \wedge dx^4, \quad
    H_4 = Q_4 \, dx^1 \wedge dx^2 \wedge dx^3,
  \end{gather}
  with the corresponding coupling functions denoted as $h_1$, $h_2$, $h_3$ and $h_4$.
\end{itemize}

\section{Solution in Quadratures}\label{sec:quadratures}

In a fully anisotropic $D=5$ model considered in \cite{Arefeva:2024mtl}, four Maxwell fields (one electric and three magnetic) are present in the action. For this model, one of the approaches outlined in the cited paper is as follows: as a starting point, specify metric functions $\fb(z)$, $\fg_1(z)$, $\fg_2(z)$, and $\fg_3(z)$ along with two coupling functions $f_0\big(\phi(z)\big)$ and $f_1\big(\phi(z)\big)$ as functions of $z$, where the former corresponds to the electric ansatz and the latter --- to one of the magnetic ones; then, solve for the potential $A_t(z)$ and the blackening function $g(z)$, and finally, solve for the other functions --- $\phi(z)$, $V\big(\phi(z)\big)$, $f_2\big(\phi(z)\big)$, $f_3\big(\phi(z)\big)$.

In this section we provide a procedure for obtaining quadratures for models belonging to the class discussed in Section~\ref{sec:actionmetric} and treated in the aforementioned approach. In terms of quadratures, models are distinguished by the dimension of the space and the set of fields, taken in the electric or magnetic ansatz. We claim that all such models, regardless of the dimension and configuration of the $p$-form fields, are solved by quadratures, by which we mean the following: 
\begin{itemize}
  \item the blackening function $g(z)$ is expressed in terms of integrals of the specified metric functions and some of the coupling functions (after having solved for $A_t(z)$ first):
  \be
    g(z) = \text{Function} \Big(
      \fb(z), \, \fg_i(z), \, f_p\big(\phi(z)\big), \, h_r\big(\phi(z)\big) \Big), \quad 
    i = \overline{1, \ D-2},
  \ee
  where $f_p\big(\phi(z)\big)$ and $h_r\big(\phi(z)\big)$ are sets of the specified coupling functions for the 2- and 3-form fields, see Appendix~\ref{app::B} for proof;
  \item all the other functions --- $\phi(z)$, $V\big(\phi(z)\big)$, $f_q\big(\phi(z)\big)$, $h_s\big(\phi(z)\big)$ --- are found as functions of $z$ (where $q \ne p, \ s \ne r$) and expressed in terms of integrals of the specified functions.
\end{itemize}

\subsection{Stress-Energy Tensor Patterns}\label{subsec:Tmunupatterns}

To obtain solutions in quadratures, it proves crucial to recognize a certain pattern in the Einstein equations for models in consideration. For this purpose, it is best to express the metric \eqref{metric} directly in its components, absorbing the common factor and the minus sign before the first term:
\be
  ds^2 = g_{00}(z) dt^2 
  + \sum^{D-2}_{i=1} g_{ii}(z) dx^2_i 
  + g_{zz} dz^2. \label{SET-Metric}
\ee

To arrive at formulas for the Lorentz signature case, we only need to make the following substitution:
\be
  \text{Euclidean} \rightarrow \text{Lorentz} \qquad
  g_{00}(z) \rightarrow - \, g_{tt}(z),
\ee
where $g_{tt}(z)$ is taken to be a positive function.

\subsubsection{Contribution of Different Fields to Stress-Energy Tensor $T_{\mu\nu}$} 

In this representation of the metric \eqref{SET-Metric}, the contributions of the dilaton $\phi$, the 2-form fields $F_{\cM}$ taken in a magnetic and an electric ansatz are correspondingly
\be
  \phi = \phi(z): \quad
  T^{\phi}_{zz} = \frac{\phi'^2}{4} - g_{zz} \, \frac{V(\phi)}{2}, \quad 
  T^{\phi}_{ii} = - \, \frac{g_{ii}}{g_{zz}} \, \frac{\phi'^2}{4} - g_{ii} \, \frac{V(\phi)}{2}, \quad 
  i = \overline{0, \, D-2},
  \label{3.4}
\ee
\begin{equation}
  F_{\cM} = q_{\cM} \, dx^p \wedge dx^k: \quad
  T^{\cM}_{\mu\nu} = \frac{f_{\cM}(\phi) \, q^2_{\cM}}{4 g_{pp} g_{kk}} \
  \text{diag} \Big(
    \begin{tikzpicture}[baseline=(top.base)]
      \node (top)
      {$-\,g_{00}, \, \dots, \, g_{pp}, \, \dots, \, g_{kk}, \, \dots, \, -\,g_{zz}$};
      \coordinate (Ck) at ($(top.south west)!0.36!(top.south east)$);
      \coordinate (Cl) at ($(top.south west)!0.63!(top.south east)$);

      \def\dy{2.0ex}
      \coordinate (Ak) at ($(Ck) + (0,-\dy)$); 
      \coordinate (Al) at ($(Cl) + (0,-\dy)$); 

      \draw (Ak) -- node[below]{\(\text{plus sign before }p, \, k\)} (Al);

      \draw[->] (Ak) -- (Ck);
      \draw[->] (Al) -- (Cl);
    \end{tikzpicture}
  \Big),
  \label{3.5}
\end{equation}
\begin{equation}
  F^{(el.)}_{\cM} = A'_t(z) \, dt \wedge dz:\quad
  T^{\cM\,\,(el.)}_{\mu\nu} = \frac{f^{(el.)}_{\cM}(\phi)\,A'^2_t(z)}{4 g_{00} g_{zz}}
  \
  \text{diag}\Big(
    \begin{tikzpicture}[baseline=(top.base)]
      \node (top)
      {$g_{00}, \, \dots, \, - \, g_{ii}, \, \dots, \, g_{zz}$};
      \coordinate (Ck) at ($(top.south west)!0.10!(top.south east)$);
      \coordinate (Cl) at ($(top.south west)!0.90!(top.south east)$);

      \def\dy{2.0ex}
      \coordinate (Ak) at ($(Ck) + (0,-\dy)$); 
      \coordinate (Al) at ($(Cl) + (0,-\dy)$); 

      \draw (Ak) -- node[below]{\(\text{plus sign before  }t, \, z\)} (Al);

      \draw[->] (Ak) -- (Ck);
      \draw[->] (Al) -- (Cl);
    \end{tikzpicture}
  \Big).
  \label{3.6}
\end{equation}
For the magnetic ansatz of a 3-form field:
\bea
  H_{\cN} &=& Q_{\cN}\, dx^p \wedge dx^k \wedge dx^l: \nn \\
  T^{\cN}_{\mu\nu} &=& \frac{h_{\cN}(\phi)\,Q^2_{\cN}}{4 g_{pp} g_{kk} g_{ll}} \
  \text{diag}\Big(
    \begin{tikzpicture}[baseline=(top.base)]
    \node (top)
    {$- \, g_{00}, \, \dots, \, g_{pp}, \, \dots, \, g_{kk}, \, \dots, \, g_{ll}, \, \dots, \, - \, g_{zz}$};
    \coordinate (C1) at ($(top.south west)!0.30!(top.south east)$);
    \coordinate (C2) at ($(top.south west)!0.50!(top.south east)$);
    \coordinate (C3) at ($(top.south west)!0.70!(top.south east)$);

    \def\dy{2.0ex}
    \coordinate (A1) at ($(C1) + (0,-\dy)$); 
    \coordinate (A2) at ($(C2) + (0,-\dy)$); 
    \coordinate (A3) at ($(C3) + (0,-\dy)$); 

    \draw (A1) -- (A3)
      node[midway,below]{\(\text{plus sign before } p,\,k,\,l\)};

    \draw[->] (A1) -- (C1);
    \draw[->] (A2) -- (C2);
    \draw[->] (A3) -- (C3);
    \end{tikzpicture}
  \Big).
  \label{3.7}
\eea

\paragraph{Revealing a Pattern in $T_{\mu\nu}/g_{\mu\nu}$}
$\,$\\
\par 
The form of the obtained expressions \eqref{3.4}--\eqref{3.7} prompts us to consider the expressions for $T_{\mu\nu}/g_{\mu\nu}$ for each of the field:
\be
  \phi = \phi(z): \quad
  \frac{T^{\phi}_{\mu\nu}}{g_{\mu\nu}} 
  = \frac{V(\phi)}{2} \
  \text{diag}\Big(
    \bm{-}, \, \dots, \, \bm{-}, \, \dots, \, \bm{-}
  \Big)
  + \frac{\phi'^2}{4 g_{zz}} \
  \text{diag}\Big(
    \bm{-}, \ \dots, \ \bm{-}, \ \bm{+}
  \Big), \label{Tg-phi}
\ee
\begin{equation}
  F_{\cM} = q_{\cM}\, dx^p \wedge dx^k:\quad
  \frac{T^{\cM}_{\mu\nu}}{g_{\mu\nu}} 
  = \frac{f_{\cM}(\phi)\,q^2_{\cM}}{4 g_{pp} g_{kk}} \
  \text{diag}\Big(
    \begin{tikzpicture}[baseline=(top.base)]
      \node (top)
      {$\bm{-}, \, \dots, \, \bm{+}, \, \dots, \, \bm{+}, \, \dots, \, \bm{-}$};
      \coordinate (Ck) at ($(top.south west)!0.35!(top.south east)$);
      \coordinate (Cl) at ($(top.south west)!0.65!(top.south east)$);

      \def\dy{2.0ex}
      \coordinate (Ak) at ($(Ck) + (0,-\dy)$); 
      \coordinate (Al) at ($(Cl) + (0,-\dy)$); 

      \draw (Ak) -- node[below]{\(\text{plus sign before  }p, \, k\)} (Al);

      \draw[->] (Ak) -- (Ck);
      \draw[->] (Al) -- (Cl);
    \end{tikzpicture}
  \Big), \label{Tg-F}
\end{equation}
\begin{equation}
  F^{(el.)}_{\cM} = A'_t(z) \, dt \wedge dz:\quad
  \frac{T^{\cM\,\,(el.)}_{\mu\nu}}{g_{\mu\nu}} 
  = \frac{f^{(el.)}_{\cM}(\phi)\,A'^2_t(z)}{4 g_{00} g_{zz}} \
  \text{diag}\Big(
    \begin{tikzpicture}[baseline=(top.base)]
      \node (top)
      {$\bm{+}, \, \dots, \, \bm{-}, \, \dots, \, \bm{+}$};
      \coordinate (Ck) at ($(top.south west)!0.10!(top.south east)$);
      \coordinate (Cl) at ($(top.south west)!0.90!(top.south east)$);

      \def\dy{2.0ex}
      \coordinate (Ak) at ($(Ck) + (0,-\dy)$); 
      \coordinate (Al) at ($(Cl) + (0,-\dy)$); 

      \draw (Ak) -- node[below]{\(\text{plus sign before  }t, \, z\)} (Al);

      \draw[->] (Ak) -- (Ck);
      \draw[->] (Al) -- (Cl);
    \end{tikzpicture}
  \Big). \label{Tg-Fel}
\end{equation}
For the magnetic ansatz of a 3-form field, the pattern is easily generalized:
\bea
  H_{\cN} &=& Q_{\cN}\, dx^p \wedge dx^k \wedge dx^l: \quad \nn \\
  \frac{T^{\cN}_{\mu\nu}}{g_{\mu\nu}} 
  &=& \frac{h_{\cN}(\phi)\,Q^2_{\cN}}{4 g_{pp} g_{kk} g_{ll}} \
  \text{diag}\Big(
    \begin{tikzpicture}[baseline=(top.base)]
      \node (top)
      {$\bm{-}, \, \dots, \, \bm{+}, \, \dots, \, \bm{+}, \, \dots, \, \bm{+}, \, \dots, \, \bm{-}$};
      \coordinate (C1) at ($(top.south west)!0.27!(top.south east)$);
      \coordinate (C2) at ($(top.south west)!0.50!(top.south east)$);
      \coordinate (C3) at ($(top.south west)!0.73!(top.south east)$);

      \def\dy{2.0ex}
      \coordinate (A1) at ($(C1) + (0,-\dy)$); 
      \coordinate (A2) at ($(C2) + (0,-\dy)$); 
      \coordinate (A3) at ($(C3) + (0,-\dy)$); 

      \draw (A1) -- (A3)
      node[midway,below]{\(\text{plus sign before } p,\,k,\,l\)};

      \draw[->] (A1) -- (C1);
      \draw[->] (A2) -- (C2);
      \draw[->] (A3) -- (C3);
    \end{tikzpicture}
  \Big). \label{Tg-H}
\eea

The case of a magnetic ansatz for a $p$-form field is easily obtained via generalization. We should also note that the expression $T_{\mu\nu}/g_{\mu\nu}$ is undefined for non-diagonal components, for which both $T_{\mu\nu}$ and $g_{\mu\nu}$ are zero.

\subsection{Sign Table}\label{subsec::signtables}

The highlighted pattern in $T_{\mu\nu}/g_{\mu\nu}$ gives rise to a certain table. Let us pick a particular model to illustrate it:
\bea
    &\text{Model:} \qquad D = 6, \quad \fF_3 = \fq_3 \, dx^3 \wedge dx^4, \quad 
    H_1 = Q_1 \, dx^2 \wedge dx^3 \wedge dx^4,& \label{6D-M2M3-Ansatz1}
    \\
    &g_{\mu\nu} = \text{diag}\Big(-g_{tt}(z), \ g_{x_1 x_1}(z), \ g_{x_2 x_2}(z), \ g_{x_3 x_3}(z), \ g_{x_4 x_4}(z), \ g_{zz}(z)\Big),&
\eea
whose metric we will further specify and consider as Model III in Sec.~\ref{sec:model3}.

\begin{table}[h]
  \centering
  \begin{tabular}{ScSr||Sc|Sc|Sc|Sc||Sl}
    & & $\displaystyle \frac{\phi'^2}{4g_{zz}}$ & $\displaystyle \frac{V(\phi)}{2}$ & $\displaystyle\frac{h_1(\phi) \, Q_1^2}{4 g_{x_2x_2} g_{x_3x_3}g_{x_4x_4}}$ & $\displaystyle\frac{\ff_3(\phi) \, \fq_3^2}{4 g_{x_3x_3}g_{x_4x_4}}$ & \\ \hlineb{1pt}
    $\bm{(z):}$ & $\displaystyle\frac{T_{zz}}{g_{zz}}$ 
    & $\bm{+}$ & $\bm{-}$ & $\bm{-}$ & $\bm{-}$ &
    $\displaystyle=\frac{G_{zz}}{g_{zz}}$ 
    \\ \hlineb{1pt}
    $\bm{(x_4):}$ & $\displaystyle\frac{T_{x_4 x_4}}{g_{x_4 x_4}}$ 
    & $\bm{-}$ & $\bm{-}$ & $\bm{+}$ & $\bm{+}$ &  
    $\displaystyle=\frac{G_{x_4 x_4}}{g_{x_4 x_4}}$ 
    \\ \hlineb{1pt}
    $\bm{(x_3):}$ & $\displaystyle\frac{T_{x_3 x_3}}{g_{x_3 x_3}}$ 
    & $\bm{-}$ & $\bm{-}$ & $\bm{+}$ & $\bm{+}$ &  
    $\displaystyle=\frac{G_{x_3 x_3}}{g_{x_3 x_3}}$ 
    \\ \hlineb{1pt}
    $\bm{(x_2):}$ & $\displaystyle\frac{T_{x_2 x_2}}{g_{x_2 x_2}}$
    & $\bm{-}$ & $\bm{-}$ & $\bm{+}$ & $\bm{-}$ & 
    $\displaystyle=\frac{G_{x_2 x_2}}{g_{x_2 x_2}}$
    \\ \hlineb{1pt}
    $\bm{(x_1):}$  & $\displaystyle\frac{T_{x_1 x_1}}{g_{x_1 x_1}}$
    & $\bm{-}$ & $\bm{-}$ & $\bm{-}$ & $\bm{-}$ & 
    $\displaystyle=\frac{G_{x_1 x_1}}{g_{x_1 x_1}}$
    \\ \hlineb{1pt}
    $\bm{(t):}$  & $\displaystyle-\,\frac{T_{tt}}{g_{tt}}$ 
    & $\bm{-}$ & $\bm{-}$ & $\bm{-}$ & $\bm{-}$ & 
    $\displaystyle=-\,\frac{G_{tt}}{g_{tt}}$ \\ 
  \end{tabular}
  \caption{Sign table for the terms contributing to the components of $T_{\mu\nu}/g_{\mu\nu}$ for the case of $D = 6$ with the fields of 2- and 3-forms taken in the magnetic ansatz \eqref{6D-M2M3-Ansatz1}.}
  \label{Table-6D-MM-34}
\end{table}

For this model, the sign table is the Table~\ref{Table-6D-MM-34}. The rows indicate which sign each header term must be equipped with to write down a diagonal component of $T_{\mu\nu}/g_{\mu\nu}$. In the rightmost column we include the corresponding ratio of the Einstein equations RHS and the corresponding metric component $g_{\mu\nu}$ to match the leftmost column. For example, in Table~\ref{Table-6D-MM-34}:
\be
  \frac{T_{x_2x_2}}{g_{x_2x_2}} \equiv
  - \, \frac{\phi'^2}{4g_{zz}}
  - \frac{V(\phi)}{2}
  + \frac{h_1(\phi) \, Q_1^2}{4 g_{x_2x_2} g_{x_3x_3}g_{x_4x_4}}
  - \frac{\ff_3(\phi) \, \fq_3^2}{4 g_{x_3x_3}g_{x_4x_4}}
  = \frac{G_{x_2x_2}}{g_{x_2x_2}}.
\ee

\subsection{Quadrature Derivation}\label{sec:quadraturederivation}

The instrumental way to find quadratures for models via the sign tables is based on the observation that we can try to non-degenerately transform the set of the original Einstein equations \eqref{einstein-eq} to a set of equations with isolated matter terms on the RHS, which appear in the headers of the sign table. 

To illustrate it, let us consider the model introduced above (Table~\ref{Table-6D-MM-34}) and transform the Einstein equations by taking linear combinations of the rows:
\bea
  \bm{(x_1)-(t)} \qquad && \Gg{x_1}+\Gg{t}
  = 0, 
  \label{TransfEOM-6D-1}\\
  \bm{(x_3)-(x_2)} \qquad && \Gg{x_3}-\Gg{x_2} 
  = \displaystyle \frac{\ff_3(\phi) \, \fq_3^2}{2\, g_{x_3x_3}g_{x_4x_4}},
  \label{TransfEOM-6D-2}\\
  \bm{(x_2)-(x_1)} \qquad && \Gg{x_2}-\Gg{x_1}
  = \displaystyle\frac{h_1(\phi) \, Q_1^2}{2\, g_{x_2x_2} g_{x_3x_3}g_{x_4x_4}} ,
  \label{TransfEOM-6D-3}\\
  \bm{(z)+(x_4)} \qquad &&\Gg{z}+\Gg{x_4}
  = -\,V(\phi),
  \label{TransfEOM-6D-4}\\
  \bm{(z)-(t)} \qquad && \Gg{z} + \Gg{t}
  = \frac{\phi'^2}{2\,g_{zz}},
  \label{TransfEOM-6D-5}\\
  \bm{(x_4)-(x_3)} \qquad && \Gg{x_4}-\Gg{x_3}
  = 0, 
  \label{TransfEOM-6D-6}
\eea
where we managed to fully isolate each of the header terms of Table~\ref{Table-6D-MM-34}. 

One can make certain remarks from these transformed Einstein equations regarding how to solve the model in the approach outlined in the introduction of the present section. Particularly:
\begin{itemize}
  \item there are two differential equations, which include metric functions only --- \eqref{TransfEOM-6D-1} and \eqref{TransfEOM-6D-6}; these can be thought of as the equations to solve for $g(z)$;
  \item the equation \eqref{TransfEOM-6D-6} is a 1st-order DE with respect to $g(z)$, see Appendix~\ref{app::B}; we are thus limited in our ability to impose the two standard boundary conditions $g(0) = 1, \ g(z_h) = 0$; such a problem is usually neutralized when the metric ansatz is specified: setting $g_{x_4x_4}(z)$ to be a scalar multiple of $g_{x_3x_3}(z)$ makes the equation \eqref{TransfEOM-6D-6} be satisfied identically, see \eqref{A:4};
  \item with the previous point in mind, \eqref{TransfEOM-6D-1} is a 2nd-order DE used to solve for $g(z)$, see Appendix~\ref{app::B}; it turns out that this equation allows for a solution in quadratures, see Sections~\ref{subsec::solex1} and~\ref{subsubsec::4Maxwell};
  \item having obtained the solution for $g(z)$, we substitute it into \eqref{TransfEOM-6D-2}, \eqref{TransfEOM-6D-3}, and \eqref{TransfEOM-6D-4}, and algebraically solve for $\ff_3(\phi)$, $\bldf_1(\phi)$, and $V(\phi)$, respectively, to find these functions as functions of $z$;
  \item the equation \eqref{TransfEOM-6D-5} does not contain the blackening function $g(z)$, see the formula \eqref{Gtt-Gzz}, and can be solved for $\phi(z)$ by a simple quadrature
  \be
    \phi(z) = \pm\bigintsss \sqrt{2 g_{zz} \bigg( \Gg{z} + \Gg{t} \bigg)} \, dz.
  \ee
\end{itemize}

For the models we are restricted to, the LHS of the transformed Einstein equations contain the following combinations of $G_{\mu\mu}/g_{\mu\mu}$ (considering the Lorentz signature):
\bea
  &\displaystyle\bm{[\alpha,\,\beta\ne z,\,t]:}\quad \frac{G_{\alpha\alpha}}{g_{\alpha\alpha}}-\frac{G_{\beta\beta}}{g_{\beta\beta}}, 
  \qquad \quad
  \bm{[\alpha\ne z]:}\quad \frac{G_{\alpha\alpha}}{g_{\alpha\alpha}}+\frac{G_{tt}}{g_{tt}},&
  \nn \\ \nn \\
  &\displaystyle\bm{[\alpha\ne z]:}\quad \frac{G_{\alpha\alpha}}{g_{\alpha\alpha}}+\frac{G_{zz}}{g_{zz}}.&
  \nn
\eea

It is not straightforward to see how one can obtain a solution for $g(z)$ from the equation \eqref{TransfEOM-6D-1}, especially for an arbitrary metric ansatz. We need the following formulas to derive the quadratures (Euclidean signature, see Appendix~\ref{app::A} for proof):
\bea
  \displaystyle 
  \bm{[\alpha,\,\beta\ne z]}\quad
  \frac{G_{\alpha\alpha}}{g_{\alpha\alpha}}-\frac{G_{\beta\beta}}{g_{\beta\beta}}
  &=&
  \frac{1}{2\,\sqrt{g_D}}
  \Bigg(
    \frac{\sqrt{g_D}}{g_{zz}}
    \frac{d}{dz}\ln{\frac{g_{\beta\beta}}{g_{\alpha\alpha}}}
  \Bigg)', \label{Gaa-Gbb}
  \\
  \displaystyle 
  \bm{[\alpha \ne z]}\quad
  \frac{G_{\alpha\alpha}}{g_{\alpha\alpha}}+\frac{G_{zz}}{g_{zz}}
  &=&
  \frac{1}{2\,\sqrt{g_D}} 
  \Bigg(
    \frac{\sqrt{g_D}}{g_{zz}}
    \frac{d}{dz}\ln{\frac{g_D}{g_{\alpha\alpha}\,g_{zz}}}
  \Bigg)', \label{Gaa+Gzz}
  \\
  \displaystyle
  g_{zz}\left(\frac{G_{00}}{g_{00}}-\frac{G_{zz}}{g_{zz}}\right)
  &=&
  \sum^{D-2}_{i=1} \frac{1}{2} \, \frac{g''_{x_ix_i}}{g_{x_ix_i}}
  - \frac{1}{4} \, \frac{g'^2_{x_ix_i}}{g^2_{x_ix_i}}
  - \frac{1}{4} \, \frac{g'_{x_ix_i}}{g_{x_ix_i}} \, \frac{\left(g_{00}g_{zz}\right)'}{g_{00}g_{zz}}. \label{Gtt-Gzz}
\eea

We now show how these formulas are applied in particular cases to illustrate their significance.

\subsubsection{Blackening Function $g(z)$ in Quadratures: Example One}\label{subsec::solex1}

In the model, that we have been considering all along (Table~\ref{Table-6D-MM-34}), the following equation has been shown to arise:
\be
  \bm{(x_1)-(t)} \quad \Gg{x_1} + \Gg{t} = 0.
\ee

Let us show how the formula \eqref{Gaa-Gbb} helps us solve for the blackening function $g(z)$ by quadratures. The procedure to obtain the solution is as follows:
\bea
  \displaystyle \frac{G_{x_1 x_1}}{g_{x_1 x_1}}+\frac{G_{tt}}{g_{tt}}
  &\overset{\eqref{Gaa-Gbb}}{=}&
  \frac{1}{2\,\sqrt{-g_D}} 
  \Bigg(
    \fg_1(z)\prod_{\gamma\ne z, t}\sqrt{g_{\gamma\gamma}(z)} \,
    \frac{d}{dz}\,\frac{g(z)}{\fg_1(z)}
  \Bigg)'=0 
  \nn \\
  &\Longrightarrow& \quad
  \displaystyle \fg_1(z)\prod_{\gamma\ne z, t}\sqrt{g_{\gamma\gamma}(z)}
  \
  \frac{d}{dz} \, \frac{g(z)}{\fg_1(z)} \equiv C_1 
  \nn \\
  &\Longrightarrow& \quad
  \displaystyle g(z) = \fg_1(z) \Bigg[
    C_1 \bigintssss_{z_0}^z \frac{d\xi}{\displaystyle \fg_1(\xi) \prod_{\gamma\ne z, t} \sqrt{g_{\gamma\gamma}(\xi)}} + C_2
  \Bigg],
\eea
where $C_1$, $C_2$ are constants of integration and $z_0 \ge 0$ is an arbitrary boundary of integration.

\subsubsection{Blackening Function $g(z)$ in Quadratures: Example Two, Einstein-dilaton-four-Maxwell Model}\label{subsubsec::4Maxwell}

In the holographic approach to the QGP study a general model was considered in \cite{Arefeva:2024mtl}. The action there contains four Maxwell fields, one electric and three magnetic ones:
\begin{gather}
  S = \displaystyle \int d^5 x \, \sqrt{-\,g_5} 
  \left[ R 
    - \sum^{3}_{\cM=0} \cfrac{1}{4} \, f_{\cM}(\phi) F^2_{\cM}
    - \cfrac{1}{2} \, \partial_{\mu} \phi \, \partial^{\,\mu} \phi 
    - V(\phi) \right], \\
  F_0 = A'_t(z) \, dz \wedge dt, 
  \quad
  \phi = \phi(z), \\
  F_1 = q_1 \, dx^2 \wedge dx^3, \quad 
  F_2 = q_2 \, dx^1 \wedge dx^3, \quad
  F_3 = q_3 \, dx^1 \wedge dx^2.
\end{gather}

The metric is factorized to be of the form
\be
  ds^2 = \fB^2(z) \left[
    - \, g(z) \, dt^2 
    + \fg_{1}(z) \, dx_1^2
    + \fg_{2}(z) \, dx_2^2
    + \fg_{3}(z) \, dx_3^2
    + \cfrac{dz^2}{g(z)} \right].
\ee

In the paper one of the approaches, proposed to solve the model, consists of these steps: at first, specify the metric functions $\fB^2(z)$, $\fg_{1}(z)$, $\fg_{2}(z)$, $\fg_{3}(z)$, the coupling function $f_0\big(\phi(z)\big)$ and one of the coupling functions $f_i\big(\phi(z)\big)$; then, solve for $A_t(z)$, $g(z)$, $\phi(z)$, $V\big(\phi(z)\big)$ and the rest coupling functions $f_{j\ne i}\big(\phi(z)\big)$. In this approach the authors derive an equation for the blackening function by choosing $f_1\big(\phi(z)\big)$ as the one to be specified; we generalize this equation:
\bea
  g'' &+& g' \left(
    \cfrac{3 \fB'}{\fB} + \cfrac{3 \fg_i'}{2 \fg_i} 
    - \sum_{j=1,j\ne i}^3 \cfrac{\fg_j'}{2 \fg_j} \right) \nn \\
  &-& g \left[ 
    - \left( \cfrac{\fg_i''}{\fg_i} - \cfrac{\fg_i'^2}{2 \fg_i} \right)
    + \sum_{j=1,j\ne i}^3 \left(
      \cfrac{\fg_j''}{\fg_j} - \cfrac{\fg_j'^2}{2 \fg_j} \right)
    + \cfrac{3 \fB'}{\fB} \left( 
      - \, \cfrac{\fg_i'}{\fg_i} + \sum_{j=1,j\ne i}^3 \cfrac{\fg_j'}{\fg_j} \right)
    + \prod_{j=1,j\ne i}^3 \cfrac{\fg_j'}{\fg_j}
  \right] \nn\\
  &-& \cfrac{f_0 A_t'^2}{\fB^2} 
  - \cfrac{2 f_i q_i^2}{\fB^2} \prod_{j=1,j\ne i}^3 \cfrac{1}{\fg_j} = 0, \quad
  i = 1, 2, 3, \label{eq:4.06}
\eea
for which it is hard to identify whether $g(z)$ can be solved by quadratures. Let us treat this model with the tools developed above. First, we write down the sign Table~\ref{Table-5D-2M-2} for the model. Second, we transform the Einstein equations with the help of the sign table to arrive at the following form:
\bea
  \bm{(z)-(t)} \qquad & \displaystyle 
  \Gg{z} + \Gg{t}& = \frac{\phi'^2}{2\,g_{zz}},
  \\
  \bm{(z)+(x_1)} \qquad & \displaystyle 
  \Gg{z} + \Gg{x_1} & = -\,V(\phi)-\frac{f_1(\phi)\,q^2_1}{2 g_{x_2x_2} g_{x_3x_3}},
  \nn\\
  \\
  \begin{aligned}
    \bm{\big((x_2)+(x_3)}& \\
    -\,\bm{(x_1)-(t)\big)}&
  \end{aligned}
  \qquad & \displaystyle 
  \Gg{x_2} + \Gg{x_3} - \Gg{x_1} + \Gg{t} & =
  \frac{f_1(\phi)\,q^2_1}{g_{x_2x_2}g_{x_3x_3}}+\frac{f_0(\phi) A'^2_t(z)}{2\,g_{tt}g_{zz}},
  \label{EOM-g-4Maxwell}
  \nn \\
  \\
  \bm{(x_3)-(x_1)} \qquad & \displaystyle \frac{G_{x_3x_3}}{g_{x_3x_3}} - \frac{G_{x_1x_1}}{g_{x_1x_1}} & = \frac{f_1(\phi)\,q^2_1}{2 g_{x_2x_2} g_{x_3x_3}} -\frac{f_3(\phi)\,q^2_3}{2 g_{x_1x_1} g_{x_2x_2}},
  \nn \\ \label{3:35}
  \\
  \bm{(x_2)-(x_1)} \qquad & \displaystyle \frac{G_{x_2x_2}}{g_{x_2x_2}}-\frac{G_{x_1x_1}}{g_{x_1x_1}} & = \frac{f_1(\phi)\,q^2_1}{2 g_{x_2x_2} g_{x_3x_3}} -\frac{f_2(\phi)\,q^2_2}{2 g_{x_1x_1} g_{x_3x_3}},
  \nn \\ \label{3:36}
\eea
where the equation \eqref{EOM-g-4Maxwell} is precisely the equation \eqref{eq:4.06} where $f_1\big(\phi(z)\big)$ is specified. 

\begin{table}[b!]
  \centering
  \begin{tabular}{Sr||Sc|Sc|Sc|Sc|Sc|Sc||Sl}
    & $\cfrac{\phi'^2}{4g_{zz}}$ 
    & \hspace{-3pt}$\cfrac{V(\phi)}{2}$\hspace{-3pt}
    & \hspace{-3pt}$\cfrac{f_1(\phi)\,q^2_1}{4 g_{x_2x_2} g_{x_3x_3}}$\hspace{-3pt}
    & \hspace{-3pt}$\cfrac{f_2(\phi)\,q^2_2}{4 g_{x_1x_1} g_{x_3x_3}}$\hspace{-3pt}
    & \hspace{-3pt}$\cfrac{f_3(\phi)\,q^2_3}{4 g_{x_1x_1} g_{x_2x_2}}$\hspace{-3pt}
    & \hspace{-3pt}$-\,\cfrac{f_0(\phi) A'^2_t(z)}{4 g_{tt} g_{zz}}$\hspace{-3pt} & \\ \hlineb{1pt}
    $\displaystyle\frac{T_{zz}}{g_{zz}}$ 
    & $\bm{+}$ & $\bm{-}$ & $\bm{-}$ & $\bm{-}$ & $\bm{-}$ & $\bm{+}$ 
    & \hspace{-3pt}$\displaystyle=\frac{G_{zz}}{g_{zz}}$ 
    \\ \hlineb{1pt}
    $\displaystyle\frac{T_{x_3 x_3}}{g_{x_3 x_3}}$ 
    & $\bm{-}$ & $\bm{-}$ & $\bm{+}$ & $\bm{+}$ & $\bm{-}$ & $\bm{-}$ 
    & \hspace{-3pt}$\displaystyle=\frac{G_{x_3 x_3}}{g_{x_3 x_3}}$ 
    \\ \hlineb{1pt}
    $\displaystyle\frac{T_{x_2 x_2}}{g_{x_2 x_2}}$
    & $\bm{-}$ & $\bm{-}$ & $\bm{+}$ & $\bm{-}$ & $\bm{+}$ & $\bm{-}$ 
    & \hspace{-3pt}$\displaystyle=\frac{G_{x_2 x_2}}{g_{x_2 x_2}}$
    \\ \hlineb{1pt}
    $\displaystyle\frac{T_{x_1 x_1}}{g_{x_1 x_1}}$
    & $\bm{-}$ & $\bm{-}$ & $\bm{-}$ & $\bm{+}$ & $\bm{+}$ & $\bm{-}$ 
    & \hspace{-3pt}$\displaystyle=\frac{G_{x_1 x_1}}{g_{x_1 x_1}}$
    \\ \hlineb{1pt}
    $\displaystyle-\,\frac{T_{tt}}{g_{tt}}$ 
    & $\bm{-}$ & $\bm{-}$ & $\bm{-}$ & $\bm{-}$ & $\bm{-}$ & $\bm{+}$ 
    & \hspace{-3pt}$\displaystyle=-\,\frac{G_{tt}}{g_{tt}}$ 
    \\ 
  \end{tabular}
  \caption{Sign table for the terms contributing to the components of $T_{\mu\nu}/g_{\mu\nu}$ for the Einstein-dilaton-four-Maxwell model.}
  \label{Table-5D-2M-2}
\end{table}

By subtracting the equations \eqref{3:35} and \eqref{3:36}, multiplied by two, from \eqref{EOM-g-4Maxwell} we get the two other variations of \eqref{eq:4.06}, corresponding to $i=2, \, 3$:
\bea
  \begin{aligned}
    \bm{\big((x_1)+(x_3)}& \\
    -\,\bm{(x_2)-(t)\big)}&
  \end{aligned}
  \qquad & \displaystyle 
  \Gg{x_1} + \Gg{x_3} - \Gg{x_2}  + \Gg{t} & =
  \frac{f_2(\phi)\,q^2_2}{g_{x_1x_1}g_{x_3x_3}}+\frac{f_0(\phi) A'^2_t(z)}{2\,g_{tt}g_{zz}},
  \nn \\ \label{3:37} \\
  \begin{aligned}
    \bm{\big((x_1)+(x_2)}& \\
    -\,\bm{(x_3)-(t)\big)}&
  \end{aligned}
  \qquad & \displaystyle 
  \Gg{x_1} + \Gg{x_2} - \Gg{x_3} + \Gg{t} & =
  \frac{f_3(\phi)\,q^2_3}{g_{x_1x_1}g_{x_2x_2}}+\frac{f_0(\phi) A'^2_t(z)}{2\,g_{tt}g_{zz}}.
  \nn \\ \label{3:38}
\eea

Let us now solve for the blackening function $g(z)$. At first, we need to find the solution for the potential $A_t(z)$.

\paragraph{Solution for the Electric Potential $A_t(z)$}
$\,$\\
\par 
The Einstein equations must be joined by the Maxwell equation \eqref{maxwell-1-5d} equipped with two boundary conditions to solve for $A_t(z)$:
\bea
  &0 = \cfrac{d}{dz} \left(
    \cfrac{\prod_{\gamma\ne z, t}\sqrt{g_{\gamma\gamma}(z)}}{\fB^2(z)}\,f_{0}\big(\phi(z)\big) A'_t(z)
  \right),& \\
  &A_t(0) = \mu, \quad A_t(z_h) = 0,&
\eea
where $z_h$ is the brane's horizon and $\mu$ is the chemical potential.

In the following notation
\be
  \fs(z) = \prod_{\gamma\ne z, t}\sqrt{g_{\gamma\gamma}(z)}, \quad
  I(z) = \int^{z}_{0}
  \frac{\fB^2(\xi)}{\fs(\xi)\,f_0(\phi(\xi))} \, d\xi, \label{3:39}
\ee
the solution for $A_t(z)$ reads
\be
  A_t(z) = \mu \left( 1 - \frac{I(z)}{I(z_h)} \right).
\ee

\paragraph{Solution for the Blackening Function $g(z)$}
$\,$\\
\par We are now ready to find the blackening function $g(z)$. Let us introduce the following additional notation and use the formula \eqref{Gaa-Gbb} to simplify the equations \eqref{EOM-g-4Maxwell}, \eqref{3:37}, and \eqref{3:38}, which we index by $i=1, \, 2, \, 3$:
\be\label{3.41}
  \fG_i = \cfrac{1}{\fg_i} \prod_{j=1,j\ne i}^3 \fg_j
  \quad\Longrightarrow\quad
  \cfrac{1}{2}\,\cfrac{1}{\fB^2\fs}\,
  \Bigg(
    \fs \fG_i \left(\frac{g}{\fG_i}\right)'
  \Bigg)'
  = \cfrac{f_i q^2_i}{\fB^4} \prod_{j=1, \, j\ne i}^3 \cfrac{1}{\fg_j}
  + \cfrac{f_0A'^2_t}{2\fB^4}. 
\ee

In this form, the equations \eqref{EOM-g-4Maxwell}, \eqref{3:37}, and \eqref{3:38} are evidently linear 2nd-order differential equations with respect to $g(z)$. This fact allows us to search for the solution by variation of constants $C_1$ and $C_2$ in the solution for the homogeneous equation

\begin{align}
  \Bigg(
    \fs \fG_i \left(\frac{g}{\fG_i}\right)'
  \Bigg)' = 0 
  \quad\Longrightarrow\quad 
  g(z) = \fG_i(z) \left[C_1 \int^{z}_{z_0}
  \frac{d\xi}{\fs(\xi)\,\fG_i(\xi)} + C_2\right],
  \ \  i=1, \, 2, \, 3,
  \label{3.42}
\end{align}
where $z_0 \ge 0$ is a fixed arbitrary integration boundary and $C_1$, $C_2$ are parameters of integration.

We then finally obtain the solution for the full non-homogeneous equation \eqref{3.41} by performing the variation of constants in the solution \eqref{3.42}, imposing $g(z_h) = 0$ right away:
\begin{align}
  g(z) = \fG_i(z) \, \bigg[
    C &\int^{z}_{z_h}
    \frac{d\xi}{\fs(\xi)\,\fG_i(\xi)}
    + 2\,q^2_i \int^{z}_{z_h}
    \frac{d\xi}{\fs(\xi)\,\fG_i(\xi)}
    \int^{\xi}_{z_h} \frac{\fs(\eta)}{ \fB^2(\eta)} \, f_i(\phi(\eta))\prod_{j=1, \, j\ne i}^3 \cfrac{1}{\fg_j(\eta)}\,d\eta
    \nn\\
    &+ \bigg(\frac{\mu}{I(z_h)}\bigg)^2
    \int^{z}_{z_h}
    \frac{d\xi}{\fs(\xi)\,\fG_i(\xi)}
    \int^{\xi}_{z_h}
    \frac{\fB^2(\eta)}{\fs(\eta)\,f_0(\phi(\eta))} \,d\eta
  \bigg],  \,  i=1, \, 2, \, 3.
  \label{four-Maxwell-quadratures}
\end{align}

In this solution, there is still one integration parameter left, $C$, which is determined by imposing the last boundary condition \[\lim_{z \to 0+}g(z)=1.\]

However, in this general form \eqref{four-Maxwell-quadratures}, it is not possible to write down the explicit solution for the parameter $C$ as it is not always possible to impose the last boundary condition. Hence, the search for $C$ can only be performed when we specify a model, i.e., when all the functions in the integrands are given explicitly. In Sec.~\ref{sec:model1} and Sec.~\ref{sec:model2} we consider particular cases of the model, choosing specific metric ansatzes and retaining only two of the four Maxwell fields.

\section{Model I: Two Maxwell Fields in D = 5 with Lifshitz- and Gauss-Type Anisotropies}\label{sec:model1}

We now move on to the particular models. Let us first consider a $D=5$ space and two Maxwell fields taken in the magnetic ansatz:
\begin{gather}
  S = \int d^5 x \, \sqrt{- g_5} \left[ R 
    - \cfrac{1}{4} \, f_{1}(\phi) F^2_{1} 
    - \cfrac{1}{4} \, f_{3}(\phi) F^2_{3}
    - \cfrac{1}{2} \, \partial_{\mu} \phi \, \partial^{\,\mu} \phi 
    - V(\phi) \right], \\
  F_3 = q_3 \, dx^1 \wedge dx^2,
  \quad
  F_1 = q_1 \, dx^2 \wedge dx^3, \label{magn-ans-5D}
\end{gather}
where $q_1, \, q_3$ are constants.

In the metric the warp factor is set to be trivial $\fb(z)\equiv1$, the same Lifshitz-type anisotropic factors are placed before $dx^2_2$ and $dx^2_3$, while $dx^2_3$ is additionally equipped with a Gauss-type anisotropic factor:
\begin{gather}
  ds^2 = \cfrac{L^2}{z^2} \Bigg( 
    - g(z)dt^2 + dx_1^2 
    + \bigg(\frac{z}{L}\bigg)^{2-\frac{2}{\nu}} dx_2^2
    + e^{c_Bz^2}\bigg(\frac{z}{L}\bigg)^{2-\frac{2}{\nu}} dx_3^2
    + \frac{dz^2}{g(z)}
  \Bigg), \label{ansatz-5D}
\end{gather}
where $\nu$, $c_B$ are anisotropy coefficients.

Maxwell equations for the two fields taken in the magnetic ansatz hold identically, leaving us with five Einstein equations and one equation for the dilaton $\phi$. We can further get rid of the dilaton equation because here it is, in fact, a consequence of the Einstein equations \cite{Arefeva:2024mtl}. Using the model's sign table, we arrive at such EOM:
\begin{gather}
  g''(z) - \left(1 + \cfrac{2}{\nu} + c_B z^2 \right) \cfrac{g'(z)}{z}
  + 2 c_B \left( \cfrac{2}{\nu} - c_B z^2 \right) g(z) = 0, \label{model1-EOM-g} \\
  f_1 \bigl(\phi(z)\bigr) = - \, 2 \left(\cfrac{L}{z}\right)^{4/\nu} \cfrac{e^{c_B z^2}}{L^2 q_1^2} \left( 1 - \cfrac{1}{\nu} \right) \left[
    g'(z) \, z - \left(2 + \cfrac{2}{\nu} - c_B z^2 \right) 
    g(z) \right], \label{model1-EOM-f1} \\
  f_3 \bigl(\phi(z)\bigr) = 2 \left(\cfrac{L}{z}\right)^{2/\nu} \cfrac{c_B}{q_3^2} \left[
    g'(z) \, z - \left(\cfrac{2}{\nu} - c_B z^2 \right) 
    g(z) \right], \label{model1-EOM-f3} \\
  \phi'(z)^2 = \left( 1 - \cfrac{1}{\nu} \right) \cfrac{4}{\nu z^2}
  - 2 c_B \left( 3 - \cfrac{2}{\nu} \right)
  - 2 c_B^2 z^2 , \\
  V \bigl(\phi(z)\bigr) = \cfrac{2 z}{L^2} \left[ 
    \left( 1  + \cfrac{1}{2\nu} - c_B z^2 \right) g'(z)
    - \left( 
      2 + \cfrac{3}{\nu} + \cfrac{1}{\nu^2} - \left( 1 + \cfrac{5}{2\nu} \right) c_B z^2 + c_B^2 z^4 
    \right) 
    \cfrac{g(z)}{z} \right].
\end{gather}

The equation \eqref{model1-EOM-g} is obtained from the model's sign table and is an explicit form of the following equation
\be
  \bm{(x_3)+(x_1)-(x_2)-(t)} \qquad 
  \cfrac{G_{x_3 x_3}}{g_{x_3 x_3}} 
  + \cfrac{G_{x_1 x_1}}{g_{x_1 x_1}} 
  - \cfrac{G_{x_2 x_2}}{g_{x_2 x_2}} 
  + \cfrac{G_{tt}}{g_{tt}} = 0,
\ee
while the boundary conditions are
\be
  g(0) = 1, \quad g(z_h) = 0.
\ee
Using \eqref{Gaa-Gbb} we arrive at the following solution:
\be
  g(z) = C e^{c_B z^2}\int^{z_h}_z e^{-\frac{3c_B}{2} \xi^2} \xi^{1 + \frac{2}{\nu}} \ d\xi,
\ee
where we have already imposed the condition $g(z_h)=0$. The constant $C$ is fixed by the last boundary condition $g(0)=1$. This is possible if and only if the integral is convergent, which leads to the following restriction:
\be
    1 + \frac{2}{\nu} > -\,1 
    \quad \Longleftrightarrow \quad 
    1 + \frac{1}{\nu} > 0
    \quad \Longleftrightarrow \quad 
    \nu \in (-\,\infty; -\,1) \cup (0; +\,\infty).
    \label{model1-nu-constraint}
\ee
After imposing the last boundary condition explicitly, we get
\be\label{4:13}
  \displaystyle 
  g(z) = e^{c_B z^2}
  \frac{\int^{z_h}_z e^{-\frac{3c_B}{2} \xi^2} \xi^{1 + \frac{2}{\nu}} \ d\xi}{\int^{z_h}_0 e^{-\frac{3c_B}{2} \xi^2} \xi^{1 + \frac{2}{\nu}} \ d\xi}, \quad 1 + \frac{1}{\nu} > 0.
\ee
The solution in the case $c_B = 0$ is of a particularly simple form
\be
  c_B = 0: \quad
  g(z) = 1 - \left(\cfrac{z}{z_h}\right)^{2+\frac{2}{\nu}}, \quad 1 + \frac{1}{\nu} > 0.
\ee

\subsection{Third Law of Thermodynamics}\label{ModelI-3rdlaw}
Now that we have solved for $g(z)$, see \eqref{4:13}, we can determine, when the model satisfies the third law of thermodynamics. For the black brane solution the entropy density and the temperature are calculated to be
\bea\label{4:15}
    s(z_h) = \cfrac{1}{4} \left( \cfrac{L}{z_h} \right)^{1+\frac{2}{\nu}} e^{\frac{1}{2}c_Bz_h^2},
    \qquad
    T(z_h) =
    \begin{cases}
    \displaystyle \frac{1}{4\pi}\,\frac{e^{-\frac{c_B}{2} z_h^2} z_h^{1 + \frac{2}{\nu}}}{\int^{z_h}_0 e^{-\frac{3c_B}{2} \xi^2} \xi^{1 + \frac{2}{\nu}} \ d\xi},
    \\
    \displaystyle 
    c_B = 0 \ \to \ \cfrac{1}{2 \pi z_h} \left( 1 + \frac{1}{\nu} \right).
    \end{cases} 
\eea
For the third law of thermodynamics, we shall find when the following holds:
\be\label{4:16}
   s(T) \to 0 \quad \text{as} \quad T \to 0.
\ee

Since the model depends on the anisotropy parameters $(\nu, \, c_B)$, we shall fully describe the parameter regime that supports the third law of thermodynamics. We note that the entropy density also depends on the AdS radius $L$, which is irrelevant to the question of whether or not $s(T)$ decays to zero at small temperatures.

\begin{figure}[b!]
  \centering
  \includegraphics[width=0.32\textwidth]{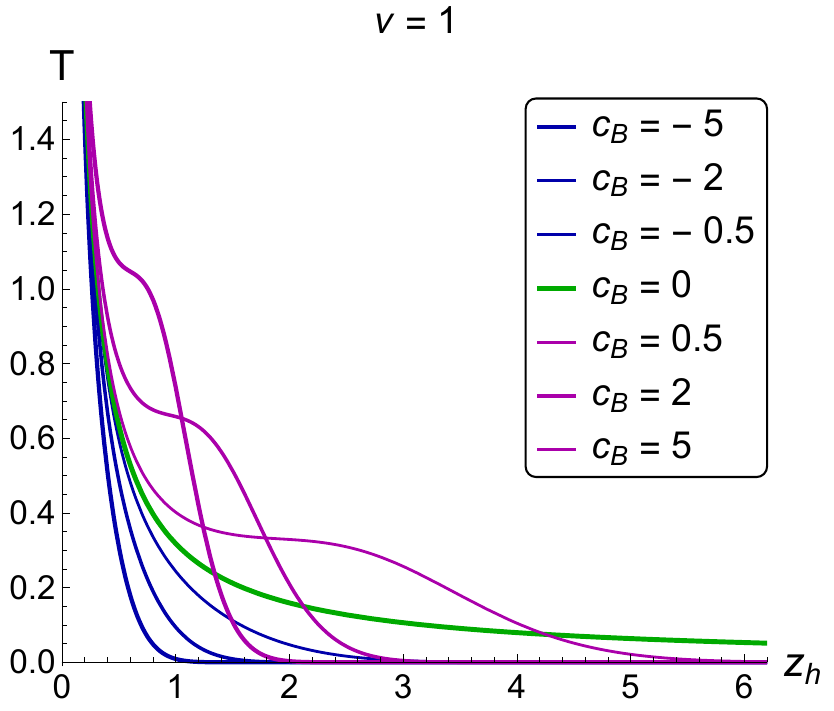} \hfill
  \includegraphics[width=0.32\textwidth]{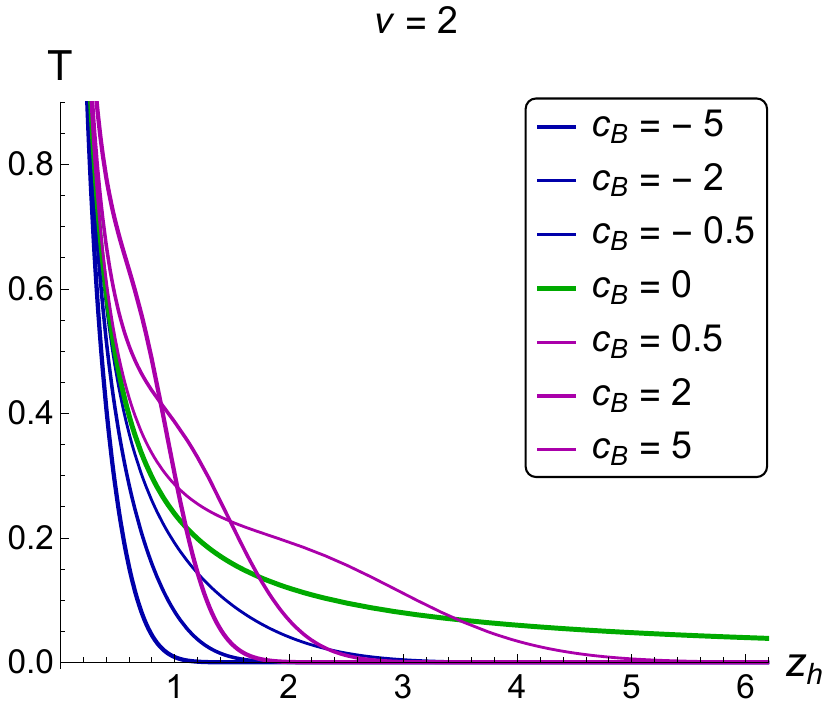} \hfill 
  \includegraphics[width=0.32\textwidth]{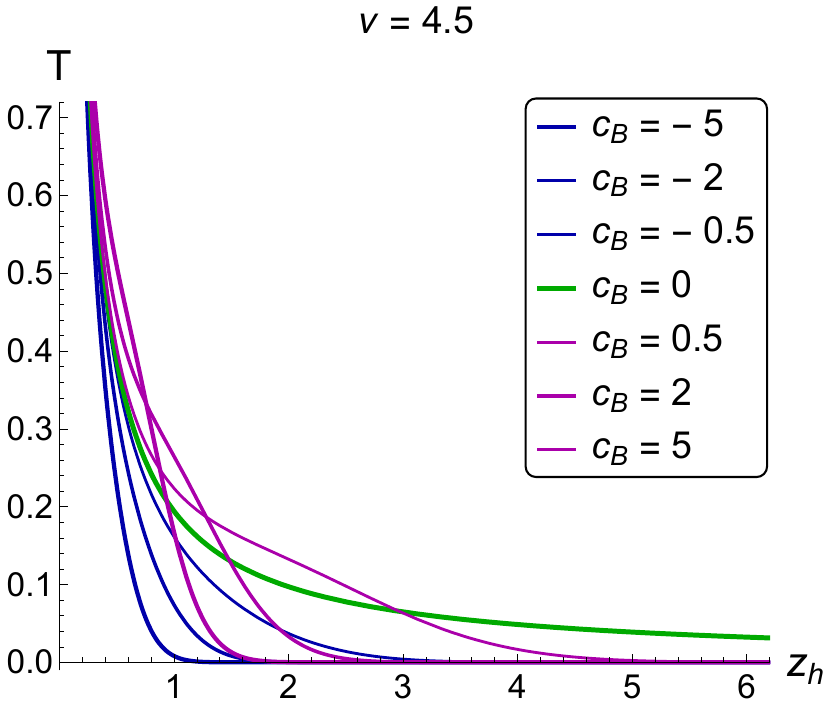} \hfill \\
  A \hspace{140pt} B \hspace{140pt} C
  \caption{Temperature $T(z_h)$ for various values of $c_B$; for $\nu=1$ (A), $\nu=2$ (B), and $\nu=4.5$ (C).\\ \,}
  \label{ModelI-Tzh}
  \centering
  \includegraphics[width=0.32\textwidth]{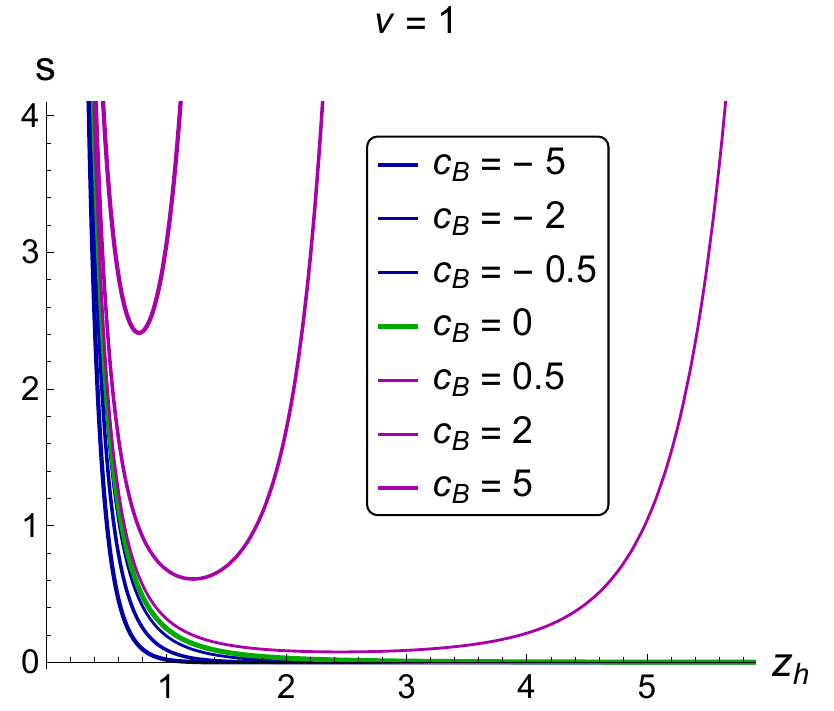} \hfill
  \includegraphics[width=0.32\textwidth]{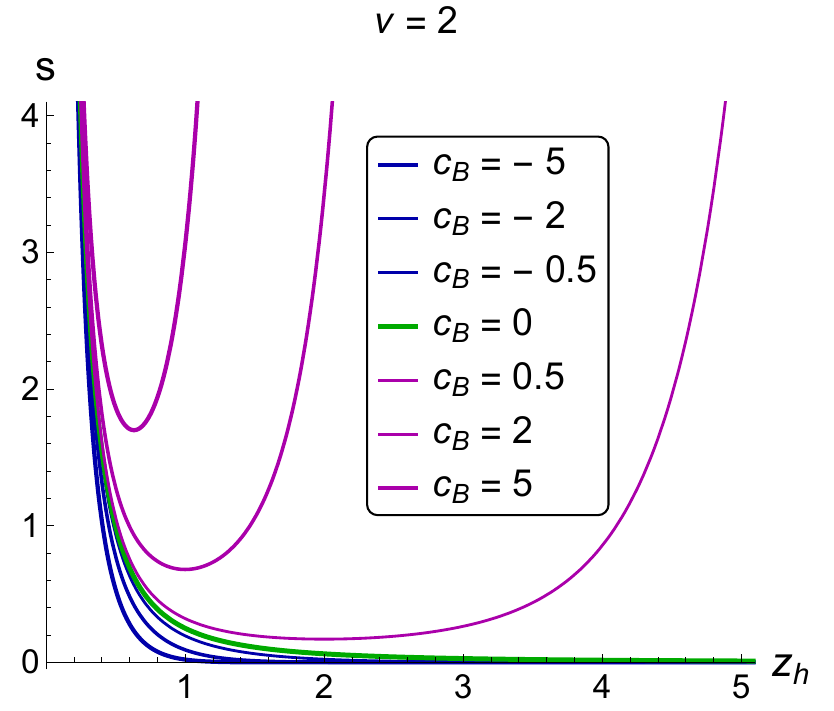} \hfill 
  \includegraphics[width=0.32\textwidth]{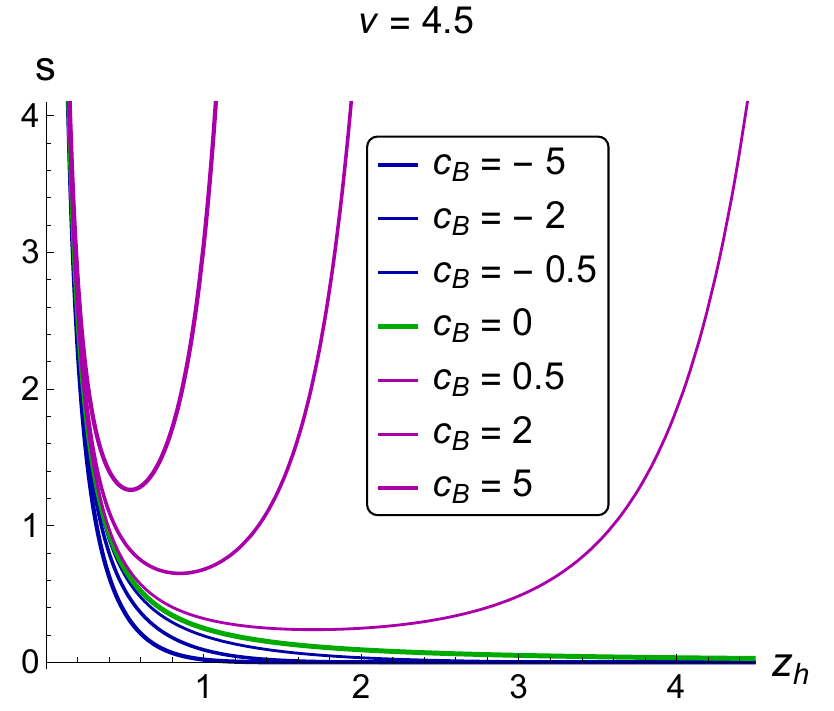} \hfill \\
  A \hspace{140pt} B \hspace{140pt} C
  \caption{Entropy density $s(z_h)$ for various values of $c_B$; for $\nu = 1$ (A), $\nu = 2$ (B), and $\nu = 4.5$ (C); $L = 1$.\\ \,}
  \label{ModelI-szh}
  \centering
  \includegraphics[width=0.32\textwidth]{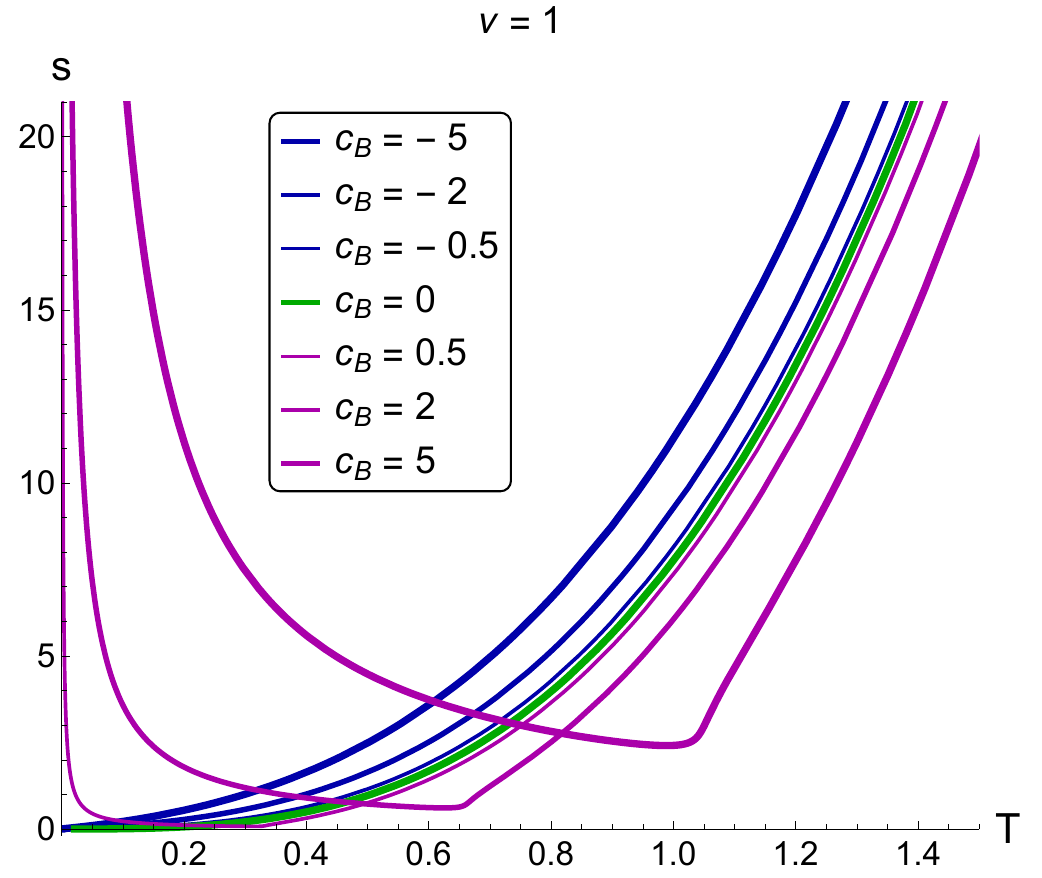} \hfill
  \includegraphics[width=0.32\textwidth]{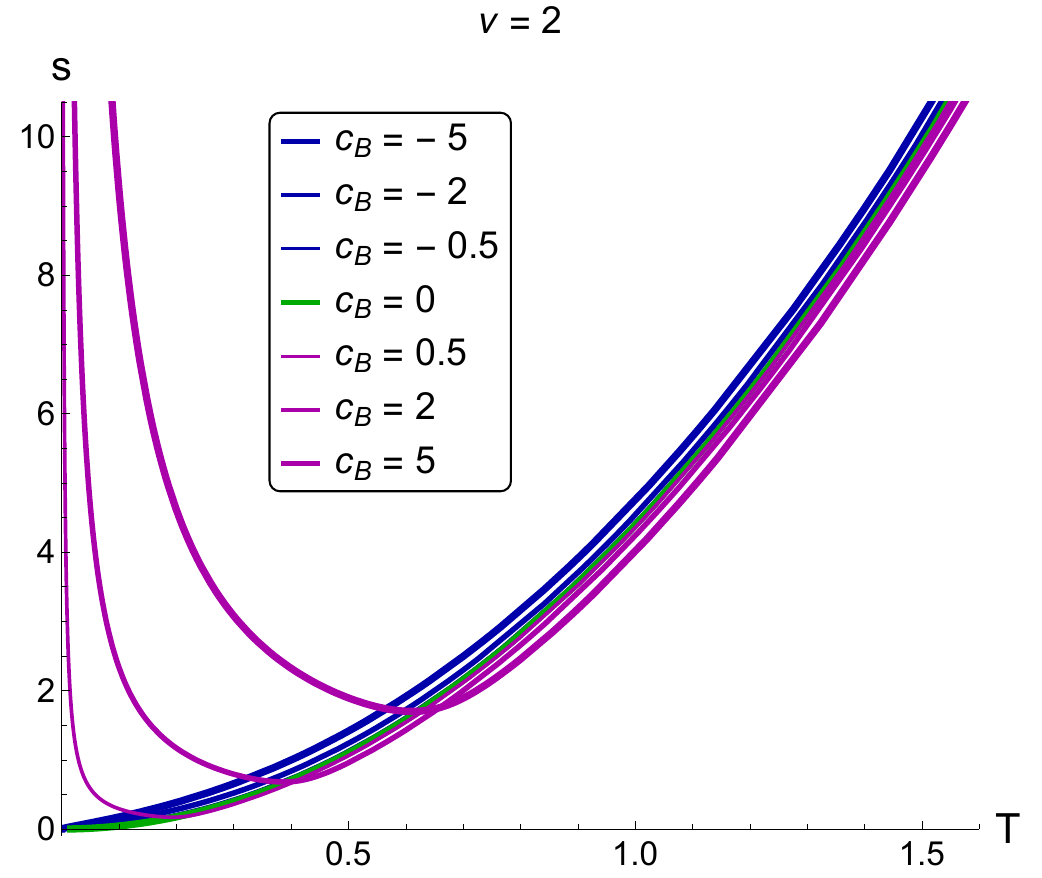} \hfill 
  \includegraphics[width=0.32\textwidth]{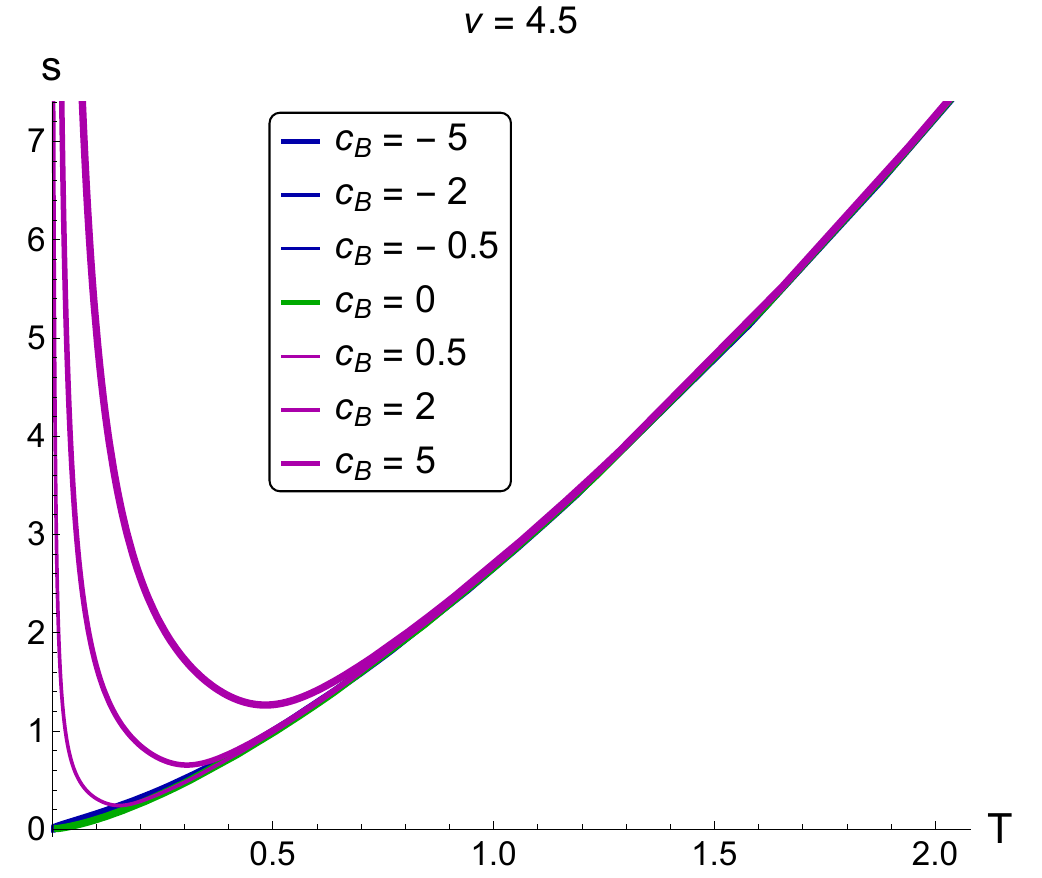} \hfill \\
  A \hspace{140pt} B \hspace{140pt} C
  \caption{Entropy density $s(T)$ for various values of $c_B$; for $\nu = 1$ (A), $\nu = 2$ (B), and $\nu = 4.5$ (C); $L = 1$.}
  \label{ModelI-sT}
\end{figure}

Both $s$ and $T$ are found as functions of the horizon $z_h$, and it proves extremely difficult to obtain an analytic expression for $s=s(T)$. However, we only need to examine the behavior of $s$ at small temperatures, as required by the third law of thermodynamics \eqref{4:16}. We can observe that the regime $T \to 0$ is possible only for $z_h \to + \, \infty$ because $T=T(z_h)$ never evaluates to zero, see \eqref{4:15}, and at small $z_h$ the temperature diverges, as is seen in the following asymptotic behavior:
\be
    T(z_h) \sim \cfrac{1}{2 \pi z_h} \left( 1 + \frac{1}{\nu} \right) \quad \text{as} \quad z_h \to +\,\infty,
\ee
which we obtain by expanding the integrand in the formula for $T(z_h)$, see \eqref{4:15}, and integrating term-by-term. This behavior is confirmed with Fig.~\ref{ModelI-Tzh}.

Since we can only have arbitrarily small $T$ at large $z_h$, we focus our examination of $s(z_h)$ and $T(z_h)$ on the regime $z_h \to +\,\infty$. The sign of the parameter $c_B$ is essential for understanding the behavior of both $s(z_h)$ and $T(z_h)$. For $c_B=0$ we obtain the following:
\be
    c_B = 0: \qquad 
    s(T) = \cfrac{1}{4} \left( \cfrac{2\pi L T}{1+\frac{1}{\nu}} \right)^{1+\frac{2}{\nu}},
\ee
so that the third law holds if and only if
\be
    1+\frac{2}{\nu} > 0.
\ee

If $c_B > 0$, the temperature decays to zero as $z_h \to +\infty$ because the denominator converges to a finite nonzero value, while the numerator decays to zero. The entropy density, however, blows up (see Fig.~\ref{ModelI-szh}):
\be
    c_B > 0: \qquad 
    s(z_h) \to +\,\infty, \quad T(z_h) \to 0
    \quad \text{as} \quad z_h \to +\,\infty.
\ee

If $c_B < 0$, the entropy density tends to zero. For $T(z_h)$ both the numerator and denominator diverge, which is why we use the L'Hôpital rule to obtain that $T(z_h)$ also decays to zero:
\be
  \lim_{z_h\to+\infty}T(z_h) = 
  \lim_{z_h\to+\infty} \left[
    \frac{\exp\left(c_B z_h^2\right)}{4\pi}
    \left(\frac{1+2/\nu}{z_h} - c_B z_h\right) 
  \right] = 0,
\ee
hence the third law is satisfied.

To sum it up, the third law holds if and only if either of these two conditions is true, keeping in mind the constraint in \eqref{4:13}:
\be\label{4:22}
  (1)\quad c_B = 0, \ 1 + \frac{2}{\nu} > 0, \ 1 + \frac{1}{\nu} > 0, \qquad 
  (2)\quad c_B < 0, \ 1 + \frac{1}{\nu} > 0.
\ee
Otherwise, the $s(T)$ either diverges for $c_B>0$ and $c_B = 0$, $1+2/\nu < 0$, or is constant for $c_B = 0$, $\nu = - \, 2$. We can observe $s(T)$ in different parameter regimes in Fig.~\ref{ModelI-sT}.

It is possible to obtain an asymptotic expression for $s(T)$ in the relevant regime $T\to0$. For that we first need to find the asymptotic expansion for $T(z_h)$, $z_h \to +\,\infty$, then invert it to obtain an explicit asymptotic expansion for $z_h(T)$, $T \to 0$, and substitute it into the formula for $s(z_h)$. The result reads:
\bea
  c_B > 0: && s(T) = \frac{(3\,c_B/2)^{1+\frac{1}{\nu}}\, L^{1+\frac{2}{\nu}}}{8 \pi \, \Gamma\left(1+\frac{1}{\nu}\right)}  \cdot \frac{1}{T} + O\left(\frac{T^2}{\sqrt{\ln(1/T)}}\right), \\
  c_B < 0: && s(T) = \frac{\sqrt{\pi}\,|c_B|^{1/4+1/\nu} \, L^{1+\frac{2}{\nu}}}{\sqrt{12}} \cdot \sqrt{T} \left(\ln \left(\frac{1}{4\pi\,T}\right)\right)^{-3/4-1/\nu} \times \nn \\
  && \phantom{s(T)} \times \left[1 + O\left(\frac{\ln\ln\left(1/T\right)}{\ln\left(1/T\right)}\right)\right].
\eea

\subsection{Null Energy Condition}\label{sec:ModelINEC}

We now impose NEC on the model, as it is understood in \cite{Chu:2019uoh}. Specifically, for a diagonal metric of Euclidean signature:
\be
  \forall \mu: \qquad R^{\mu}_{\mu} - R^0_0 \ge 0 
  \quad\Longrightarrow\quad
  \cfrac{R_{\mu\mu}}{g_{\mu\mu}} - \cfrac{R_{00}}{g_{00}} =  \cfrac{G_{\mu\mu}}{g_{\mu\mu}} - \cfrac{G_{00}}{g_{00}} \ge 0. \label{NEC-02}
\ee

Here we notice the linear combinations of $G_{\mu\mu}/g_{\mu\mu}$ we encountered earlier in Section~\ref{sec:quadraturederivation} when discussing sign tables. Using this model's sign table, we write down the Einstein equations in the following form:
\begin{gather}
    \cfrac{G_{x_3 x_3}}{g_{x_3 x_3}}+\cfrac{G_{x_1 x_1}}{g_{x_1 x_1}} - \cfrac{G_{x_2 x_2}}{g_{x_2 x_2}} + \cfrac{G_{tt}}{g_{tt}} = 0, 
    \label{einsten-ratios-01} \\
    \cfrac{G_{x_1 x_1}}{g_{x_1 x_1}}+\cfrac{G_{tt}}{g_{tt}} = \cfrac{f_3(\phi)\,q^2_3}{2\,g_{x_1x_1}\,g_{x_2x_2}},
    \qquad
    \cfrac{G_{x_3 x_3}}{g_{x_3 x_3}}+\cfrac{G_{tt}}{g_{tt}} = \cfrac{f_1(\phi)\,q^2_1}{2\,g_{x_2x_2}\,g_{x_3x_3}},
    \label{einsten-ratios-03} \\
    \cfrac{G_{x_2x_2}}{g_{x_2x_2}}+\cfrac{G_{zz}}{g_{zz}} = - \, V(\phi), \qquad
    \cfrac{G_{tt}}{g_{tt}} + \cfrac{G_{zz}}{g_{zz}} =\cfrac{\phi'^2}{2\,g_{zz}}.
    \label{einsten-ratios-05}
\end{gather}
In Lorentz signature null energy conditions read:
\begin{equation}
  \cfrac{G_{zz}}{g_{zz}}+\cfrac{G_{tt}}{g_{tt}} \ge 0, \quad
  \cfrac{G_{x_1x_1}}{g_{x_1x_1}}+\cfrac{G_{tt}}{g_{tt}} \ge 0, \quad
  \cfrac{G_{x_2x_2}}{g_{x_2x_2}}+\cfrac{G_{tt}}{g_{tt}} \ge 0, \quad
  \cfrac{G_{x_3x_3}}{g_{x_3x_3}}+\cfrac{G_{tt}}{g_{tt}} \ge 0. \label{Model1-NEC-general}
\end{equation}
Using \eqref{einsten-ratios-01}--\eqref{einsten-ratios-05}, we conclude that the NEC \eqref{Model1-NEC-general} are equivalent to
\be\label{4:30}
  \phi'^2 \ge 0, \quad f_3(\phi) \ge 0, \quad f_1(\phi) \ge 0.
\ee

The complete solution for the model depends on various parameters. We now investigate which parameter regimes ensure that the model satisfies the NEC. 

First, we notice that the NEC must hold for the coupling functions at $z=z_h$, determined by the equations \eqref{model1-EOM-f1}--\eqref{model1-EOM-f3}:
\bea
  \displaystyle f_3 \left(\phi(z_h)\right)
  =&\left(\cfrac{L}{z_h}\right)^{2/\nu} \cfrac{2 c_B z_h}{q_3^2} \, g'(z_h) & \ge 0, \\
  \displaystyle f_1 \left(\phi(z_h)\right)
  =&\left(\cfrac{L}{z_h}\right)^{4/\nu} \cfrac{2e^{c_B z_h^2} z_h}{L^2 q_1^2} \, g'(z_h)\,\cfrac{1-\nu}{\nu} & \ge 0.
\eea

They lead us to two more conditions in addition to \eqref{model1-nu-constraint}, keeping in mind $g'(z_h)<0$, see Appendix~\ref{app::C}:
\be
  c_B \le 0, \quad \cfrac{1}{\nu} \le 1.
\ee
Thus for $\nu$ we have two possible regimes: $\nu < -\,1$ and $\nu \ge 1$. The second regime $\nu \ge 1$ gives us non-negative coupling functions. This can be seen if one considers the sign of each term in \eqref{model1-EOM-f1} and \eqref{model1-EOM-f3}, keeping in mind that $g'(z)<0$ for $c_B \le 0$, see \eqref{C:9}:
\be
  \displaystyle c_B\le0, \ \nu \ge 1 \quad \longrightarrow \quad 
  f_3\left(\phi(z)\right), \ f_1\left(\phi(z)\right)\Big|_{(0, z_h)} \ge 0.
\ee

\begin{figure}[t!]
  \centering
  \includegraphics[width=0.49\textwidth]{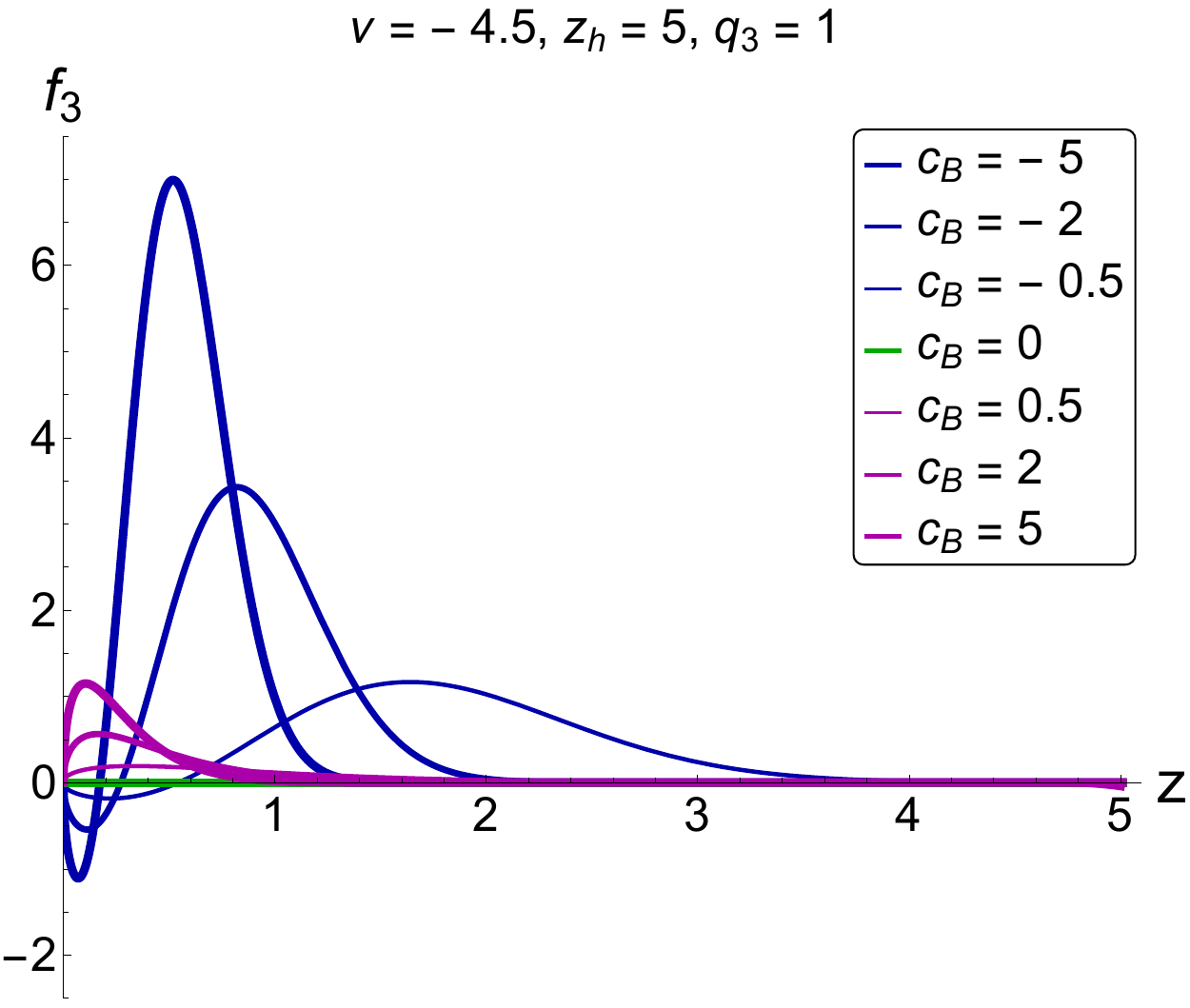} \hfill
  \includegraphics[width=0.49\textwidth]{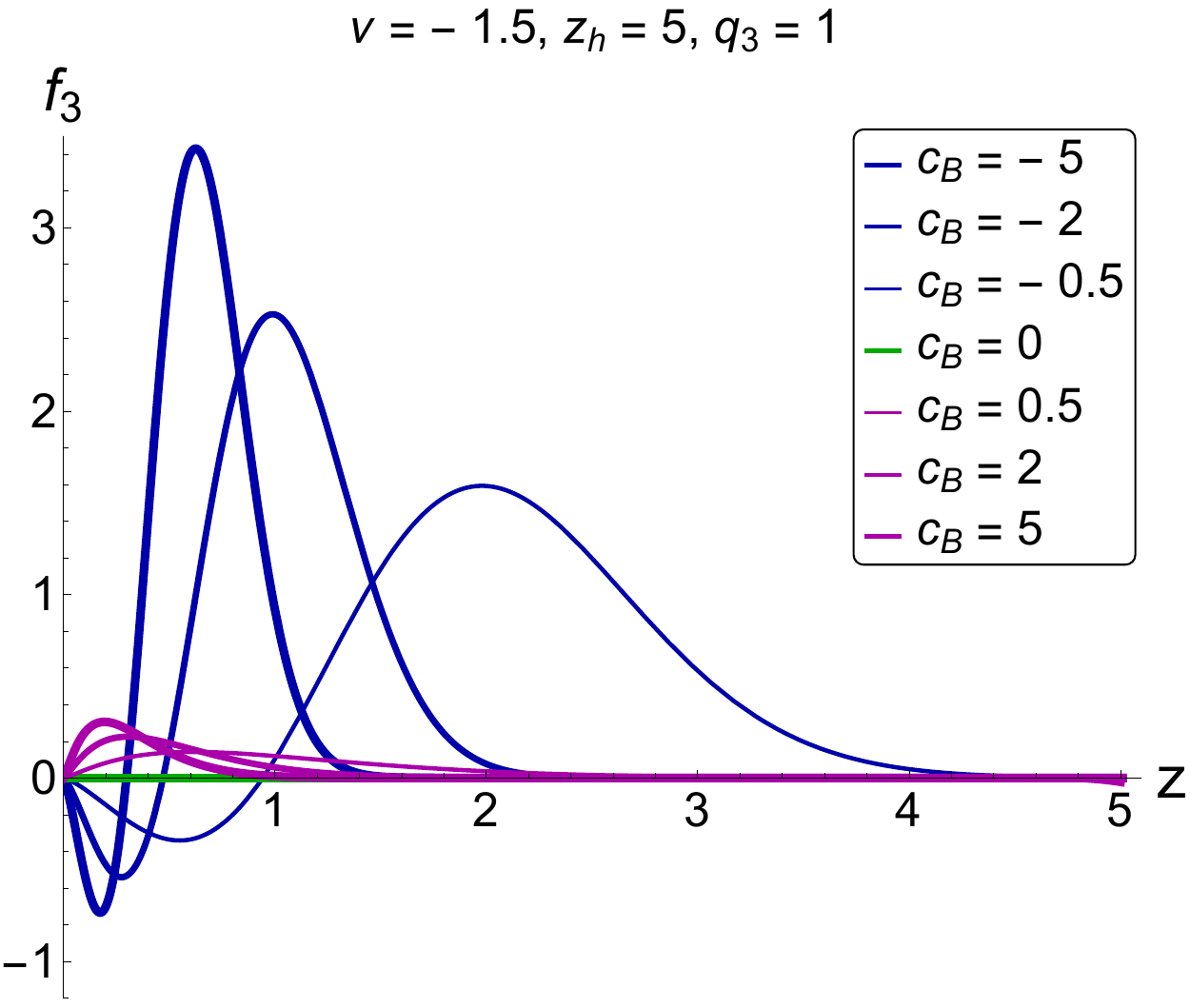} \hfill \\
  A \hspace{200pt} B
  \caption{Coupling function $f_3(z)$ for various values of $c_B$; $\nu = - \, 4.5$ (A) and $\nu = - \, 1.5$ (B); $L = 1$, $q_3 = 1$.}
  \label{model1-negative-f3}
\end{figure}

However, for some particular values $\nu < -\,1$ the coupling function $f_3(z)$ is negative for $c_B < 0$ in the vicinity of $z = 0$, see Fig.~\ref{model1-negative-f3}.

Second, let us consider the condition on the dilaton. The analysis of this condition can be reduced to that of a bi-quadratic equation
\be
  \displaystyle \phi'(z)^2
  = - \, \frac{2 c^2_B}{z^2} \left(
  z^4 + \frac{3\nu-2}{c_B\nu} \, z^2 + \frac{2-2\nu}{c^2_B \nu^2}
  \right) \ge 0.
\ee
Since the restriction $c_B\le0$ has been already established, the results read
\bea
    c_B = 0: &&
    \phi'(z)^2 = \left(1-\frac{1}{\nu}\right) \frac{4}{\nu z^2} \ge 0
    \quad\Rightarrow\quad \nu \ge 1,
    \\
    c_B < 0, \ \nu \ge 1: &&
    \phi'(z)^2 \ge 0 \quad\text{for} \quad 
    0 \ < \ z \ \le \ z_{-}, \label{model1-nec-cbnu1}
    \\
    c_B < 0, \ \nu < -\,1: &&
    \phi'(z)^2 \ge 0 \quad\text{for}\quad 0 < \text{min}(z_-, \, z_+)
    \le z \le 
    \text{max}(z_-, \, z_+),
\eea
where
\be
    z_{+} = \sqrt{\frac{2-3\nu+\sqrt{9\nu^2 - 4\nu - 4}}{2c_B\nu}}, 
    \qquad 
    z_{-} = \sqrt{\frac{2-3\nu-\sqrt{9\nu^2 - 4\nu - 4}}{2c_B\nu}}.
\ee

The regime $c_B \le 0$, $\nu < -\,1$ is therefore forbidden as $\phi'(z)^2<0$ in the vicinity of $z=0$. Therefore we have managed to fully describe what the NEC is equivalent to:
\be\label{4:40}
    \text{NEC} \quad
    \begin{cases}
        c_B = 0, \ \nu \ge 1,
        \\
        c_B < 0, \ \nu \ge 1, \ z_h \le z_{-}.
    \end{cases}
\ee

The upper boundary for the horizon $z_h$ appears due to the fact that the condition $\phi'(z)^2 \ge 0$ is satisfied for the regime $c_B < 0$, $\nu \ge 1$ only on the interval $z \in [0, \ z_{-}]$, see \eqref{model1-nec-cbnu1}, thus limiting which values $z_h$ can take for the NEC to be satisfied:
\be
  z_h \le z_h^{max} \equiv z_{-}.
\ee

\subsection{NEC and the Third Law of Thermodynamics}\label{Model1-NEC-3Law}

Recall the result for the third law of thermodynamics: it is only satisfied in either of these two cases
\be\label{4:41}
  (1)\quad c_B = 0, \ 1 + \frac{2}{\nu} > 0, \ 1 + \frac{1}{\nu} > 0, \qquad 
  (2)\quad c_B < 0, \ 1 + \frac{1}{\nu} > 0.
\ee

For the first case when NEC is satisfied, see \eqref{4:40}, we have
\be
    c_B = 0, \ \nu \ge 1,
\ee
which also support the third law of thermodynamics \eqref{4:41}.

For the second case when NEC is satisfied, see \eqref{4:40}, we have
\be
    c_B < 0, \ \nu \ge 1, \ z_h \le z_h^{max}.
\ee

\begin{figure}[b!]
  \centering
  \includegraphics[width=0.32\textwidth]{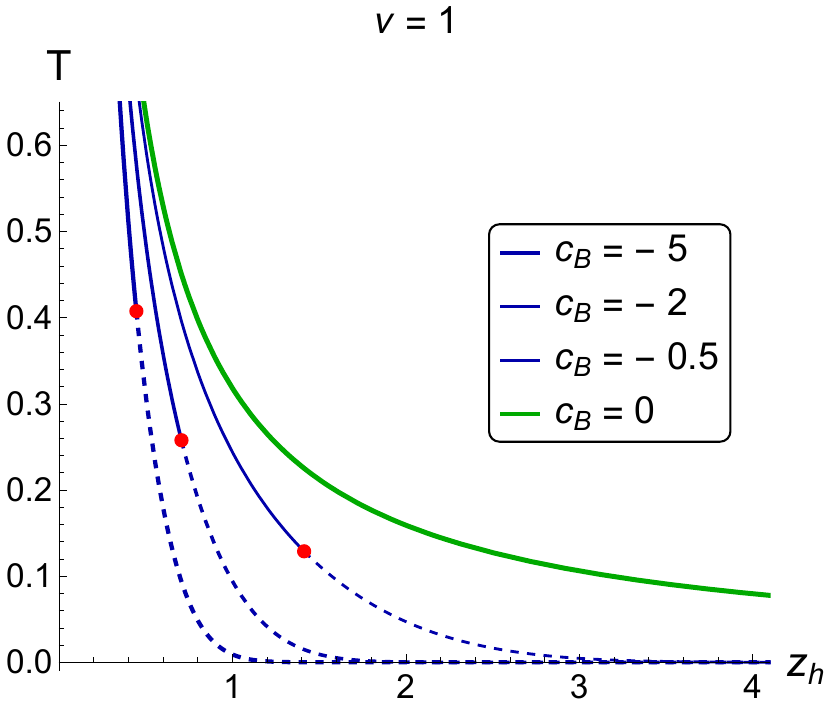} \hfill
  \includegraphics[width=0.32\textwidth]{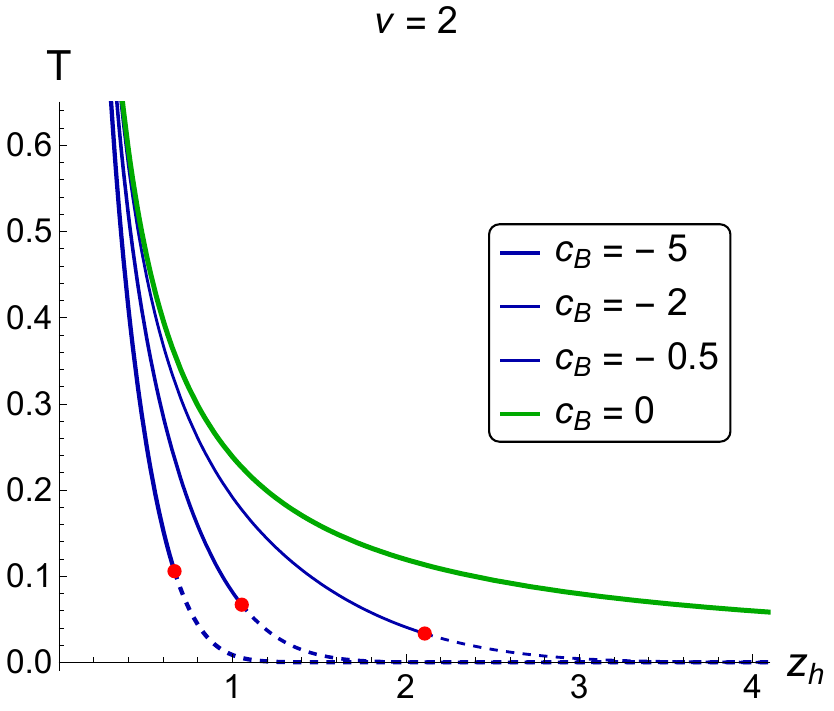} \hfill 
  \includegraphics[width=0.32\textwidth]{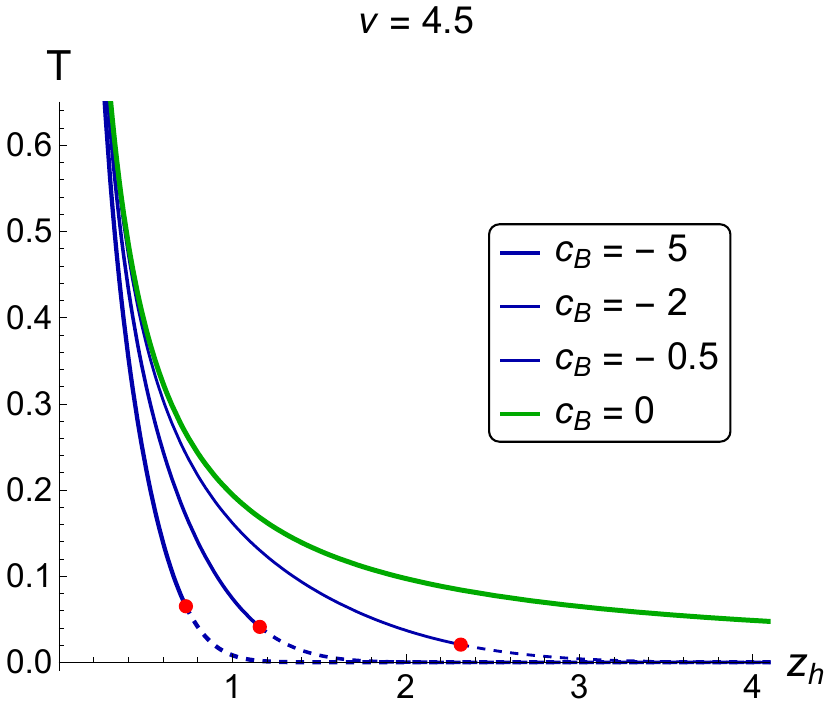} \hfill \\
  A \hspace{140pt} B \hspace{140pt} C
  \caption{Temperature $T(z_h)$ for various values of $c_B$; for $\nu = 1$ (A), $\nu = 2$ (B), and $\nu = 4.5$ (C); the red points mark $T_{min}$ from \eqref{3.69}; the dashed lines represent regimes forbidden by NEC.\\ \,}
  \label{model1-Tcapped}
  \centering
  \includegraphics[width=0.32\textwidth]{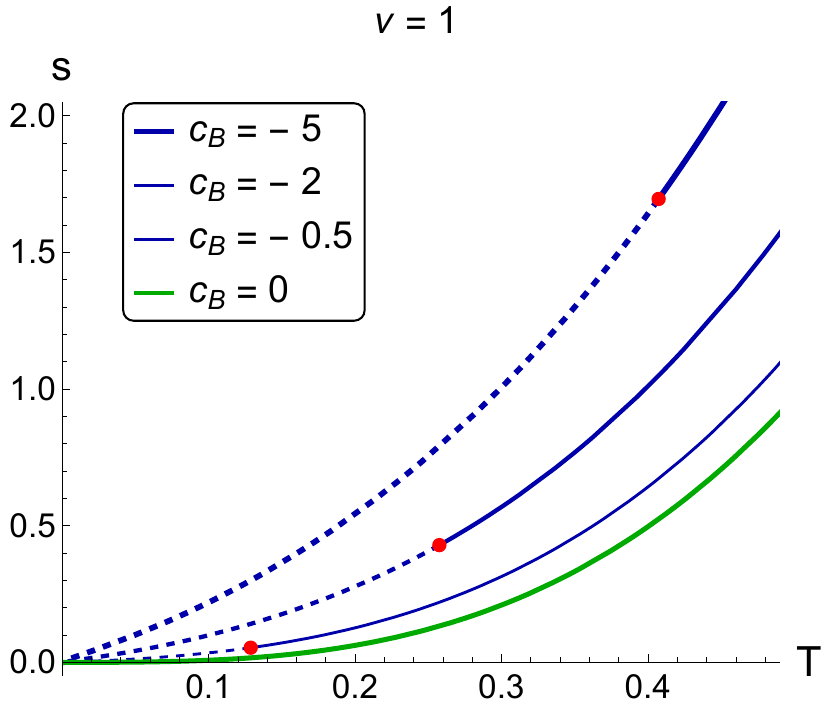} \hfill
  \includegraphics[width=0.32\textwidth]{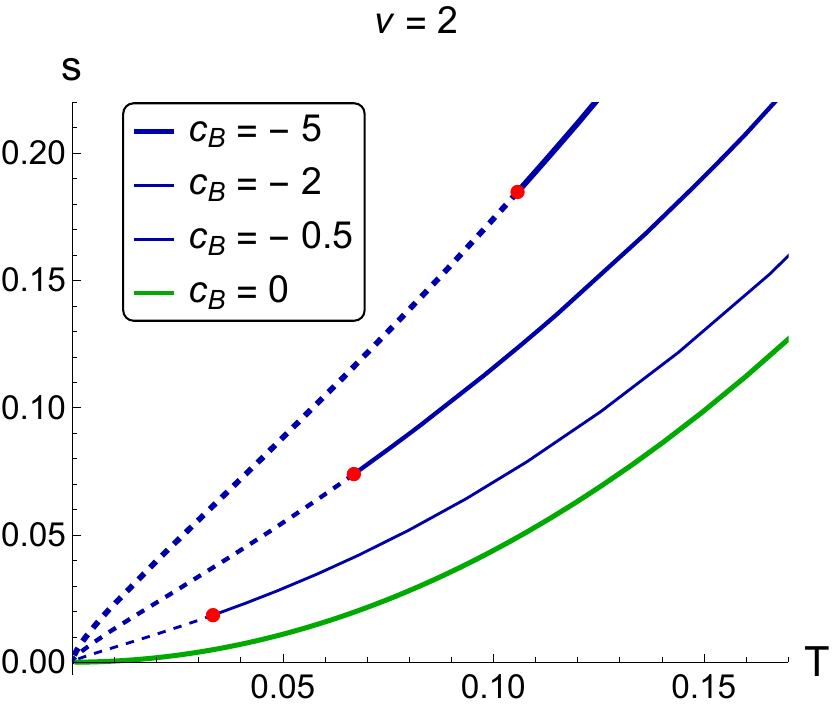} \hfill 
  \includegraphics[width=0.32\textwidth]{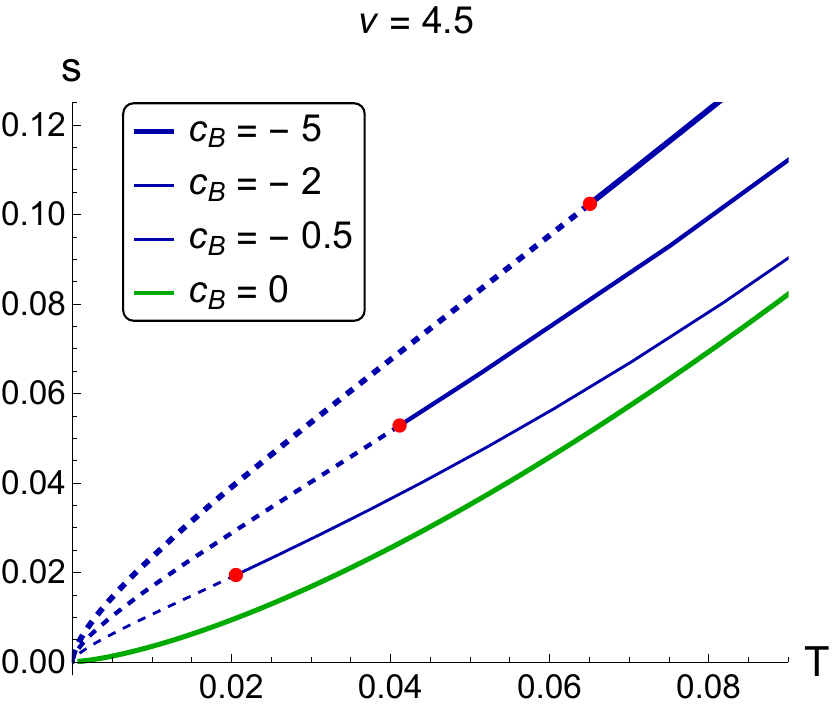} \hfill \\
  A \hspace{140pt} B \hspace{140pt} C
  \caption{Entropy density $s(T)$ for various values of $c_B$; for $\nu=1$ (A), $\nu=2$ (B), and $\nu=4.5$ (C); the red points mark $T_{min}$ from \eqref{3.69}; the dashed lines represent regimes forbidden by NEC.}
  \label{model1-sTcapped}
\end{figure}

In the regime $c_B < 0$, $\nu \ge 1$, the temperature $T(z_h)$ decays to zero monotonically. Therefore, there is a boundary below for $T(z_h)$, see Fig.~\ref{model1-Tcapped}:

\be
  z_h \le z^{max}_h=\sqrt{\frac{2-3\nu-\sqrt{9\nu^2 - 4\nu - 4}}{2c_B\nu}}
  \quad\Longrightarrow\quad
  T \ge T_{min} = T(z^{max}_h). \label{3.69}
\ee

As a consequence, we can no longer claim fulfillment of the third law of thermodynamics for $c_B < 0$, $\nu \ge 1$, since the temperature cannot tend to zero, see Fig.~\ref{model1-sTcapped}. Therefore, the case $c_B = 0$ and $\nu \ge 1$ is the only one that satisfies both the third law and NEC.

For this model we obtain that the third law of thermodynamics and NEC are independent conditions --- neither leads to the other. For example, the regime $c_B = 0$, $\nu < - \,2$ is inconsistent with NEC while supporting the third law of thermodynamics. For $c_B < 0$, the NEC leads to an upper boundary on the horizon $z_h \le z_h^{max}$, which is inconsistent with the premise $T\to0$.

\section{Model II: Two Maxwell Fields in D = 5 with Two Lifshitz-Type Anisotropies and Non-Trivial Warp Factor}\label{sec:model2}

The next model we consider is a $D = 5$ model, that differs from Model I in that its metric ansatz contains the second anisotropy factor in a Lifshitz form, not in a Gauss form $\exp(c_Bz^2)$:
\be
    ds^2=\frac{L^2}{z^2}\,\fb(z)
    \left( 
    -\,g(z)dt^2 + dx_1^2 
    +\bigg(\frac{z}{L}\bigg)^{2-\frac{2}{\nu_1}} dx_2^2
    +\bigg(\frac{z}{L}\bigg)^{2-\frac{2}{\nu_2}} dx_3^2
    +\frac{dz^2}{g(z)}
    \right).
\ee

There are three possible configurations of two magnetic Maxwell fields:
\bea\label{5:2}
    \textbf{i}\quad&F_2=q_2 \, dx^1\wedge dx^3, \qquad F_3=q_3 \, dx^1\wedge dx^2,&
    \\
    \textbf{ii}\quad&F_1=q_1 \, dx^2\wedge dx^3, \qquad F_3=q_3 \, dx^1\wedge dx^2,&
    \\
    \textbf{iii}\quad&F_1=q_1 \, dx^2\wedge dx^3, \qquad F_2=q_2 \, dx^1\wedge dx^3.&
    \label{5:4}
\eea

Here is the action for the case $\textbf{ii}$:
\be
    \textbf{ii} \qquad S = \int d^5 x \, \sqrt{- g_5} \left[R - \cfrac{1}{4} \, f_{1}(\phi) F^2_{1} - \cfrac{1}{4} \, f_{3}(\phi) F^2_{3}
    - \cfrac{1}{2} \, \partial_{\mu} \phi \, \partial^{\,\mu} \phi - V(\phi)\right],
\ee
while in other cases the action can be written down in a similar way.

For simplicity, let us insert the sign table (Table~\ref{Table-5D-2M-1}) that includes all the three distinct magnetic ansatzes. A particular case ($\textbf{i}$, $\textbf{ii}$, or $\textbf{iii}$) is obtained by removing a column corresponding to the Maxwell field not included in the case.

\begin{table}[t!]
\centering
    \begin{tabular}{ScSr||Sc|Sc|Sc|Sc|Sc||Sl}
         & & $\displaystyle\frac{\phi'^2}{4g_{zz}}$ & $\displaystyle \frac{V(\phi)}{2}$ & $\displaystyle\frac{f_1(\phi)\,q^2_1}{4\,g_{x_2x_2}\,g_{x_3x_3}}$ & $\displaystyle\frac{f_2(\phi)\,q^2_2}{4\,g_{x_1x_1}\,g_{x_3x_3}}$ & $\displaystyle\frac{f_3(\phi)\,q^2_3}{4\,g_{x_1x_1}\,g_{x_2x_2}}$ & \\ \hlineb{1pt}
         $\bm{(z):}$ & $\displaystyle\frac{T_{zz}}{g_{zz}}$ 
         & $\bm{+}$ & $\bm{-}$ & $\bm{-}$ & $\bm{-}$ & $\bm{-}$ &
         $\displaystyle=\frac{G_{zz}}{g_{zz}}$ 
         \\ \hlineb{1pt}
         $\bm{(x_3):}$ & $\displaystyle\frac{T_{x_3 x_3}}{g_{x_3 x_3}}$ 
         & $\bm{-}$ & $\bm{-}$ & $\bm{+}$ & $\bm{+}$ & $\bm{-}$ &  
         $\displaystyle=\frac{G_{x_3 x_3}}{g_{x_3 x_3}}$ 
         \\ \hlineb{1pt}
         $\bm{(x_2):}$ & $\displaystyle\frac{T_{x_2 x_2}}{g_{x_2 x_2}}$
         & $\bm{-}$ & $\bm{-}$ & $\bm{+}$ & $\bm{-}$ & $\bm{+}$ &
         $\displaystyle=\frac{G_{x_2 x_2}}{g_{x_2 x_2}}$
         \\ \hlineb{1pt}
         $\bm{(x_1):}$  & $\displaystyle\frac{T_{x_1 x_1}}{g_{x_1 x_1}}$
         & $\bm{-}$ & $\bm{-}$ & $\bm{-}$ & $\bm{+}$ & $\bm{+}$ &
         $\displaystyle=\frac{G_{x_1 x_1}}{g_{x_1 x_1}}$
         \\ \hlineb{1pt}
         $\bm{(t):}$  & $\displaystyle-\,\frac{T_{tt}}{g_{tt}}$ 
         & $\bm{-}$ & $\bm{-}$ & $\bm{-}$ & $\bm{-}$ & $\bm{-}$ &
         $\displaystyle=-\,\frac{G_{tt}}{g_{tt}}$ 
         \\ 
    \end{tabular}
  \caption{Sign table for the terms contributing to the components of $T_{\mu\nu}/g_{\mu\nu}$ for $D=5$ Model II. Note that this table includes all the three possible magnetic Maxwell fields, thus being of use for all three distinct cases.}
\label{Table-5D-2M-1}
\end{table}

The following equations for $g(z)$ are obtained in each case
\bea
    \textbf{i} \quad &&\frac{G_{x_2 x_2}}{g_{x_2 x_2}}+\frac{G_{x_3 x_3}}{g_{x_3 x_3}} -\frac{G_{x_1 x_1}}{g_{x_1 x_1}} + \frac{G_{tt}}{g_{tt}}=0,
    \\
    \textbf{ii} \quad &&\frac{G_{x_1 x_1}}{g_{x_1 x_1}}+\frac{G_{x_3 x_3}}{g_{x_3 x_3}} - \frac{G_{x_2 x_2}}{g_{x_2 x_2}} + \frac{G_{tt}}{g_{tt}}=0,
    \\
    \textbf{iii} \quad &&\frac{G_{x_1 x_1}}{g_{x_1 x_1}} + \frac{G_{x_2 x_2}}{g_{x_2 x_2}} - \frac{G_{x_3 x_3}}{g_{x_3 x_3}} + \frac{G_{tt}}{g_{tt}}=0.
\eea

These are second-order DEs governing the behavior of the blackening function $g(z)$. We can write down a general solution, parametrized by a single constant $\alpha_a$:
\begin{gather}
    g_{a}(z)= C_a \, z^{\alpha_a} \displaystyle \int_{z}^{z_h} \displaystyle\fb(\xi)^{-\frac{3}{2}}\xi^{1+\frac{1}{\nu_1}+\frac{1}{\nu_2}-\alpha_a} d\xi, \quad 
    a = \mathbf{i, \ ii, \ iii}, \label{model3-gGeneral}
    \\
    \alpha_\textbf{i}=4-\frac{2}{\nu_1}-\frac{2}{\nu_2}, \qquad
    \alpha_\textbf{ii}=\frac{2}{\nu_1}-\frac{2}{\nu_2}, \qquad
    \alpha_\textbf{iii}=\frac{2}{\nu_2}-\frac{2}{\nu_1}.
\end{gather}

We choose a warp factor of considerable generality $\fb(z)=e^{cz^n}$, $n>0$, $c\in\mathbb{R}$. This includes the trivial warp factor $\fb(z)\equiv 1$, which is reproduced by setting $c=0$. The solutions obtained for the blackening function are
\bea
  g_\textbf{i}(z)&=&
  \begin{cases}
    \displaystyle 1-\frac{I(z)}{I(z_h)}, \text{   where   }I(z)=\int^{z}_0 \xi^3\,\exp\left(-\frac{3}{2}c\xi^2\right)\,d\xi, \ \frac{1}{\nu_1}+\frac{1}{\nu_2}=2, \\
    \displaystyle 8 z^{8} \int_{z}^{z_h}\xi^{-9}\,\exp\left(-\frac{3}{2}c\xi^2\right)\,d\xi, \ \frac{1}{\nu_1}+\frac{1}{\nu_2}=-\,2,
  \end{cases} \label{model3-gSol1} \\
  g_\textbf{ii}(z)&=&
  \begin{cases}
    \displaystyle 1-\frac{I(z)}{I(z_h)}, \text{   where   }I(z)=\int^{z}_0 \xi^{1+\frac{2}{\nu_1}}\,\exp\left(-\frac{3}{2}c\xi^2\right)\,d\xi, \ \nu_1=\nu_2, \ 1+\frac{1}{\nu_1}>0, \\
    \displaystyle 4\left(1+\frac{1}{\nu_1}\right) z^{4+\frac{4}{\nu_1}} \int_{z}^{z_h}\xi^{-5-\frac{4}{\nu_1}}\,\exp\left(-\frac{3}{2}c\xi^2\right)\,d\xi, \ \frac{1}{\nu_1}+\frac{1}{\nu_2}=-\,2, \ 1+\frac{1}{\nu_1}>0,
  \end{cases} \label{model3-gSol2} \\
  g_\textbf{iii}(z)&=&g_\textbf{ii}(z), \ \mbox{where} \ \nu_1\leftrightarrow\nu_2.
\eea

Note that we have omitted the full process of obtaining these solutions, as it included an extensive analysis of when the remaining boundary condition $g(0)=1$ can be imposed in \eqref{model3-gGeneral}. The very process of imposing this boundary condition leads to the restrictions on $\nu_1$ and $\nu_2$ presented in \eqref{model3-gSol1}--\eqref{model3-gSol2}. What is particularly interesting is that in both \eqref{model3-gSol1} and \eqref{model3-gSol2}, the second solutions are constructed with the integrals that are divergent at $z \to 0\,+$. This divergence is suppressed by the $z^8$, $z^{4+4/\nu_1}$ factors before the integrals, so that $g(z) \to 1$ at $z \to 0\,+$.

The case \textbf{iii} is the same as \textbf{ii} up to the interchange $\nu_2 \leftrightarrow \nu_1$, $f_2 \leftrightarrow f_3$, $q_2 \leftrightarrow q_3$. We therefore generally omit formulas for \textbf{iii}. The special case of the trivial warp-factor $\fb(z)\equiv1$ is reproduced from the formulas above by the straightforward setting $c=0$:
\bea
    g_\textbf{i}(z)&=&
    \begin{cases}
        \displaystyle 1-\left(\frac{z}{z_h}\right)^4, \quad \frac{1}{\nu_1}+\frac{1}{\nu_2}=2,
        \\
        \displaystyle 1-\left(\frac{z}{z_h}\right)^8, \quad \frac{1}{\nu_1}+\frac{1}{\nu_2}=-\,2,
    \end{cases}
    \\
    g_\textbf{ii}(z)&=&
    \begin{cases}
        \displaystyle 1-\left(\frac{z}{z_h}\right)^{2+\frac{2}{\nu_1}}, \quad \nu_1=\nu_2, \ 1+\frac{1}{\nu_1}>0,
        \\
        \displaystyle 1-\left(\frac{z}{z_h}\right)^{4+\frac{4}{\nu_1}}, \quad \frac{1}{\nu_1}+\frac{1}{\nu_2}=-\,2, \ 1+\frac{1}{\nu_1}>0.
    \end{cases}
\eea

\subsection{Third Law of Thermodynamics}

We now move on to a careful examination of the model thermodynamics. Let us first calculate the entropy density, that is the same for all the cases \textbf{i}, \textbf{ii}, and \textbf{iii}:
\be
    s(z_h) = \frac{1}{4}\sqrt{g_{x_1x_1}g_{x_2x_2}g_{x_3x_3}} = \frac{1}{4} \left(\frac{L}{z_h}\right)^{1 + \frac{1}{\nu_1} + \frac{1}{\nu_2}} \ \exp\left(\frac{3 c z^n_h}{2}\right).
\ee

Since the solutions above are equipped with restrictions on the parameters $(\nu_1, \, \nu_2)$, we can further simplify the expression for the entropy density, depending on the case:
\begin{gather}
    \displaystyle 
    s_{div}(z_h)\equiv s_{div, \, \textbf{i}}(z_h)=s_{div, \, \textbf{ii}}(z_h) = \frac{1}{4} \frac{z_h}{L} \ \exp\left(\frac{3 c z^n_h}{2}\right),
    \\
    \displaystyle s_{conv, \, \textbf{i}}(z_h) = \frac{\left(L/z_h\right)^{3}}{4} \ \exp\left(\frac{3 c z^n_h}{2}\right), \ 
    s_{conv, \, \textbf{ii}}(z_h) = \frac{\left(L/z_h\right)^{1+\frac{2}{\nu_1}}}{4} \ \exp\left(\frac{3 c z^n_h}{2}\right), \label{6:18}
\end{gather}
where \textit{div} and \textit{conv} in the subscripts denote the case of the divergent and convergent integral, respectively; \textbf{i} and \textbf{ii} specify the particular case of the field configuration.

The solutions with a diverging integral are the easiest to consider. Let us compute the temperature:
\be
    T_{div, \, \textbf{i}}(z_h) = \frac{2}{\pi z_h} \exp\left(-\,\frac{3 c z^n_h}{2}\right), \qquad
    T_{div, \, \textbf{ii}}(z_h) = \frac{1+1/\nu_1}{\pi z_h} \exp\left(-\,\frac{3 c z^n_h}{2}\right).
\ee
We then multiply $s(z_h)$ and $T(z_h)$ to obtain an expression independent of $z_h$ 
\be
    s_{div}(z_h) \ T_{div, \, \textbf{i}}(z_h) = \frac{1}{2\pi L}, \qquad s_{div}(z_h) \ T_{div, \, \textbf{ii}}(z_h) = \frac{1+1/\nu_1}{4\pi L},
\ee
therefore the entropy density can be expressed as a function of the temperature:
\be\label{6:221}
    s_{div, \, \textbf{i}}(T) = \frac{1}{2\pi\,L}\cdot\frac{1}{T}, \qquad s_{div, \, \textbf{ii}}(T) = \frac{1+1/\nu_1}{4\pi\,L}\cdot\frac{1}{T},
\ee
so that the third law of thermodynamics clearly fails:
\be
    s_{div, \, \textbf{i}}(T), \ s_{div, \, \textbf{ii}}(T) \sim \frac{1}{T} \to \infty \quad\text{   as   }\quad T \to 0.
\ee

\paragraph{Solutions with a Converging Integral}
$\,$\\
\par 
The analysis for the convergent solutions is more complicated. Let us compute the temperature:
\begin{gather}
    T_{conv, \, \textbf{i}}(z_h) = \frac{1}{4\pi}\,\frac{z_h^3\,\exp\left(-\frac{3 c z^n_h}{2}\right)}{\int^{z_h}_0 \xi^3\,\exp\left(-\frac{3 c \xi^n}{2}\right)\,d\xi}, \quad
    T_{conv, \, \textbf{ii}}(z_h) = \frac{1}{4\pi}\,\frac{z_h^{1+2/\nu_1}\,\exp\left(-\frac{3 c z^n_h}{2}\right)}{\int^{z_h}_0 \xi^{1+2/\nu_1}\,\exp\left(-\frac{3 c \xi^n}{2}\right)\,d\xi}.
\end{gather}

These formulas are similar to the one for Model I, see \eqref{4:15}, and the analysis is similar, see Section~\ref{ModelI-3rdlaw}. Temperature can only tend to zero at $z_h \to +\,\infty$. For $c=0$ we have
\begin{gather}
  s_{conv, \, \textbf{i}}(T) =
  \frac{1}{4}(\pi L\,T)^3, \qquad
  s_{conv, \, \textbf{ii}}(T) =
  \frac{1}{4}\left(\frac{2\pi L\,T}{1+\frac{1}{\nu_1}}\right)^{1+\frac{2}{\nu_1}},
\end{gather}
so that the third law holds for \textbf{i}, while we additionally need
\be
    1 + \frac{2}{\nu_1} > 0
\ee
for the third law to hold for \textbf{ii}.

\begin{figure}[b!]
  \centering
  \includegraphics[width=0.49\textwidth]{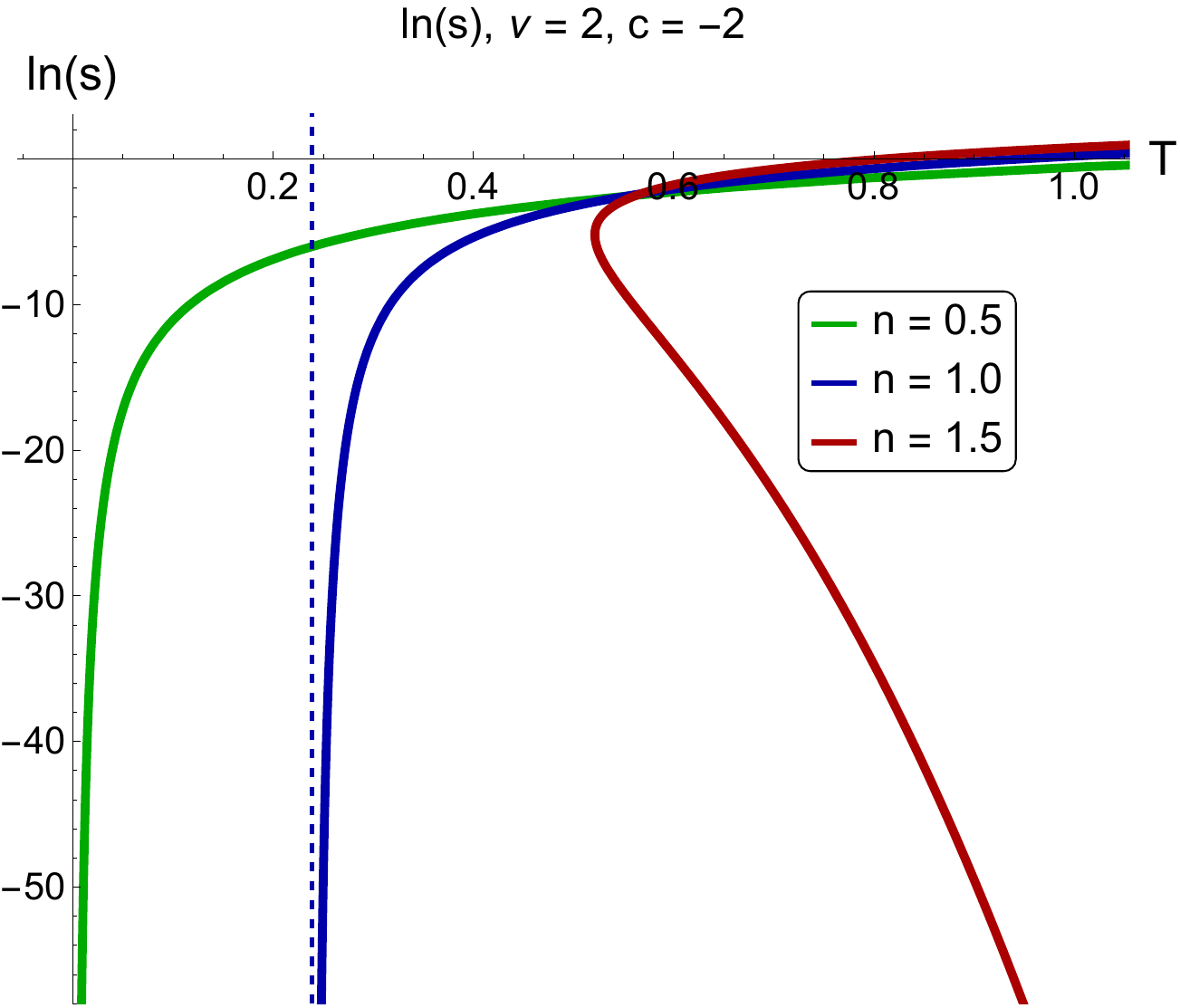} \hfill
  \includegraphics[width=0.49\textwidth]{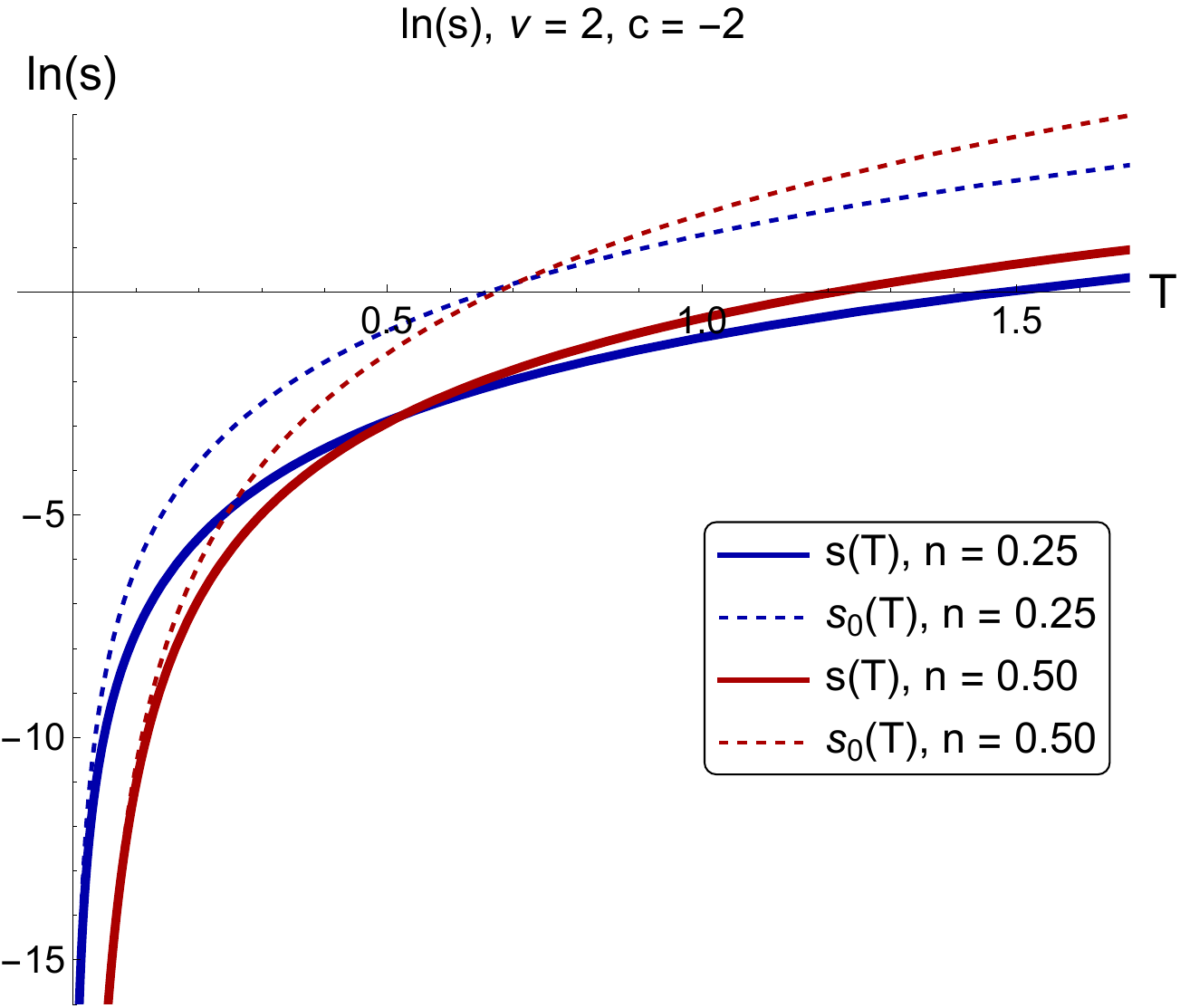} \\
  A \hspace{200pt} B
  \caption{Plots for the entropy density $s(T)$ for the cases \textbf{ii} and \textbf{iii}, $\nu_1=\nu_2=2$ and $c=-\,2$. The plot (A) demonstrates that for some $n$ values the temperature $T$ might never tend to zero: for instance, for $n=1$ the temperature $T$ tends to a finite value for large $z_h$, which is highlighted with a vertical asymptotic. The plot (B) compares the leading term $s_0(T)$, see the asymptotic expansion \eqref{5:31}, with the entropy density $s(T)$ itself. \\ \,}
  \label{ModelII-1}
  \centering
  \includegraphics[width=0.5\textwidth]{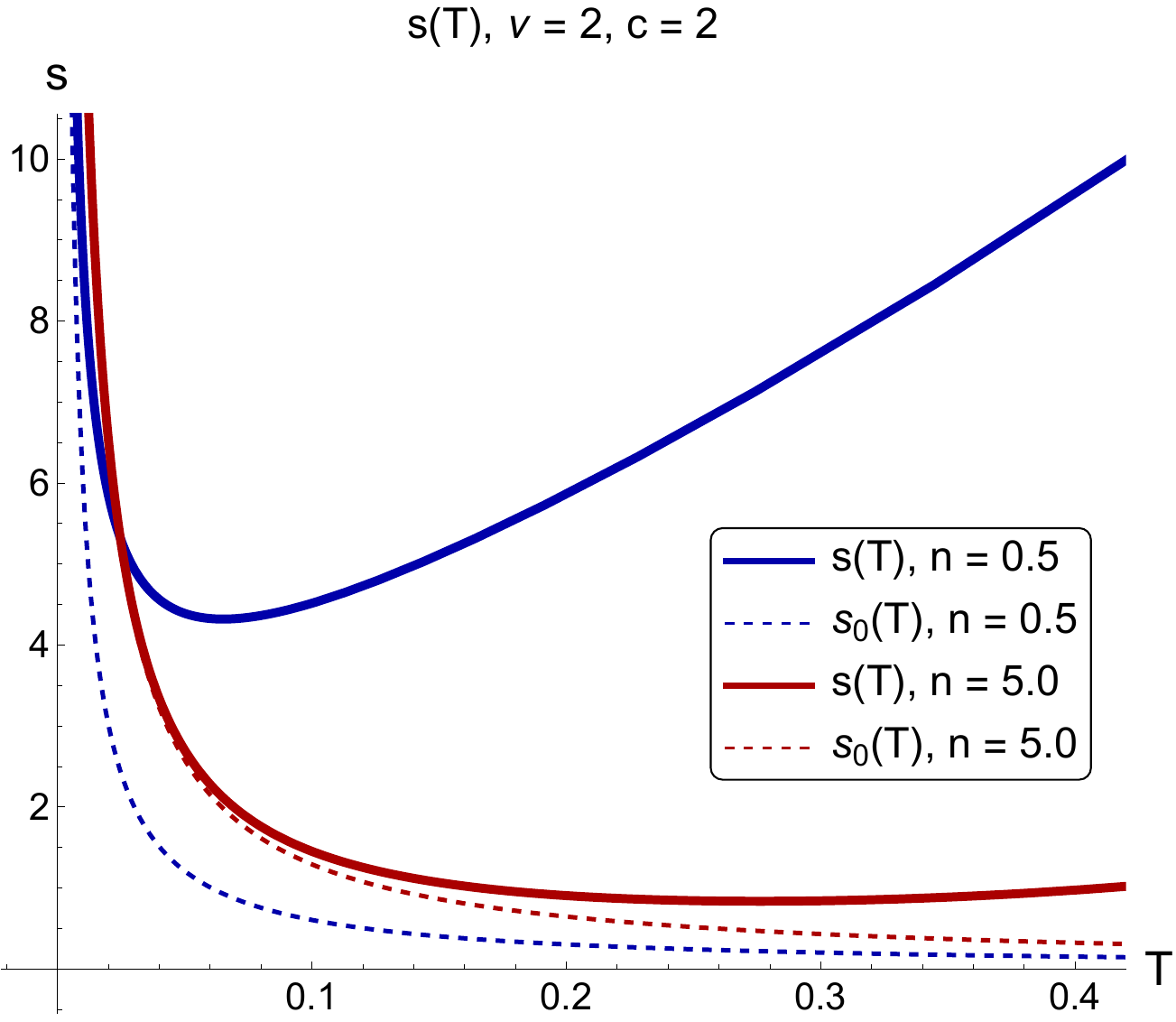}
  \caption{Plot for the entropy density $s(T)$ for the cases \textbf{ii} and \textbf{iii}, $\nu_1=\nu_2=2$ and $c=2$, when $s(T)$ blows up at small temperatures. The plot compares the leading term $s_0(T)$, see the asymptotic expansion \eqref{5:33}, with the entropy density $s(T)$ itself.
  }
  \label{ModelII-2}
\end{figure}

Similarly to Section~\ref{ModelI-3rdlaw}, the case $c>0$ leads the following:
\be
    s_{conv, \, \textbf{i}/\textbf{ii}}(z_h) \to +\,\infty, \quad T_{conv, \, \textbf{i}/\textbf{ii}}(z_h) \to 0
    \quad \mbox{as} \quad z_h \to +\,\infty.
\ee
But for the case $c < 0$, the L'Hôpital rule yields a nontrivial result:
\bea
    \lim_{z_h\to+\infty}
    T_{conv, \, \textbf{i}}(z_h) &=&
    \lim_{z_h\to+\infty}
    \frac{1}{4\pi}\left(\frac{3}{z_h} - \frac{3 c n}{2} z^{n-1}_h \right), \\
    \lim_{z_h\to+\infty}
    T_{conv, \, \textbf{ii}}(z_h) &=& 
    \lim_{z_h\to+\infty}
    \frac{1}{4\pi}\left(\frac{1+2/\nu_1}{z_h} - \frac{3 c n}{2} z^{n-1}_h \right),
\eea
from which we conclude, keeping in mind that the entropy density decays to zero \eqref{6:18}, that the third law holds if and only if $0<n<1$, because otherwise the temperature does not tend to zero for large $z_h$, see Fig.~\ref{ModelII-1}.

Asymptotic formulas have been obtained: 
\begin{itemize}
  \item in the regime $c<0$, $0<n<1$ that supports the third law for both \textbf{i} and \textbf{ii}:
  \begin{gather}
    s_{conv, \, \textbf{i}}(T) =
    \frac{L^3\,e^{\frac{4-n}{n-1}}}{4} 
    \left(\frac{3 n |c|}{8\pi\,T}\right)^{-\frac{3}{1-n}} 
    \exp\left(-\frac{3|c|}{2} \left(\frac{3 n |c|}{8\pi\,T}\right)^{n/(1-n)}\right) 
    \left[1 + O\left(T^{\,n/(1-n)}\right)\right],
    \\
    s_{conv, \, \textbf{ii}}(T) =
    \frac{L^{1+\frac{2}{\nu_1}}\,e^{\frac{2+2/\nu_1-n}{n-1}}}{4}
    \left(\frac{3 n |c|}{8\pi\,T}\right)^{-\frac{1+2/\nu_1}{1-n}}
    \exp\left(-\frac{3|c|}{2} \left(\frac{3 n |c|}{8\pi\,T}\right)^{n/(1-n)}\right)  
    \left[1 + O\left(T^{\,n/(1-n)}\right)\right]; \label{5:31}
  \end{gather}
  \item the regime $c>0$ leads to the entropy blowing up:
  \bea
    s_{conv, \, \textbf{i}}(T) &=& \frac{L^3}{16\pi} \frac{n\,(3c/2)^{\frac{4}{n}}}{\Gamma\left(\frac{4}{n}\right)} \cdot \frac{1}{T} +  O\left(\left(\ln\frac{1}{T}\right)^{(1-n)/n}\right), \label{5:32}
    \\
    s_{conv, \, \textbf{ii}}(T) &=& \frac{L^{1+\frac{2}{\nu_1}}}{16\pi} \frac{n\,(3c/2)^{\frac{2+2/\nu_1}{n}}}{\Gamma\left(\frac{2+2/\nu_1}{n}\right)} \cdot \frac{1}{T} +  O\left(\left(\ln\frac{1}{T}\right)^{(1-n)/n}\right). \label{5:33}
  \eea
\end{itemize}

\vspace{-10pt}

We plot the leading terms of the asymptotic formulas \eqref{5:31} and \eqref{5:33} in Fig.~\ref{ModelII-1} and Fig.~\ref{ModelII-2}, respectively. In Table~\ref{tab::m2-3rdlaw} we summarize all the results for Model II.

\begin{table}[h]
\centering
\begin{tabular}{|cccc|}
\hline
\multicolumn{4}{|c|}{\textbf{i}} \\ \hline
\multicolumn{1}{|c|}{\multirow{2}{*}{Divergent}} & \multicolumn{3}{c|}{Convergent} \\ \cline{2-4} 
\multicolumn{1}{|c|}{} & \multicolumn{1}{c|}{$c>0$} & \multicolumn{1}{c|}{$c=0$} & $c<0$ \\ \hline
\multicolumn{1}{|c|}{fails} & \multicolumn{1}{c|}{fails} & \multicolumn{1}{c|}{holds} & \begin{tabular}{@{}c@{}}
    holds if $0<n<1$ \\
    otherwise $T=0$ never approached
  \end{tabular} \\ \hline
\multicolumn{1}{|c|}{$\displaystyle s \sim \frac{1}{T}$} & \multicolumn{1}{c|}{$\displaystyle s \sim \frac{1}{T}$} & \multicolumn{1}{c|}{$s\sim T^{\,3}$} & \multicolumn{1}{l|}{$s\sim T^{\frac{3}{1-n}}
    \exp\left[-\frac{3 |c|}{2}
    \left(\frac{8\pi \, T}{3 |c| n}\right)^{n/(n-1)}\right]$} \\ \hline
\multicolumn{4}{c}{} \\ \hline
\multicolumn{4}{|c|}{\textbf{ii} and \textbf{iii}} \\ \hline
\multicolumn{1}{|c|}{\multirow{2}{*}{Divergent}} & \multicolumn{3}{c|}{Convergent} \\ \cline{2-4} 
\multicolumn{1}{|c|}{} & \multicolumn{1}{c|}{$c>0$} & \multicolumn{1}{c|}{$c=0$} & $c<0$ \\ \hline
\multicolumn{1}{|c|}{fails} & \multicolumn{1}{c|}{fails} & \multicolumn{1}{c|}{holds if $1+\frac{2}{\nu_1}>0$} & \begin{tabular}{@{}c@{}}
    holds if $0<n<1$ \\
    otherwise $T=0$ never approached
  \end{tabular} \\ \hline
\multicolumn{1}{|c|}{$\displaystyle s \sim \frac{1}{T}$} & \multicolumn{1}{c|}{$\displaystyle s \sim \frac{1}{T}$} & \multicolumn{1}{c|}{$s\sim T^{\,1+\frac{2}{\nu_1}}$} & \multicolumn{1}{l|}{$s\sim T^{\frac{1+\frac{2}{\nu_1}}{1-n}}
    \exp\left[-\frac{3 |c|}{2}
    \left(\frac{8\pi \, T}{3 |c| n}\right)^{n/(n-1)}\right]$} \\ \hline
\end{tabular}
\caption{Third law of thermodynamics satisfaction for Model II. ``Divergent'' and ``Convergent'' refer to the type of the solution obtained, see \eqref{model3-gSol1}--\eqref{model3-gSol2}. The table is broken up into two parts with headers \textbf{i}, \textbf{ii} and \textbf{iii} representing which field configuration case the part of the table concerns. Each part should be read from the top down.}
\label{tab::m2-3rdlaw}
\end{table}

\subsection{Null Energy Condition}

For NEC in the case \textbf{ii} we can use the intermediate result \eqref{4:30} of Model I consideration in Section~\ref{sec:ModelINEC}, as these models are the same in terms of the dimension of the space and the  set of the 2-form fields. For the other cases, the resulting conditions are similar:
\bea
    \textbf{i}\qquad&&\phi'^2 \ge 0, \quad f_2(\phi) \ge 0, \quad f_3(\phi) \ge 0,
    \\
    \textbf{ii}\qquad&&\phi'^2 \ge 0, \quad f_1(\phi) \ge 0, \quad f_3(\phi) \ge 0,
    \\
    \textbf{iii}\qquad&&\phi'^2 \ge 0, \quad f_1(\phi) \ge 0, \quad f_2(\phi) \ge 0.
\eea

First, we impose $f_i\big(\phi(z_h)\big)\ge0$, which removes the solutions with divergent integrals and further constraints anisotropy parameters $\nu_1$, $\nu_2$:
\bea
    \textbf{i}\quad&& g_\textbf{i}(z)=
    \displaystyle 1-\frac{I(z)}{I(z_h)}, \ \mbox{where} \ I(z)=\int^{z}_0 \xi^3\,\exp\left(-\,\frac{3}{2}\,c\,\xi^n\right) d\xi, \ \nu_1=\nu_2=1,
    \\
    &&\mbox{for} \ c=0 \ \longrightarrow \ g_\textbf{i}(z)=\displaystyle 1-\left(\frac{z}{z_h}\right)^4, \ \nu_1=\nu_2=1;
    \\
    \textbf{ii}\quad&& g_\textbf{ii}(z)=
    \displaystyle 1-\frac{I(z)}{I(z_h)}, \mbox{where} \ I(z)=\int^{z}_0 \xi^{1+\frac{2}{\nu_1}}\,\exp\left(-\frac{3}{2}\,c\,\xi^n\right) d\xi, \ \nu_1=\nu_2, \ 
    -\,1 < \frac{1}{\nu_1} \le 1, \nn \\
    \\
    &&\mbox{for} \ c=0 \ \longrightarrow \ g_\textbf{ii}(z)=\displaystyle 1-\left(\frac{z}{z_h}\right)^{2+\frac{2}{\nu_1}}, \ \nu_1=\nu_2, \ -\,1 < \frac{1}{\nu_1} \le 1.
\eea

We see that in the case \textbf{i} anisotropy cannot be supported, when the warp factor is chosen to be $\fb(z) = e^{cz^n}$, $n > 0$, $c\in\mathbb{R}$, which includes the trivial warp factor. For \textbf{ii} and \textbf{iii} the anisotropy parameters have to be the same $\nu_1=\nu_2$, though they do not necessarily need to be equal to $1$.

The EOM for $\phi'(z)^2$ and the corresponding $f_i\big(\phi(z)\big)$ then are
\bea
    \textbf{i} \quad &&
    \begin{cases}
        \phi'(z)^2 
        = \cfrac{1}{2 z^2} \, 3 c n \big(c n z^{n} - 2 \left(1+n\right) \big) \, z^n,
        \\
        f_3\big(\phi(z)\big)=f_2\big(\phi(z)\big) \equiv 0,
    \end{cases}
    \\
    \textbf{ii} \quad &&
    \begin{cases}
        \phi '(z)^2 &= \cfrac{1}{2 z^2} \, \Bigg( 3cn(cnz^n-2(1+n))z^n + \cfrac{8}{\nu_1} \, \bigg( 1-\cfrac{1}{\nu_1} \bigg) \Bigg),
        \\ \\
        f_1\big(\phi(z)\big)&=\displaystyle \left(1-\frac{1}{\nu_1}\right)\frac{2 e^{c z^n}}{L^2 q_1^2 I(z_h)}\left(\frac{L}{z}\right)^{\frac{4}{\nu_1}} 
        \bigg[z^{2+2/\nu_1}\,\exp\left(-\,\frac{3}{2}c z^2\right) +
        \\ 
        &\phantom{=}+\displaystyle\left(-\,\frac{3 c z^n}{2} + 2 \left(1+\frac{1}{\nu_1}\right)\right) \int^{z_h}_z \xi^{1+2/\nu_1}\,\exp\left(-\,\frac{3}{2}c\xi^2\right)\,d\xi \bigg],
        \\
        f_3\big(\phi(z)\big) & \equiv 0.
    \end{cases}. \label{6:24}
\eea

Further, we need to determine which values of the parameters ($c\in\mathbb{R}$, $n>0$, $\nu_1=\nu_2$, and possibly other ones) make the model satisfy NEC.

Case \textbf{i}.
\begin{itemize}
    \item If $c=0$, then $\phi '(z)^2 \equiv 0$, thus all NEC are satisfied.
    \item If $c<0$, then (we make the substitution $c \to -\,|c|$ to make non-negativity more apparent):
    
    \be
        \phi'(z)^2 = \cfrac{3 |c| n}{z^2} \left(\cfrac{|c|nz^{n}}{2} + \left(1+n\right)\right) z^n \ge 0 \ \mbox{for} \ n>0,
    \ee
    so all NEC are satisfied.
    \item If $c > 0$, then for $n > 0$ we have the following behavior:
    \bea
        \phi'(z)^2 < 0 && \text{   for    } z \in \left(0; \left(\frac{2+2 n}{n c}\right)^{1/n}\right),
        \\
        \phi'(z)^2 \ge 0 && \text{   for    } z \ge \left(\frac{2+2 n}{n c}\right)^{1/n},
    \eea
    thus violating NEC for any solution as $\phi'(z)^2 < 0$ in the vicinity of $z=0$.
\end{itemize}

Case \textbf{ii}.
\begin{itemize}
    \item If $\nu_1 = 1$, then $f_1\big(\phi(z)\big) \equiv 0$ and the case reduces to that of \textbf{i}, whose analysis is presented above.
    \item If $\nu_1 \ne 1$, on the allowed range of $\nu_1$, then $f_1\big(\phi(z)\big) \ge 0$ on the interval for all allowed $\nu_1$ and all $c$. For brevity, we only outline the proof of this fact: removing the positive factors before the square bracket in the expression for $f_1\big(\phi(z)\big)$ in \eqref{6:24}, one can show that the coupling function is positive at the ends of the interval $(0; z_h)$, and the derivative of the expression in the square brackets does not evaluate to zero for any $z\in(0; z_h)$, thus ensuring that $f_1\big(\phi(z)\big)$ is positive on the interval $(0; z_h)$.
    
    If we set $c=0$, then:
    \bea
        f_1\big(\phi(z)\big)&=&
        \frac{4}{L^2\,q_1^2}\left(1-\frac{1}{\nu_1^2}\right)\left(\frac{L}{z}\right)^{4/\nu_1},
        \\
        \phi '(z)^2 &=& \displaystyle \frac{4}{\nu_1\,z^2}\left(1-\frac{1}{\nu_1}\right).
    \eea
    In such case, for the condition on the dilaton to be satisfied, we need to additionally impose:
    \be
        \nu_1 \ge 1.
    \ee
    
    Assuming $n>0$, $c \ne 0$, we analyze the condition on the dilaton:
    \begin{itemize}
        \item if $\nu_1 \in (-\infty;-1)$, NEC is violated in the vicinity of $z=0$:
        \bea
            &&\phi '(z)^2 >0 \text{  for  } z\in(z_0, \infty), \\
            &&\phi '(z)^2 <0 \text{  for  } z\in(0, z_0),
        \eea
        where
        \be
            z_0 = \Biggl(\frac{1}{\sqrt{3} c n}
            \left(\sqrt{3} \left(n+1\right) -\text{sgn}(c) \, \frac{1}{\nu_1}\sqrt{3 \nu_1^2 (n+1)^2-8 \nu_1+8}\right)\Biggr)^{1/n} > 0;
        \ee

        \item if $\nu_1 \in (1; +\infty)$, there are two cases depending on the sign of $c$:
        \begin{itemize}
            \item  if $c>0$, we have
            \bea
                &&\phi '(z)^2 >0 \ \mbox{for} \ z\in \big(0; \text{min}(z_1,\,z_2)\big)\cup\big(\text{max}(z_1,\,z_2), +\infty \big),
                \\
                &&\phi '(z)^2 <0 \ \mbox{for} \ z\in \big(\text{min}(z_1,\,z_2); \,\,\text{max}(z_1,\,z_2)\big),
            \eea
            where
            \be
                z_{1,\,2} = \left({\frac{3(1+n) \pm \sqrt{3}\sqrt{\frac{8-8 \nu_1 +3 (1+n)^2 \nu_1^2}{\nu_1^2}}}{3\,c\,n}}\right)^{1/n} > 0,
            \ee
            thus only allowing the solution to have $z_h \le \text{min}(z_1,\,z_2)$;
        
            \item if $c < 0$, we have
            \be
                \phi '(z)^2 >0 \text{  for  } z\in (0; +\infty),
            \ee
            thus allowing for a solution with arbitrary horizon value $z_h$, meaning NEC is satisfied for arbitrary $z_h$.
        \end{itemize}

    \end{itemize}
\end{itemize}

We can sum all the results up in a table (Tab.~\ref{tab::m2-NEC}).
\begin{table}[H]
\centering
\begin{tabular}{|ccc|ccc|}
\hline
\multicolumn{3}{|c|}{\textbf{i}} & \multicolumn{3}{c|}{\textbf{ii} and \textbf{iii}} 
\\ \hline
\multicolumn{3}{|c|}{%
  \begin{tabular}{@{}c@{}}
    $\nu_1=\nu_2=1$ \\
    otherwise NEC violated
  \end{tabular}
} & 
\multicolumn{3}{c|}{%
  \begin{tabular}{@{}c@{}}
    $\nu_1=\nu_2\ge1$ \\
    otherwise NEC violated
  \end{tabular}
} 
\\ \hline
\multicolumn{1}{|c|}{\multirow{2}{*}{$c=0$}} & \multicolumn{1}{c|}{\multirow{2}{*}{$c<0$}} & \multirow{2}{*}{$c>0$} & \multicolumn{3}{c|}{\begin{tabular}{@{}c@{}}
    $\nu_1=\nu_2=1 \quad \Rightarrow \quad \textbf{i}$ \\
    for $\nu_1=\nu_2>1$ depends on $c$
  \end{tabular}} 
\\ \cline{4-6} 
\multicolumn{1}{|c|}{} & \multicolumn{1}{c|}{} &  & \multicolumn{1}{c|}{$c=0$} & \multicolumn{1}{c|}{$c<0$} & $c>0$ \\ \hline
\multicolumn{1}{|c|}{\begin{tabular}{@{}c@{}}
    NEC \\
    satisfied
  \end{tabular}} &
\multicolumn{1}{c|}{\begin{tabular}{@{}c@{}}
    NEC \\
    satisfied
  \end{tabular}} & 
\begin{tabular}{@{}c@{}}
    NEC \\
    violated
  \end{tabular}  &
\multicolumn{1}{c|}{\begin{tabular}{@{}c@{}}
    NEC \\
    satisfied
  \end{tabular}} &
\multicolumn{1}{c|}{\begin{tabular}{@{}c@{}}
    NEC \\
    satisfied
  \end{tabular}} &
\begin{tabular}{@{}c@{}}
    NEC satisfied if \\
    $z_h \le \text{min}(z_1,\,z_2)$
  \end{tabular}\\ \hline
\end{tabular}
\caption{NEC satisfaction for Model II. Here we summarize the NEC for the three cases \textbf{i}, \textbf{ii}, and \textbf{iii}. The table should be read from the top: first, there is a restriction on $\nu_1$, $\nu_2$, and then, the cases diverge depending on the sign of $c$. An additional restriction, that is, an upper boundary on the horizon, $z_h$, appears.}
\label{tab::m2-NEC}
\end{table}

\subsection{NEC and the Third Law of Thermodynamics}\label{Model3-NEC-3Law}

The solutions with a divergent integral, see \eqref{model3-gSol1}--\eqref{model3-gSol2}, fall out of consideration for both the third law of thermodynamics and the NEC. 

For the other solutions, the case $c=0$ satisfies the NEC if and only if 
\bea
    c = 0 \qquad 
    \begin{cases}
        \textbf{i} \phantom{i, iii } \quad \nu_1=\nu_2=1, \\
        \textbf{ii, iii} \quad \nu_1=\nu_2 \ge 1,
    \end{cases} 
\eea
and this regime also supports the third law of thermodynamics, see Table~\ref{tab::m2-3rdlaw}. However, the NEC is too restrictive in this case because, for example, \textbf{i} satisfies the third law for all $\nu_1, \, \nu_2$, which are subject to the constraint in \eqref{model3-gSol1}.

The case $c>0$ violates NEC for \textbf{i} and leads to an upper boundary for \textbf{ii} and \textbf{iii}, thus making it impossible for $T(z_h)$ to attain arbitrarily small values:
\be
    c > 0 \qquad \textbf{ii},\,\textbf{iii} \quad T(z_h) \not\to 0,
\ee
and the third law of thermodynamics, when considered on its own, is not satisfied in the regime $c>0$, see Table~\ref{tab::m2-3rdlaw}, because the entropy density diverges as the temperature decays to zero.

The case $c < 0$ satisfies the NEC if and only if
\bea\label{6:58}
    c < 0 \qquad 
    \begin{cases}
        \textbf{i} \phantom{i, iii } \quad \nu_1=\nu_2=1, \\
        \textbf{ii, iii} \quad \nu_1=\nu_2 \ge 1,
    \end{cases} 
\eea
while the third law of thermodynamics only requires the following constraint for it to be satisfied, see Table~\ref{tab::m2-3rdlaw}:
\be 
    0 < n < 1.
\ee

Therefore, the NEC and the third law are independent conditions in this case, though they do have an intersection: in \eqref{6:58} add $0<n<1$ to both cases

To sum it up, here are all the cases that support both the NEC and the third law of thermodynamics:
\bea
    c = 0 &\qquad&
    \begin{cases}
        \textbf{i} \phantom{i, iii } \quad \nu_1=\nu_2=1, \\
        \textbf{ii, iii} \quad \nu_1=\nu_2 \ge 1,
    \end{cases} \label{5:60} \\
    c < 0 &\qquad&
    \begin{cases}
        \textbf{i} \phantom{i, iii } \quad \nu_1=\nu_2=1, \ \ 0 < n < 1, \\
        \textbf{ii, iii} \quad \nu_1=\nu_2 \ge 1, \ \ 0 < n < 1.
    \end{cases} \label{5:61} 
\eea

\section{Model III: Magnetic 2- and 3-form Fields in D = 6 with Two Lifshitz-Type Anisotropies}\label{sec:model3}

Let us now consider a $D = 6$ model that includes one 2-form field and one 3-form field, both taken in magnetic ansatz:
\begin{gather}
    S = \int d^{\,6} x \, \sqrt{- g_6} \left[R - \cfrac{1}{4} \, \ff_3(\phi) \fF_3^2 - \cfrac{1}{12} \, h_1(\phi) H_1^2
    - \cfrac{1}{2} \, \partial_{\mu} \phi \, \partial^{\,\mu} \phi - V(\phi)\right],
    \\
    \fF_3=\fq_3 \, dx^3 \wedge dx^4, \qquad 
    H_1=Q_1 \, dx^2 \wedge dx^3 \wedge dx^4.
\end{gather}

We specify the following metric ansatz
\begin{gather}
  ds^2=\cfrac{L^2\fb(z)}{z^2} \left[
    - \, g(z) dt^2 + dx^2_1
    + {\left( \cfrac{z}{L} \right)^{2-\frac{2}{\nu_1}}}\!dx^2_2
    + {\left( \cfrac{z}{L} \right)^{2-\frac{2}{\nu_2}}}\!dx^2_3
    + {\left( \cfrac{z}{L} \right)^{2-\frac{2}{\nu_2}}}\!dx^2_4
    + \cfrac{dz^2}{g(z)} \right].
  \label{kappa-nu-metric}
\end{gather}

Following the developed procedure for solving such models, see Section~\ref{sec:quadratures}, we use the model's sign Table~\ref{Table-6D-MM-34} to obtain the solutions for all unknown functions. First, for the blackening function $g(z)$ we have
\be
  \displaystyle g(z) = C_1 \int^{z_h}_z \cfrac{\xi^{1+\frac{1}{\nu_1}
  + \frac{2}{\nu_2}}}{\fb^2(\xi)} \ d\xi,
\ee
where $C_1$ is a constant, fixed by the remaining boundary condition $g(0)=1$. The trivial warp factor $\fb(z) \equiv 1$ and the boundary condition at $z = 0$ lead to a restriction on the anisotropy parameters $\nu_1$, $\nu_2$. The blackening function solution is
\be
  g(z) = 1 - \bigg(\frac{z}{z_h}\bigg)^{2+\frac{1}{\nu_1}+\frac{2}{\nu_2}},
  \quad 
  2+\frac{1}{\nu_1}+\frac{2}{\nu_2}>0.
  \label{gz6Dbb1}
\ee

The trivial warp factor allows us to find explicit expressions for the coupling functions' solutions as functions of the dilaton. This contrasts to the general case, when $f_i(\phi)$ are obtained only implicitly (by finding $f_i=f_i\big(\phi(z)\big)$ and $\phi=\phi(z)$, and then plotting one against the other to visualize $f_i=f_i(\phi)$). The final results are
\bea
  \displaystyle g(z) &=& 1 - \bigg(\frac{z}{z_h}\bigg)^{2+\frac{1}{\nu_1}+\frac{2}{\nu_2}},
  \quad 
  2+\frac{1}{\nu_1}+\frac{2}{\nu_2}>0,
  \\
  \phi(z)
  &=&\phi_0\pm\log\bigg(\frac{z}{L}\bigg)\,\sqrt{\frac{2}{\nu_1^2}\,(\nu_1-1)+\frac{4}{\nu_2^2}\,(\nu_2-1)}, \label{6D-phi-sol} \\
  h_1(\phi)
  &=&\frac{2}{L^2\,Q_1^2}
  \left(1-\frac{1}{\nu_1}\right)
  \left(2+\frac{1}{\nu_1}+\frac{2}{\nu_2}\right)
  \exp\Bigg[
  \bigg(\frac{2}{\nu_1}+\frac{4}{\nu_2}\bigg)
  \,\frac{\mp(\phi-\phi_0)}{\sqrt{\frac{2}{\nu_1^2}(\nu_1-1)+\frac{4}{\nu_2^2}(\nu_2-1)}}\Bigg], \nn \\ \label{Model2-h1sol} 
  \\ 
  \ff_3(\phi)
  &=&\frac{2}{L^2\,\fq_3^2}
  \left(\frac{1}{\nu_1}-\frac{1}{\nu_2}\right)
  \left(2+\frac{1}{\nu_1}+\frac{2}{\nu_2}\right)
  \exp\Bigg[\frac{4}{\nu_2}\,\frac{\mp(\phi-\phi_0)}{\sqrt{\frac{2}{\nu_1^2}(\nu_1-1)+\frac{4}{\nu_2^2}(\nu_2-1)}}\Bigg], \label{Model2-f3sol} \\
  V(\phi)
  &\equiv&-\,\frac{1}{L^2}\bigg(4+\frac{4}{\nu_1}+\frac{6}{\nu_2}+\frac{1}{\nu_1^2}+\frac{2}{\nu_2^2}+\frac{3}{\nu_1\nu_2}\bigg). \label{model2-Vsol}
\eea
In particular, this model reproduces the isotropic black hole in $AdS_{6}$ \cite{Ramallo:2013bua}, for which one has to set $\nu_1=\nu_2=1$:
\be
    \displaystyle ds^2=\cfrac{L^2}{z^2} \left[- \, g(z) dt^2 + d\vec{x}^2+\frac{dz^2}{g(z)} \right], \quad 
    g(z)=1-\left(\frac{z}{z_h}\right)^5.
\ee

This also ``switches off'' the fields of the 2- and 3-forms (see factors $(1-1/\nu_1)$ and $(1/\nu_1-1/\nu_2)$ in \eqref{Model2-h1sol} and \eqref{Model2-f3sol}, respectively). Additionally, the dilaton potential reduces to the double value of the cosmological constant of the $AdS_{6}$ space:
\be
    h_1(\phi)=\ff_3(\phi) \equiv 0, \quad \phi(z) \equiv \phi_0, \quad
    \displaystyle V(\phi) \equiv -\,\frac{20}{L^2}.
\ee

\subsection{Third Law of Thermodynamics}

Temperature and entropy density read
\be
    \displaystyle T(z_h)=\frac{2+\frac{1}{\nu_1}+\frac{2}{\nu_2}}{4\,\pi\,z_h}, 
    \qquad
    s(z_h)=\frac{1}{4}\left(\frac{L}{z_h}\right)^{1+\frac{1}{\nu_1}+\frac{2}{\nu_2}},
\ee
thus allowing us to get an expression for the entropy density as a function of the temperature
\be
    s(T)=4^{\frac{1}{\nu_1}+\frac{2}{\nu_2}}
    \left(\frac{L\,\pi\,T}{2+\frac{1}{\nu_1}+\frac{2}{\nu_2}}\right)^{1+\frac{1}{\nu_1}+\frac{2}{\nu_2}}
    \sim
    T^{\,1+\frac{1}{\nu_1}+\frac{2}{\nu_2}}. \label{6:14}
\ee
In consequence, the third law of thermodynamics holds if and only if
\be
    1+\frac{1}{\nu_1}+\frac{2}{\nu_2} > 0,
\ee
which must be joined by the constraint  in \eqref{gz6Dbb1}:
\be
    2 + \frac{1}{\nu_1} + \frac{2}{\nu_2} > 0.
\ee
In Fig.~\ref{kappanuplane} we graphically show the regions of the $(\nu_1,\,\nu_2)$ plane, where the third law~holds.

\subsection{Null Energy Condition}
For the NEC we perform the same steps as in the previous model. Interestingly, in this case the NEC does not lead to both coupling functions being non-negative independently:
\begin{equation}
    \phi'^2 \ge 0, \quad
    h_1(\phi) \ge 0, \quad
    \displaystyle\frac{h_1(\phi) \, Q_1^2}{g_{x_2x_2}}+\displaystyle \ff_3(\phi) \, \fq_3^2\ge 0.
    \label{5.15}
\end{equation}

Substitution of the solutions \eqref{6D-phi-sol}--\eqref{Model2-f3sol} into the NEC \eqref{5.15} leads to inequalities for the Lifshitz parameters $(\nu_1, \nu_2)$. Careful handling of the NEC gives the following conditions on the anisotropy parameters:
\bea
    &\displaystyle \frac{1}{\nu_1}\,\left(1-\frac{1}{\nu_1}\right)+\frac{2}{\nu_2}\,\left(1-\frac{1}{\nu_2}\right) \ge 0,&
    \label{5:16}
    \\
    &\displaystyle \left(1-\frac{1}{\nu_1}\right)
    \left(2+\frac{1}{\nu_1}+\frac{2}{\nu_2}\right) \ge 0,
    \quad 
    \left(1-\frac{1}{\nu_2}\right)
    \left(2+\frac{1}{\nu_1}+\frac{2}{\nu_2}\right) \ge 0,&
    \label{5:17}
\eea
which must be joined by the constraint  in \eqref{gz6Dbb1}:

\be
    2 + \frac{1}{\nu_1} + \frac{2}{\nu_2} > 0.
    \label{5:18}
\ee

\begin{figure}[b!]
  \centering
  \includegraphics[width=0.49\textwidth]{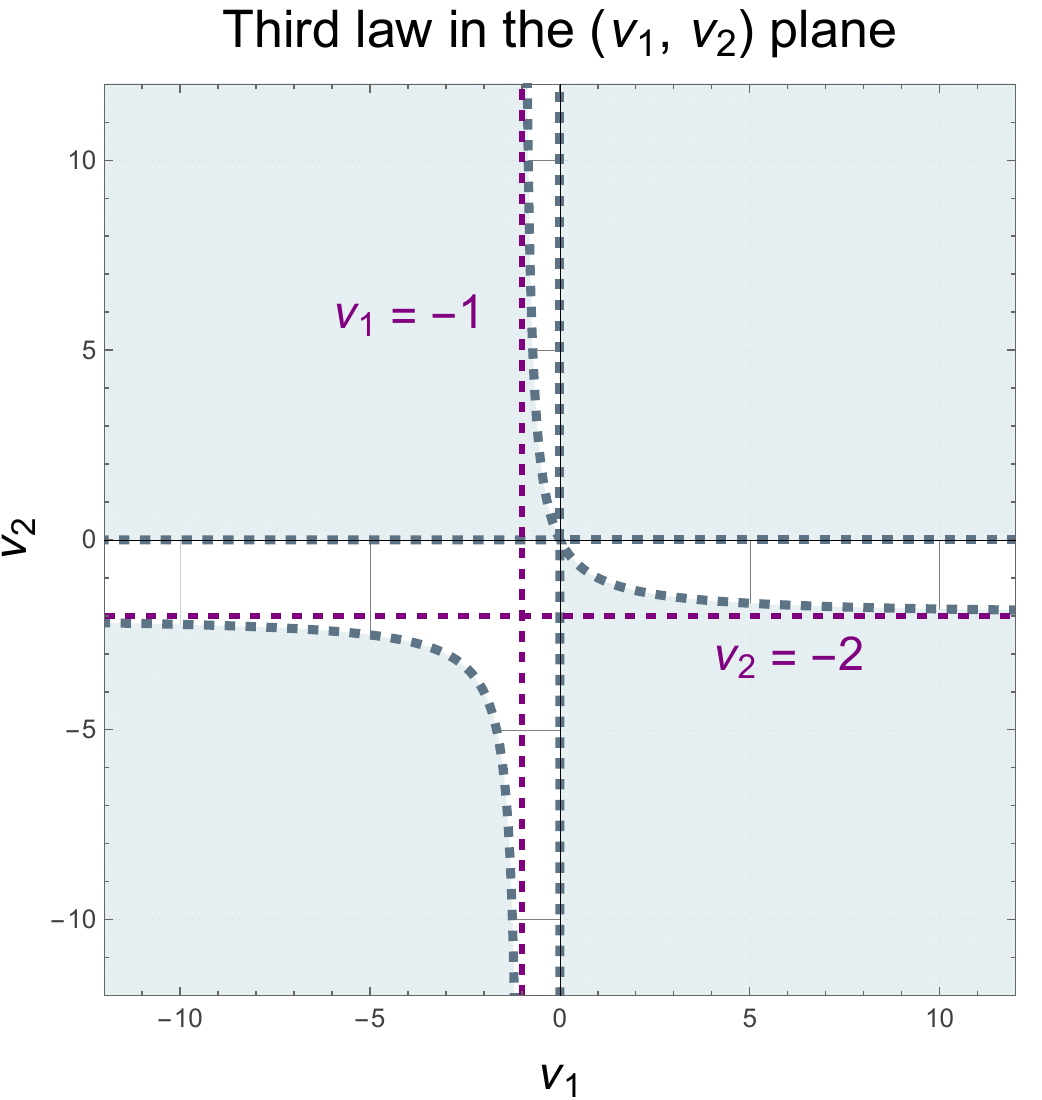} \hfill
  \includegraphics[width=0.49\textwidth]{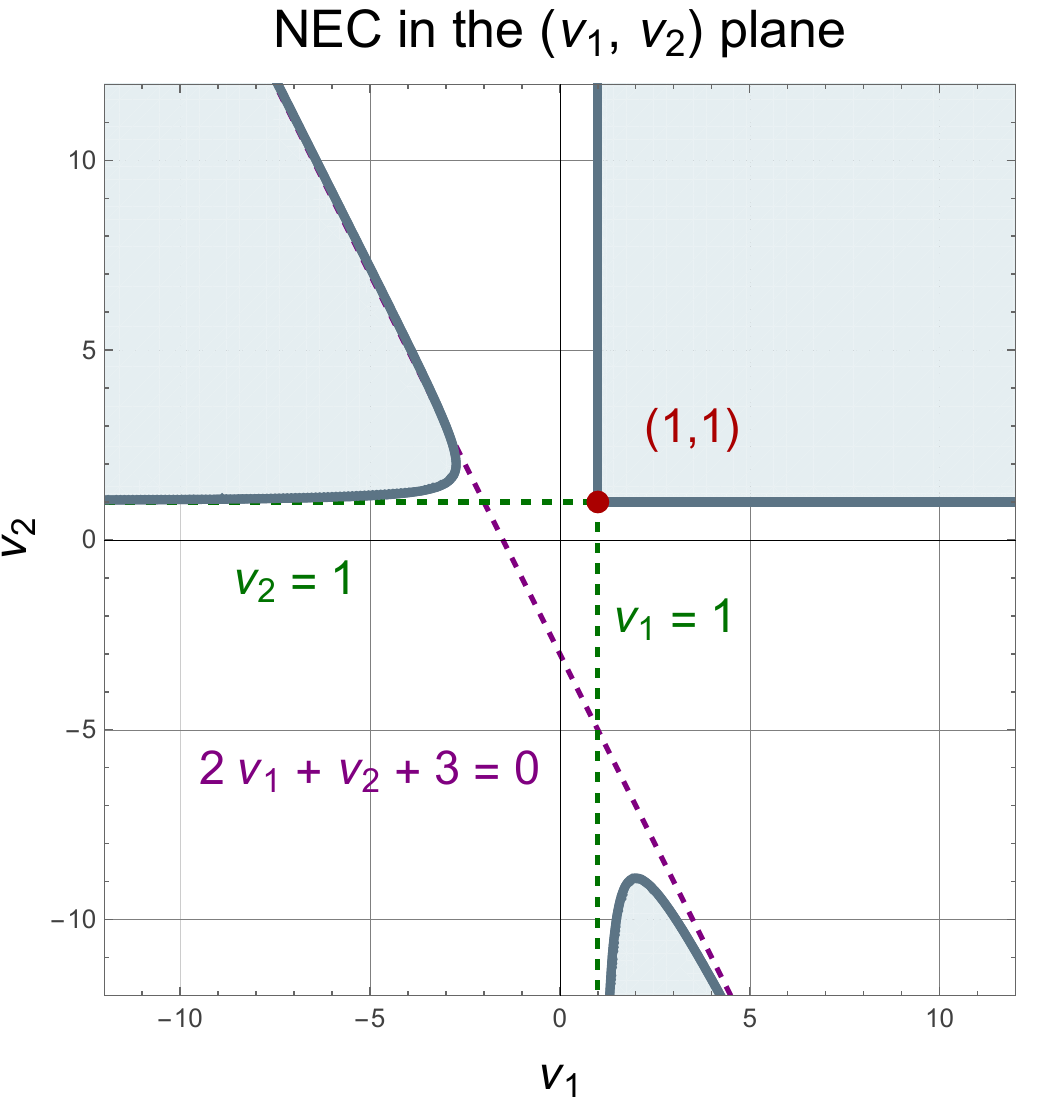} \\
  A \hspace{190pt} B
  \caption{(A) Third law of thermodynamics for Model III: the shaded region (without the dashed blue line) represents the values of $\nu_1$ and $\nu_2$ for which the third law of thermodynamics holds; the asymptotes provided are $\nu_1=-1$, $\nu_2=-\,2$. (B) NEC for Model III: the shaded region (blue with the boundary) represents the values of $\nu_1$ and $\nu_2$ for which the NEC \eqref{5:16}--\eqref{5:18} holds; the asymptotes provided for the two regions outside the first quadrant are $\nu_1=1$, $\nu_2=1$, and $2 \nu_1 + \nu_2 + 3 = 0$.
  }
  \label{kappanuplane}
\end{figure}

The solution of this system of inequalities is easier to present graphically by shading the region of the $(\nu_1, \, \nu_2)$ plane, for which the NEC holds, see Figure~\ref{kappanuplane}.

For non-negative $\nu_1$ and $\nu_2$, the NEC is simply equivalent to
\be
 \nu_1 \ge 1, \quad \nu_2 \ge 1.
\ee

We can also note that the 3-form's coupling function $h_1(\phi)$ is always non-negative, while the 2-form's function $\ff_3(\phi)$ might be negative, see \eqref{Model2-h1sol} and \eqref{Model2-f3sol}:
\bea
  \text{sgn}\big(h_1(\phi)\big)&=&\text{sgn}
  \left(1-\frac{1}{\nu_1}\right) = 0, \ +\,1, \\
  \text{sgn}\big(\ff_3(\phi)\big)&=&\text{sgn}
  \left(\frac{1}{\nu_1}-\frac{1}{\nu_2}\right) = -\,1, \ 0, \ +\,1, \label{6:23}
\eea
for which one has to use \eqref{5:18} to simplify \eqref{5:17}. For a negative $\ff_3(\phi)$, take for example $(\nu_1, \, \nu_2) = (2, \, 1)$, which clearly satisfies the NEC, see Fig.~\ref{kappanuplane}.

\subsection{NEC and the Third Law of Thermodynamics}\label{Model2-NEC-3Law}

Let us now make a note about how the NEC and the third law of thermodynamics are related for this model. We obtained the following criterion for the third law:
\be
    1 + \frac{1}{\nu_1} + \frac{2}{\nu_2} > 0, \qquad 2 + \frac{1}{\nu_1} + \frac{2}{\nu_2} > 0.
\ee

The second constraint is redundant in this case, so that the third law of thermodynamics holds if
\be
    1 + \frac{1}{\nu_1} + \frac{2}{\nu_2} > 0.
\ee

We can show that the third law of thermodynamics holds for all the regimes supported by the NEC. For that, let us add a positive term to both sides of \eqref{5:16}:
\be
    \left[\frac{1}{\nu_1}\,\left(1-\frac{1}{\nu_1}\right)+\frac{2}{\nu_2}\,\left(1-\frac{1}{\nu_2}\right)\right] 
    + 1 + \frac{1}{\nu^2_1} + \frac{2}{\nu^2_2} \ge 1 + \frac{1}{\nu^2_1} + \frac{2}{\nu^2_2},
\ee
to obtain after simplifying
\be
    1 + \frac{1}{\nu_1} + \frac{2}{\nu_2} \ge 1 + \frac{1}{\nu^2_1} + \frac{2}{\nu^2_2} > 0,
\ee
meaning that 
\be
    1 + \frac{1}{\nu_1} + \frac{2}{\nu_2} > 0
\ee
as a consequence of imposing the NEC. 

Therefore, if NEC holds, the third law of thermodynamics is satisfied automatically. The opposite, however, is not true, which is easily confirmed with the help of Figure~\ref{kappanuplane}.


\section{Discussion}\label{sec:discussion}

We now briefly discuss each of the considered models and outline the obtained results. \\

Model I is a particular case of the already published setting \cite{Arefeva:2023jjh}. Although the cited paper provides a detailed description of a more sophisticated model, it does not contain the results we have obtained in this work:
\begin{itemize}
    \item the NEC is proven to be equivalent to requiring that $c_B \le 0$, $\nu \ge 1$, and additionally the horizon is bounded above for $c_B < 0$: $z_h \le z^{max}_h$, see \eqref{3.69}, while for $c_B = 0$, the horizon is unbounded;
    \item for $c_B < 0$, $\nu \ge 1$, the upper boundary $z_h^{max}$ limits the range of temperatures attainable by the model $T_{min} = T(z^{max}_h)>0$;
    \item full description of the parameters, for which the third law of thermodynamics is satisfied, was given in \eqref{4:22}, and it has been established, that the third law and the NEC are independent conditions, see Sec.~\ref{Model1-NEC-3Law};
    \item the only regime that satisfies both conditions is $c_B = 0$, $\nu \ge 1$.
\end{itemize}
$\,$

Model II is a novel $D=5$ model, deformed by two Lifshitz factors and equipped with two magnetic Maxwell fields. This model was considered as part of an attempt to encode the magnetic anisotropy in Model I with a Lifshitz factor, as opposed to the Gauss-type one. We considered all the three possible sets of the magnetic Maxwell fields denoted as \textbf{i}, \textbf{ii}, and \textbf{iii}, see \eqref{5:2}--\eqref{5:4}. The warp factor was taken to be considerably general: $\exp(cz^n)$, where $c \in \mathbb{R}$, $n>0$. The results we obtained are:  
\begin{itemize}
    \item for the chosen warp factor $\fb(z)$, the parameters $\nu_1$ and $\nu_2$ become connected via explicit and simple algebraic constraints, see \eqref{model3-gSol1}--\eqref{model3-gSol2};
    \item the NEC was examined in full detail for all the values of $\nu_1$ and $\nu_2$, see Tab.~\ref{tab::m2-NEC}; notably, some of the NEC restrictions leads to $\nu_1=\nu_2$ for all three cases, while for the case \textbf{i} the restriction is stronger $\nu_1 = \nu_2 = 1$, leading to isotropy;
    \item an upper boundary for the horizon $z^{max}_h$ appears for cases \textbf{ii} and \textbf{iii}, when $\nu_1=\nu_2>1$, $c>0$, similarly to Model I,
    \item the third law of thermodynamics was fully examined, see Tab.~\ref{tab::m2-3rdlaw}, and it has been established, that the third law and the NEC are independent conditions, see Sec.~\ref{Model3-NEC-3Law},
    \item the regimes that satisfy both conditions were found, see \eqref{5:60}--\eqref{5:61}.
\end{itemize}
$\,$

Model III is a novel $D = 6$ model, deformed by two Lifshitz factors and equipped with a magnetic $3$-form, along with a magnetic Maxwell field. We solved the model for $\fb(z) = 1$ and obtained the following:
\begin{itemize}
    \item the dilaton potential $V(\phi)$ is constant and reduces to the twice cosmological constant of the $AdS_6$ space when $\nu_1 = \nu_2 = 1$; the coupling functions are found explicitly as functions of the dilaton \eqref{Model2-h1sol}--\eqref{Model2-f3sol}; the entropy density is found explicitly as a function of the temperature $s=s(T)$, see \eqref{6:14};
    \item as opposed to Models I and II, the NEC for this model does not imply non-negativity of coupling functions --- the 2-form coupling function $\ff_3(\phi)$ can be negative, see \eqref{6:23};
    \item the NEC and the third law of thermodynamics are described fully with Fig.~\ref{kappanuplane}, where the regions $(\nu_1, \nu_2)$, for which the model satisfies these two conditions, are shown respectively; we show that the NEC leads to the satisfaction of the third law, while the opposite is false.
\end{itemize}
$\,$

In Section~\ref{sec:quadratures}
we highlighted the pattern arising for $T_{\mu\nu}/g_{\mu\nu}$, which leads to sign tables as a way to represent the Einstein equations, see Section~\ref{subsec::signtables}. Further, in Section~\ref{sec:quadraturederivation} we discussed how the sign tables let us significantly simplify the process of the model solving. In particular, it allows us to easily write down the equation for the blackening function $g(z)$, for which we then provide a procedure to solve in quadratures. We then demonstrated how the solution for $g(z)$ can be found within the holographic Einstein-dilaton-four-Maxwell model using the developed procedure, see \eqref{four-Maxwell-quadratures}. The algorithm can be easily applied to any model within the class of models we restricted ourselves to, upon which we expand in Appendix~\ref{app::B}.\\

Thus we have shown that the magnetic field prevents the fulfillment of the third law of thermodynamics for the asymptotically AdS$_5$ Einstein-dilaton-Maxwell black branes with the metric deformation introduced via the warp-factor $\fb = e^{c z^2}$, which serves as a backbone for the HQCD considerations of heavy quarks media \cite{Arefeva:2018hyo,He:2020fdi,Shukla:2023pbp,Arefeva:2023jjh,Rannu:2024vrq,Arefeva:2025okg,Arefeva:2026yms}. The third law of thermodynamics requires $\fb = e^{c z^n}$, $0 < n < 1$, for nonzero parameter $c$. This result agrees with the $\nu_1 = \nu_2$ and more strict $\nu_1 = \nu_2 = 1$ limits on the metric anisotropies, as $\nu_1 \ne \nu_2$ should be supported by the magnetic field. The values $\nu_1, \nu_2 \ge 1$, usually considered within the HQCD models, are confirmed as the requirements of the solution physicality.

In the complementary paper \cite{AGN}, a more general class of $D$-dimensional black brane solutions with Lifshitz-like asymptotics that satisfy the classical third law has been found.

\section*{Acknowledgment}

We are grateful to  Anastasia Golubtsova, Valeriya  Nerovnova, Pavel Slepov and Igor Volovich for useful discussions. The work of I.A. was performed at the Steklov  Mathematical Institute and supported by the Russian Science Foundation grand 24-11-00039. V. Z. was supported by a scholarship from the Theoretical Physics and Mathematics Advancement Foundation ``BASIS."

\appendix
\section{General EOM for 6D Models}\label{app::6D-EOM}

Here we explicitly write down the EOM for the 6D model that includes all the possible magnetic 2- and 3-form fields, see Sec.~\ref{subsec::6D-notation} for the notation introduced for these fields. The metric is of the most general form
\begin{gather}
  ds^2 = - \, g_{00}(z) dt^2 + g_{11}(z) dx_1^2 + g_{22}(z) dx_2^2 + g_{33}(z) dx_3^2 + g_{44}(z) dx_4^2 + g_{55}(z) dx_5^2,
\end{gather}
such that the dilaton EOM and the Einstein equations are:

\begin{gather}
  \begin{split}
    \phi'' \ + \ & \cfrac{\phi'}{2} \sum_{\mu}  (-1)^{\delta_{\mu 5}} \cfrac{g_{\mu\mu}'}{g_{\mu\mu}} \ - \\
    - \ & g_{55} \left(
    \sum_{\cK} \cfrac{q_{\cK}^2}{2} \, \cfrac{\partial f_{\cK}}{\partial \phi}
      \prod_{i \ne \cK} \cfrac{1}{g_{ii}}
      + \sum_{\cM} \cfrac{\fq_{\cM}^2}{2 g_{\cM\cM} g_{44}} \, \cfrac{\partial \ff_{\cM}}{\partial \phi}
      + \sum_{\cN} \cfrac{Q_{\cN}^2}{2} \, \cfrac{\partial H_{\cN}}{\partial \phi} \prod_{j\ne\cN} \cfrac{1}{g_{jj}}
      + \cfrac{\partial V(\phi)}{\partial \phi}
    \right) = 0, \\
    &\cK, \cM = \overline{1,3}, \ \cN = \overline{1,4}, \ \mu, \nu = \overline{0,5}, \ i = \overline{1,3}, \ j = \overline{1,4}, \ k,l = \overline{1,5},
  \end{split} \\
  \begin{split}
    \sum_{j} &\left( 
      \cfrac{2 g_{jj}''}{g_{jj}} - \cfrac{g_{jj}'^2}{g_{jj}^2} 
    \right)
    + \sum_{\to k < l} (-1)^{\delta_{5k}} (-1)^{\delta_{5l}} \cfrac{g_{kk}'}{g_{kk}} \, \cfrac{g_{ll}'}{g_{ll}} \ + \\
    + \ g_{55} &\left(
      \sum_{\cK} f_{\cK} q_{\cK}^2
      \prod_{i\ne\cK} \cfrac{1}{g_{ii}}
      + \sum_{\cM} \cfrac{\ff_{\cM} \fq_{\cM}^2}{g_{\cM\cM} g_{44}}
      + \sum_{\cN} H_{\cN} Q_{\cN}^2 \prod_{j\ne\cN} \cfrac{1}{g_{jj}}
      + \cfrac{\phi'^2}{g_{55}}
      + 2 V(\phi)
    \right) = 0,
  \end{split} \\
  \begin{split}
    \sum_{\mu\ne j} ( 1 - \delta_{\mu5} ) &\left( 
      \cfrac{2 g_{\mu\mu}''}{g_{\mu\mu}} - \cfrac{g_{\mu\mu}'^2}{g_{\mu\mu}^2} 
    \right)
    + \sum_{\,\mu<\nu\ne j} (-1)^{\delta_{5\mu}} (-1)^{\delta_{5\nu}} \cfrac{g_{\mu\mu}'}{g_{\mu\mu}} \, \cfrac{g_{\nu\nu}'}{g_{\nu\nu}} \ + \\
    + \, g_{55} &\left(
      \sum_{\cK} (-1)^{\delta_{\cK j}} (-1)^{1+\delta_{4j}} f_{\cK} q_{\cK}^2
      \prod_{i\ne\cK} \cfrac{1}{g_{ii}}
      + \sum_{\cM} (-1)^{\delta_{\cM j}} (-1)^{\delta_{4j}} \cfrac{\ff_{\cM} \fq_{\cM}^2}{g_{\cM\cM} g_{44}} \right. + \\
      &\left.\!+ \sum_{\cN} (-1)^{1+\delta_{\cN j}} H_{\cN} Q_{\cN}^2 \prod_{i\ne\cN} \cfrac{1}{g_{ii}}
      + \cfrac{\phi'^2}{g_{55}}
      + 2 V(\phi)
    \right) = 0, \ j = \overline{1,4},
  \end{split} \\
  \begin{split}
    \cfrac{g_{00}'}{g_{00}} \sum_{j} \cfrac{g_{jj}'}{g_{jj}}
    &+ \prod_{k< l} ( 1  - \delta_{l5} ) \, \cfrac{g_{kk}'}{g_{kk}} \, \cfrac{g_{ll}'}{g_{ll}} \, + \\
    &+ g_{55} \left(
      \sum_{\cK} f_{\cK} q_{\cK}^2
        \prod_{i\ne\cK} \cfrac{1}{g_{ii}}
        + \sum_{\cM} \cfrac{\ff_{\cM} \fq_{\cM}^2}{g_{\cM\cM} g_{44}}
      + \sum_{\cN} H_{\cN} Q_{\cN}^2 \prod_{j\ne\cN} \cfrac{1}{g_{jj}}
      - \cfrac{\phi'^2}{g_{55}}
      + 2 V(\phi)
    \right) = 0.
  \end{split}
\end{gather}

\section{Derivation of the Formula \eqref{Gaa-Gbb}}\label{app::A}
For the diagonal metric depending on one coordinate only, \eqref{SET-Metric}, we obtained the following formulas for $G_{\mu\nu}$ by manual calculations, which we omit for brevity of this appendix section:
\bea
    \bm{[\beta\ne z]:} G_{\beta\beta} &=& 
    \frac{1}{2}\frac{g_{\beta\beta}}{g_{zz}} \sum_{\alpha\ne z,\beta} \frac{g_{\alpha\alpha, z z}}{g_{\alpha\alpha}}
    -\frac{1}{4}\frac{g_{\beta\beta}}{g_{zz}} \sum_{\alpha\ne z,\beta} \frac{g_{\alpha\alpha,z}^2}{g_{\alpha\alpha}^2}
    -\frac{1}{4}\frac{g_{\beta\beta}}{g_{zz}}\frac{g_{zz,z}}{g_{zz}} \sum_{\alpha\ne z,\beta} \frac{g_{\alpha\alpha,z}}{g_{\alpha\alpha}}+
    \nn \\ 
    &+&\frac{1}{4}\frac{g_{\beta\beta}}{g_{zz}} \sum_{\mu<\nu: \ \mu,\,\nu \,\ne\,\beta, \,z}\frac{g_{\mu\mu,z}g_{\nu\nu,z}}{g_{\mu\mu}g_{\nu\nu}}, 
    \label{Gbb-general}
    \\
    G_{zz}&=&\frac{1}{4} \sum_{\mu<\nu: \ \mu,\nu\ne z} \frac{g_{\mu\mu,z}g_{\nu\nu,z}}{g_{\mu\mu}g_{\nu\nu}},
    \label{Gzz-general}
\eea
while the off-diagonal components are identically zero. For the expression \eqref{Gaa-Gbb}, we can utilize \eqref{Gbb-general} to arrive at
\bea
    g_{zz} \bigg(\frac{G_{\alpha\alpha}}{g_{\alpha\alpha}}-\frac{G_{\beta\beta}}{g_{\beta\beta}}\bigg)
    =
    \frac{1}{2}\bigg(\frac{g''_{\beta\beta}}{g_{\beta\beta}}-\frac{g''_{\alpha\alpha}}{g_{\alpha\alpha}}\bigg)
    -\frac{1}{4}\bigg(\frac{g'_{\beta\beta}}{g_{\beta\beta}}-\frac{g'_{\alpha\alpha}}{g_{\alpha\alpha}}\bigg)
    \frac{d\ln}{dz}\bigg(\frac{g_{\alpha\alpha}g_{\beta\beta}g_{zz}}{\prod_{\gamma \ne z, \alpha, \beta}g_{\gamma\gamma}}\bigg).
\eea

Here, we already notice that if $g_{\beta\beta}$ is a scalar multiple of $g_{\alpha\alpha}$, then we have:
\be
    g_{\beta\beta}(z) = \const \cdot g_{\alpha\alpha}(z) 
    \quad \Longrightarrow \quad
    \frac{G_{\alpha\alpha}}{g_{\alpha\alpha}}-\frac{G_{\beta\beta}}{g_{\beta\beta}} = 0. \label{A:4}
\ee

We would further like to factorize the first term by the factor before the logarithmic derivative in the second term. For these purposes let us represent $g_{\beta\beta}$ in the following form:
\be
    g_{\beta\beta}(z)=u(z) \, g_{\alpha\alpha}(z),
\ee
so that
\bea
    \frac{g''_{\beta\beta}}{g_{\beta\beta}}-\frac{g''_{\alpha\alpha}}{g_{\alpha\alpha}}&=&
    \frac{u''}{u}+2\,\frac{u'}{u}\frac{g'_{\alpha\alpha}}{g_{\alpha\alpha}}=
    \frac{u''}{u'}\frac{u'}{u}+2\,\frac{u'}{u}\frac{g'_{\alpha\alpha}}{g_{\alpha\alpha}}=\frac{u'}{u}\left(\frac{u''}{u'}+2\,\frac{g'_{\alpha\alpha}}{g_{\alpha\alpha}}\right) \nn
    \\
    &=&\frac{d\ln}{dz}\bigg(\frac{g_{\beta\beta}}{g_{\alpha\alpha}}\bigg) \, \frac{d\ln}{dz}\Bigg(\bigg(\frac{g_{\beta\beta}}{g_{\alpha\alpha}}\bigg)'g^2_{\alpha\alpha}\Bigg).
\eea

Therefore we can factorize the whole expression by the logarithmic derivative of $u(z)$ and further simplify
\bea
    g_{zz}\bigg(\frac{G_{\alpha\alpha}}{g_{\alpha\alpha}}-\frac{G_{\beta\beta}}{g_{\beta\beta}}\bigg)
    &=&
    \frac{1}{2}\,\frac{d\ln}{dz}\bigg(\frac{g_{\beta\beta}}{g_{\alpha\alpha}}\bigg) 
    \left[ 
    \frac{d\ln}{dz}\Bigg(\bigg(\frac{g_{\beta\beta}}{g_{\alpha\alpha}}\Bigg)'g^2_{\alpha\alpha}\bigg)
    -\frac{1}{2}\,\frac{d\ln}{dz}\bigg(\frac{g_{\alpha\alpha}g_{\beta\beta}g_{zz}}{\prod_{\gamma \ne z, \alpha, \beta}g_{\gamma\gamma}}\bigg)\right] \nn
    \\
    &=&
    \frac{1}{2}\,\frac{d\ln}{dz}\bigg(\frac{g_{\beta\beta}}{g_{\alpha\alpha}}\bigg) \frac{d\ln}{dz}
    \Bigg( \frac{\prod_\gamma \sqrt{g_{\gamma\gamma}}}{g_{zz}} \,\frac{d\ln}{dz}\bigg(\frac{g_{\beta\beta}}{g_{\alpha\alpha}}\bigg)
    \Bigg) \nn
    \\
    &=& \frac{1}{2} \, \frac{g_{zz}}{\sqrt{g_D}} \, \frac{d}{dz}
    \Bigg( \frac{\sqrt{g_D}}{g_{zz}} \, \frac{d\ln}{dz}\bigg(\frac{g_{\beta\beta}}{g_{\alpha\alpha}}\bigg) \Bigg).
\eea

Dividing both sides by $g_{zz}$, we finally obtain the formula \eqref{Gaa-Gbb}:
\be
\displaystyle 
  \bm{[\alpha,\,\beta\ne z]}\quad
  \frac{G_{\alpha\alpha}}{g_{\alpha\alpha}}-\frac{G_{\beta\beta}}{g_{\beta\beta}}
  =
  \frac{1}{2\,\sqrt{g_D}} 
  \Bigg(
    \frac{\sqrt{g_D}}{g_{zz}}
    \,\frac{d}{dz}\ln{\frac{g_{\beta\beta}}{g_{\alpha\alpha}}}
    \Bigg)'.
\ee

\section{On the Equation for $g(z)$}\label{app::B}

In Section~\ref{sec:quadratures} we outlined the procedure of simplifying the Einstein equations with the help of sign tables. In this procedure, one combines the rows in the sign table (i.e., equations $G_{\mu\mu}/g_{\mu\mu}=T_{\mu\mu}/g_{\mu\mu}$) to obtain an equation for the blackening function $g(z)$. Since the Einstein equations contain the second derivative of the blackening function, $g''(z)$, we usually try to obtain a 2nd-order DE for $g(z)$. 

Recall that in Section~\ref{sec:actionmetric} we discussed, which family of models we consider --- their actions are structurally the same, but the dimension and the set of electric and magnetic 2- and 3-form fields are arbitrary. It turns out that we can make some general statements regarding the DE for $g(z)$ for this family of models. 

Suppose that for a particular model we combine the rows of the Sign Table in such a way that an equation without the RHS arises:
\be
    \sum^{D-1}_{\mu = 0} k_\mu \, \frac{G_{\mu\mu}}{g_{\mu\mu}} 
    \equiv
    k_0 \, \frac{G_{00}}{g_{00}} +
    \sum^{D-2}_{i = 0} k_i \, \frac{G_{x_ix_i}}{g_{x_ix_i}} +
    k_z \, \frac{G_{zz}}{g_{zz}}
    = 0, \label{11:1}
\ee
where $k_{\mu}$ are the corresponding numerical factors, where we identified the $D-1$ index with $z$.

We should think of this equation as follows: in the LHS of \eqref{11:1}, we use the Einstein equations $G_{\mu\nu}=T_{\mu\nu}$ and then substitute the particular formulas for the stress-energy tensor $T_{\mu\nu}$, see \eqref{Tg-phi}--\eqref{Tg-H}:
\be
    \sum^{D-1}_{\mu = 0} k_\mu \, \frac{G_{\mu\mu}}{g_{\mu\mu}} 
    = [G_{\mu\nu}=T_{\mu\nu}] = 
    \sum^{D-1}_{\mu = 0} k_\mu \, \frac{T_{\mu\mu}}{g_{\mu\mu}} = 0. \label{11:2}
\ee

\textbf{Proposition:} The family of models we are considering is not restricted in terms of the dimension $D$ of the space and the set of electric and magnetic 2- and 3-form fields, while the dilaton $\phi=\phi(z)$ is always present. Suppose that a particular model allows for an equation in the form \eqref{11:1}. Then
\begin{enumerate}
    \item such an equation does not include the $G_{zz}/g_{zz}$ term, or equivalently $k_z \equiv k_{D-1} = 0$;
    \item the coefficients $k_{\mu}$ can all be scaled simultaneously to be integers, which allows us to solve for $g(z)$ in equation \eqref{11:1} using formula \eqref{Gaa-Gbb};
    \item if the model includes an electric field, the equation \eqref{11:1} does not contain the $G_{00}/g_{00}=-\,G_{tt}/g_{tt}$ term $k_0 \equiv k_t = 0$; moreover, this equation is a 1st order differential equation w.r.t. $g(z)$.
\end{enumerate}

\textbf{Proof:} For proof, we need to derive equations for the coefficients $k_\mu$. For that let us substitute the expression for the stress-energy tensor \eqref{Tg-phi}--\eqref{Tg-H} into the equation \eqref{11:2}:
\bea
    0 &=& \sum^{D-1}_{\mu = 0} k_\mu \, \frac{T_{\mu\mu}}{g_{\mu\mu}}
    = - \frac{V(\phi)}{2} \ \sum^{D-1}_{\mu = 0} k_\mu 
    - \frac{\phi'^2}{4\,g_{zz}} \ \left(-\,k_{D-1} + \sum^{D-2}_{\mu = 0} k_\mu \right)
    \nn \\
    &-& \frac{f_{\cM}(\phi)\,q^2_{\cM}}{4\,g_{ii}\,g_{jj}} \ \left(-k_i - k_j + \sum^{D-1}_{\mu = 0, \, \mu \ne i, j} k_\mu \right)
    \nn \\
    &-& \frac{h_{\cN}(\phi)\,Q^2_{\cN}}{4\,g_{mm}\,g_{nn}\,g_{pp}} \ \left(-k_m - k_n - k_p + \sum^{D-1}_{\mu = 0, \, \mu \ne m,n,p} k_\mu \right),
\eea
where we only included one 2-form and one 3-form field. If a particular model has a different collection of $p$-form fields, this expression is straightforwardly adjusted.

Here, we notice that each of the independent types of contributions to the $T_{\mu\mu}/g_{\mu\mu}$ 
\be
    \frac{V(\phi)}{2}, \quad
    \frac{\phi'^2(z)}{4g_{zz}}, \quad
    \frac{f_\cM(\phi)q_\cM^2}{4g_{ii}g_{jj}}, \quad 
    \frac{h_\cN(\phi)Q_\cN^2}{4g_{pp}g_{kk}g_{ll}}
    \label{11:6}
\ee
needs to have zero contribution to the whole expression \eqref{11:2}, which translates to the following conditions on the coefficients:
\bea
    \phi(z):&&
    \sum^{D-1}_{\mu = 0} k_\mu = 0,
    \quad 
    -\,k_{D-1} + \sum^{D-2}_{\mu = 0} k_\mu = 0, \label{11:3}
    \\
    F_{\cM} = q_\cM \ dx^i \wedge dx^j:&&
    -\,k_i - k_j + \sum^{D-1}_{\mu = 0, \, \mu \ne i, j} k_\mu = 0,
    \\
    H_{\cN} = Q_\cN \ dx^m \wedge dx^n \wedge dx^p:&&
    -\,k_m - k_n - k_p + \sum^{D-1}_{\mu = 0, \, \mu \ne m,n,p} k_\mu = 0.
\eea

Since we assume that the dilaton $\phi$ is always present, we can use the condition \eqref{11:3} corresponding to it to simplify the conditions corresponding to the $p$-form fields:
\bea
    \phi(z):&&
    k_{D-1} = 0, \qquad 
    \sum^{D-1}_{\mu = 0} k_\mu = \sum^{D-2}_{\mu = 0} k_\mu = 0,
    \label{11:12} \\
    \left[ \phi(z) \, + \right] \  F_{\cM} = q_\cM \ dx^i \wedge dx^j:&&
    k_i + k_j = 0, \label{11:13}
    \\
    \left[ \phi(z) \, + \right] \ H_{\cN} = Q_\cN \ dx^m \wedge dx^n \wedge dx^p:&&
    k_m + k_n + k_p = 0. \label{11:14}
\eea

Therefore, we proved the first part of the proposition.

For the second part of the proposition, we need to understand that the conditions \eqref{11:12}, \eqref{11:13}, \eqref{11:14} we derived are equivalent to solving a matrix equation. The matrix is the sign matrix part from a model's sign table. Each column's entries are summed with the corresponding weights $k_\mu$, which represents finding the contribution of each of the terms \eqref{11:6} in the header of the sign table. Each result is then equated to zero. To illustrate that, let us consider one of the cases of Model II,
\be
    \textbf{i} \qquad
    F_2=q_2 \, dx^1\wedge dx^3, \qquad F_3=q_3 \, dx^1\wedge dx^2,
\ee
whose sign table is located above, see Table~\ref{Table-5D-2M-1}. The equations on $k_{\mu}$ are equivalent to the following matrix equation:
\begin{equation}
\begin{bmatrix}
k_z \\
k_{x_3} \\
k_{x_2} \\
k_{x_1} \\
k_0
\end{bmatrix}^T
\begin{bmatrix}
\bm{+} & \bm{-} & \bm{-} & \bm{-}\\
\bm{-} & \bm{-} & \bm{+} & \bm{-}\\
\bm{-} & \bm{-} & \bm{-} & \bm{+}\\
\bm{-} & \bm{-} & \bm{+} & \bm{+}\\
\bm{-} & \bm{-} & \bm{-} & \bm{-}
\end{bmatrix}
=
\begin{bmatrix}
0 \\
0 \\
0 \\
0 \\
0
\end{bmatrix}^{T}
\Longleftrightarrow
\begin{bmatrix}
\bm{+} & \bm{-} & \bm{-} & \bm{-} & \bm{-} \\
\bm{-} & \bm{-} & \bm{-} & \bm{-} & \bm{-} \\
\bm{-} & \bm{+} & \bm{-} & \bm{+} & \bm{-} \\
\bm{-} & \bm{-} & \bm{+} & \bm{+} & \bm{-}
\end{bmatrix}
\begin{bmatrix}
k_z \\
k_{x_3} \\
k_{x_2} \\
k_{x_1} \\
k_0
\end{bmatrix}
=
\begin{bmatrix}
0 \\
0 \\
0 \\
0 \\
0
\end{bmatrix}.
\end{equation}

We can bring the matrix to its reduced row-echelon form by the process of Gaussian elimination. In this form, the entries of the reduced matrix are all rational numbers. If a nontrivial solution for $k_\mu$ exists, we can scale it to be integer-valued, so:
\be
    \sum^{D-2}_{\mu = 0} k_\mu \, \frac{G_{\mu\mu}}{g_{\mu\mu}} = 0, \quad k_\mu \in \mathbb{Z}, \label{11:18}
\ee
where we do not include the $D-1$ term as we proved its absence.

We claimed that we can solve for $g(z)$ in this equation utilizing the formula \eqref{Gaa-Gbb}. This statement is quite nontrivial as the formula \eqref{Gaa-Gbb} tells us that we can subtract pairs of $G_{\alpha\alpha}/g_{\alpha\alpha}$. What we can notice is that the condition \eqref{11:12} actually says that the sum of negative $k_\mu$ is equal to the sum of positive $k_\mu$:
\be
    \sum^{D-2}_{\mu = 0} k_\mu = 0
    \Longleftrightarrow
    \sum^{D-2}_{\mu = 0, \ k_{\mu}>0} |k_\mu| 
    =
    \sum^{D-2}_{\nu = 0, \ k_{\nu}<0} |k_\nu|, \label{B:14}
\ee
so that we can further rewrite \eqref{11:18}:
\begin{equation}
\sum^{D-2}_{\mu = 0} k_\mu \, \frac{G_{\mu\mu}}{g_{\mu\mu}} = 0, \quad k_\mu \in \mathbb{Z}
    \Longleftrightarrow
\begin{tikzpicture}[baseline=(top.base)]
  \node (top)
    {$\displaystyle \sum^{D-2}_{\mu = 0, \ k_{\mu}>0} |k_\mu| \, \frac{G_{\mu\mu}}{g_{\mu\mu}}
    =
    \sum^{D-2}_{\nu = 0, \ k_{\nu}<0} |k_\nu| \, \frac{G_{\nu\nu}}{g_{\nu\nu}}$};
  \coordinate (Ck) at ($(top.south west)!0.25!(top.south east)$);
  \coordinate (Cl) at ($(top.south west)!0.75!(top.south east)$);

  \def\dy{2.0ex}
  \coordinate (Ak) at ($(Ck) + (0,-\dy)$); 
  \coordinate (Al) at ($(Cl) + (0,-\dy)$); 

  \draw (Ak) -- node[below]{\(\text{the same number of $G_{\rho\rho}/g_{\rho\rho}$}\)} (Al);

  \draw[->] (Ak) -- (Ck);
  \draw[->] (Al) -- (Cl);
\end{tikzpicture}
. \label{11:20}
\end{equation}

We then use the formula \eqref{Gaa-Gbb} to write down \eqref{11:20} as
\be
    \displaystyle
    \frac{1}{2\,\sqrt{g_D}}
    \Bigg(
        \frac{\sqrt{g_D}}{g_{zz}}\,
        \frac{d}{dz}
        \ln{
        \frac{\prod_{k_\nu<0}g^{|k_\nu|}_{\nu\nu}}{\prod_{k_\mu>0}g^{|k_\mu|}_{\mu\mu}}
        }
    \Bigg)' = 0, \label{11:24}
\ee
which can be simplified to be, keeping in mind the definition of $\fs_D$ from \eqref{3:39} (we use the subscript $D$ to remind that this quantity depends on the dimension of the space):
\be
    \displaystyle
    \Bigg(
        g \, \fs_D 
        \,
        \frac{d}{dz}
        \ln{ \bigg( g^{k_0} \, \prod^{D-2}_{\rho=1}\fg^{k_\rho}_{\rho} \bigg) }
    \Bigg)' = 0.
    \label{11:25}
\ee

Further analysis is broken up into two cases.
\begin{itemize}
    \item First, suppose $k_0 = 0$, then \eqref{11:25} is evidently a 1st order DE for $g(z)$, whose solution is:
    \be
        g(z) = \cfrac{C}{\displaystyle \fs_D(z) \, \frac{d}{dz}
        \ln{\prod^{D-2}_{\rho=1}\fg^{k_\rho}_{\rho}(z)}},
        \label{11:26}
    \ee
    where we need to be careful as
    \be
        \frac{d}{dz}
        \ln{\prod^{D-2}_{\rho=1}\fg^{k_\rho}_{\rho}(z)}
    \ee
    could be zero, thus making the equation \eqref{11:25} be satisfied automatically.
    \item Second, suppose $k_0 \ne 0$, then \eqref{11:25} is a 2nd order DE for $g(z)$. We write down the final solution and omit the particular steps to obtain it for brevity. Introducing notation:
    \be
        \fG(z)=\prod^{D-2}_{\rho=1}\fg^{-k_\rho/k_0}_{\rho}(z)
        \ \Longrightarrow \
        g(z) = \fG(z) \,
        \left[\displaystyle
        C_1 \int_{z_0}^{z} \frac{d\xi}{\displaystyle \fs_D(\xi) \fG(\xi)} + C_2
        \right]. \label{11:32}
    \ee
\end{itemize}

For the third part of the proposition, let us utilize the condition \eqref{11:13}:
\be
    F^{(el.)} = A'_t(z) \ dx^0 \wedge dx^{D-1}:
    \qquad
    k_0 + k_{D-1} = 0,
\ee
which together with the condition $k_{D-1} = 0$ from the dilaton gives us
\be
    k_0 = k_{D-1} = 0 \quad \Longrightarrow \quad \eqref{11:1}:
    \quad
    \sum^{D-2}_{\mu = 1} k_\mu \, \frac{G_{\mu\mu}}{g_{\mu\mu}} = 0,
\ee
which is the case of a 1st order DE for $g(z)$ we discussed right above, see \eqref{11:26}.

\subsection{Models with an Inhomogeneous Second-Order DE}\label{app::B1}
If a model does not allow for a 2nd-order DE for $g(z)$ without a RHS, we can solve this equation by variation of constants in the solution \eqref{11:32} for the homogeneous equation. This is possible due to this DE being linear with respect to $g(z)$. To see this, let us write down the DE in such a form, using \eqref{11:24}--\eqref{11:25}:
\be
    - \frac{k_0}{2} \,
    \cfrac{1}{\frac{L^2}{z^2}\fb(z) \fs_D(z)} 
    \frac{d}{dz}
    \Bigg(
        \fs_D(z) \fG(z)
        \frac{d}{dz}\frac{g(z)}{\fG(z)}
    \Bigg)
    =
    \sum^{D-2}_{\mu = 0} k_\mu \, \frac{T_{\mu\mu}}{g_{\mu\mu}}, \label{B:23}
\ee
where on the LHS we have an expression that is evidently linear w.r.t. $g(z)$, while on the RHS we have gotten rid of the contributions from the dilaton by imposing \eqref{11:12} in \eqref{B:14}, so that only such expression can appear, up to a numerical factor (compare with \eqref{11:6}):
\be
    \frac{f_\cM(\phi)q_\cM^2}{4g_{ii}g_{jj}}, \quad 
    \frac{h_\cN(\phi)Q_\cN^2}{4g_{pp}g_{kk}g_{ll}}.
\ee

Since we choose either a magnetic or electric ansatz, these expressions do not include the blackening function $g(z)$. For a magnetic ansatz, $g_{ii}$ are from the components $g_{x_ix_i}$. For the electric ansatz, $g_{00}g_{zz}$ does not explicitly contain $g(z)$ as
\be
    g_{00}(z) \, g_{zz}(z)
    = \frac{L^4}{z^4} \, \fb^2(z).
\ee

Therefore, the equation \eqref{B:23} is linear w.r.t. $g(z)$ and can be solved by variation of constants in \eqref{11:32}. In Section~\ref{subsubsec::4Maxwell}, this was performed for a particular model. But here we have proven the possibility of solving for $g(z)$ for the whole family of models we have initially restricted ourselves to.

\section{On the Sign of $g'(z_h)$}\label{app::C}
We have established in the previous Appendix section that if a particular model allows for a 2nd order DE for $g(z)$ without a RHS, it is of the form
\be
    \sum^{D-2}_{\mu = 0} k_\mu \, \frac{G_{\mu\mu}}{g_{\mu\mu}} = 0, \quad k_\mu \in \mathbb{Z}, \ k_0 \ne 0,
\ee
and is solved by the following formula
\be
    g(z) = \fG(z) 
    \left[\displaystyle
    C_1  \int_{z_0}^{z} \frac{d\xi}{\displaystyle \fs_D(\xi) \fG(\xi)} + C_2
    \right], 
    \quad \fG(z)=\prod^{D-2}_{i=1}\fg^{-\frac{k_i}{k_0}}_{i}(z), 
    \ \  \fs_D(z) = \prod^{D-2}_{i=1}g^{1/2}_{x_ix_i}(z).
\ee

Since the metric function we usually choose are not zero at the horizon (or approach infinity), we can straightforwardly impose the boundary condition at the horizon
\be
    g(z_h) = 0
    \quad \Longrightarrow \quad 
    g(z) = C \cdot \fG(z)
    \displaystyle \int_{z}^{z_h} \frac{d\xi}{\displaystyle \fs_D(\xi) \fG(\xi)}, \label{C:3}
\ee
therefore the derivative is
\be
    g'(z) = C 
    \left[
    \fG'(z)
    \displaystyle \int_{z}^{z_h} \frac{d\xi}{\displaystyle \fs_D(\xi) \fG(\xi)}
    - \frac{1}{\fs_D(z)}
    \right], 
    \qquad 
    g'(z_h) = - \, \frac{C}{\fs_D(z_h)}. 
\ee

The metric functions $\fg_i$, $g_{x_ix_i}$ are usually chosen to be positive, so that
\be
    \forall z \in (0; z_h): 
    \qquad 
    \fG(z) > 0, \quad
    \displaystyle \int_{z}^{z_h} \frac{d\xi}{\displaystyle \fs_D(\xi) \fG(\xi)} > 0.
\ee

This means that if we successfully imposed the last boundary condition $g(0)=1$, which is not always possible, the parameter of integration we determined is positive, see \eqref{C:3}:
\be
    C > 0.
\ee

Therefore, given the assumptions we made, the derivative of the blackening function $g(z)$ is always negative at the horizon $z_h$:
\be
    g'(z_h) < 0.
\ee

For example, for Model I in Section~\ref{sec:model1} we have
\be
  \displaystyle 
  g(z) = e^{c_B z^2}
  \frac{\int^{z_h}_z e^{-\frac{3c_B}{2} \xi^2} \xi^{1 + \frac{2}{\nu}} \ d\xi}{\int^{z_h}_0 e^{-\frac{3c_B}{2} \xi^2} \xi^{1 + \frac{2}{\nu}} \ d\xi}, \quad 1 + \frac{1}{\nu} > 0,
\ee
so that the derivative is
\be
  \displaystyle 
  g'(z) = \frac{1}{\int^{z_h}_0 e^{-\frac{3c_B}{2} \xi^2} \xi^{1 + \frac{2}{\nu}} \ d\xi} 
  \left(
  2 c_B z \, e^{c_B z^2}
  \int^{z_h}_z e^{-\frac{3c_B}{2} \xi^2} \xi^{1 + \frac{2}{\nu}} \ d\xi
  - e^{-\frac{1}{2}c_B z^2} z^{1 + \frac{2}{\nu}} \right). \label{C:9}
\ee


\end{document}